\documentclass[a4paper,11pt]{article}
\pdfoutput=1

\usepackage{xcolor}
\usepackage{jcappub}

\newcommand{\rev}[1]{\textcolor{black}{#1}}

\AtBeginDocument{\let\citet\citep \let\citealt\citealp}

\newcommand{\Mpch}{\,h^{-1}{\rm Mpc}}
\newcommand{\Mnu}{M_\nu}
\newcommand{\Om}{\Omega_{\rm m}}
\newcommand{\Ob}{\Omega_{\rm b}}
\newcommand{\sig}{\sigma_8}
\newcommand{\zfour}{\zeta^{(4)}}                 
\newcommand{\zconn}{\zeta^{(4)}_{\rm conn}}      
\newcommand{\Ncov}{N_{\rm cov}}
\newcommand{\Nderiv}{N_{\rm deriv}}
\newcommand{\gramsci}{\textsc{Gramsci}}

\title{Climbing the $N$-point Ladder Part I: Information in the Higher-Order Configuration-Space Clustering of Dark Matter Halos}

\author[a]{Sumi Kim,}
\author[b,1]{Cristiano G. Sabiu,}
\author[a]{Inkyu Park}

\note[1]{Corresponding author.}

\affiliation[a]{Department of Physics, University of Seoul, 163 Seoulsiripdaero, Dongdaemun-gu, Seoul 02504, Republic of Korea}
\affiliation[b]{Natural Science Research Institute (NSRI), University of Seoul, Seoul 02504, Republic of Korea}

\emailAdd{sumikim1027@gmail.com}
\emailAdd{csabiu@gmail.com}
\emailAdd{icpark00@gmail.com}

\abstract{
The two-point correlation function completely describes a Gaussian random
field, but nonlinear gravitational growth, halo bias, and redshift-space
distortions drive the late-time halo field strongly non-Gaussian, moving a
substantial part of the cosmological information into higher-order
correlations. We quantify the information content of the configuration-space
two-, three-, and connected four-point correlation functions of Quijote
dark-matter haloes at $z=0$ and fixed number density. We build Fisher
forecasts for $\{\Omega_m, \Omega_b, h, n_s, \sigma_8, M_\nu\}$ in real and
redshift space from ${\sim}38{,}000$ GPU-accelerated $N$-point measurements.
Treating the statistics as a ladder,
$\mathrm{2PCF} \rightarrow +\mathrm{3PCF} \rightarrow
+\zeta^{(4)}_{\mathrm{conn}}$, we report the information gained at each rung.
The 3PCF supplies most of the accessible higher-order information: it
tightens every parameter, most strongly $\sigma_8$ and $M_\nu$, whose
degeneracy it partially breaks, by factors of $1.6$--$6.3$ over the
redshift-space $\{\xi_0,\xi_2\}$ baseline with per-parameter gains consistent with
those of the Fourier-space halo bispectrum on the same simulations. The
3PCF gains persist under conservative, modelling-motivated scale cuts:
restricted to the tree-level-validated regime with minimum triangle side
$\ge40\,h^{-1}{\rm Mpc}$, it still improves $M_\nu$ by $3.9\times$. The
connected 4PCF adds a further $\sim1.2$--$1.5\times$, an increment
insensitive to whether the quadrupole is included in the baseline but
traceable to small-scale configurations with sides
$\lesssim30\,h^{-1}{\rm Mpc}$. This rung-to-rung increment is
stable against derivative-sample noise and compression regularization,
whereas the absolute constraints remain limited by the finite simulation
ensembles and are reported as preliminary. We validate the measured 3PCF
against a tree-level perturbation-theory model, recovering a linear bias
consistent with the 2PCF. The configuration-space ladder thus offers an
independent and complementary route to the higher-order information probed
by the Fourier-space poly-spectra.
}

\begin{document}
\maketitle
\flushbottom


\section{Introduction}
\label{sec:intro}
The two-point correlation function, and its Fourier transform the power
spectrum, have long been the primary tools of large-scale-structure cosmology
\citep{Peebles1980}. They encode the complete statistical content of a Gaussian
random field, and the primordial density field is Gaussian to high precision.
However, nonlinear gravitational growth, galaxy and halo biasing, and
redshift-space distortions drive the late-time field strongly non-Gaussian,
depositing a large fraction of the cosmological information into higher-order
correlations that two-point statistics cannot reach
\citep{Fry1984, Bernardeau2002}. Extracting this information is becoming
increasingly important. Stage-IV surveys such as DESI \citep{DESI2016}, Euclid
\citep{Laureijs2011}, and SPHEREx \citep{Dore2014} are mapping the clustering of
tens of millions of objects, with a precision at which the higher-order sector
becomes measurable, and for parameters such as the summed neutrino mass
\citep{LesgourguesPastor2006} it may even be decisive.

The natural step beyond two points is the three-point function and its Fourier
counterpart, the bispectrum. The bispectrum breaks the degeneracy between linear
bias and the clustering amplitude \citep{Fry1993,Frieman1994,Sefusatti2006} and has been measured in
spectroscopic surveys \citep{GilMarin2017}. 
In configuration space the three-point function has a correspondingly
long observational history, from the hierarchical amplitudes of the APM
survey \citep{Gaztanaga1994} through measurements in the Las Campanas
\citep{Jing1998}, 2dF \citep{Jing2004}, and SDSS
\citep{Kayo2004, McBride2011} galaxy samples, together with dedicated
modelling of the galaxy 3PCF \citep{Marin2008}, to the detection of the
baryon-acoustic feature in luminous red galaxies \citep{Gaztanaga2009}, the
large-scale 3PCF of BOSS CMASS \citep{Slepian2017}, and joint two- and
three-point analyses at higher redshift with VIPERS
\citep{Moresco2017, Veropalumbo2021}, through a dedicated methodological
analysis of the halo 3PCF \citep{Veropalumbo2022}. Recently the first full-shape
joint analysis of the two- and three-point correlation functions
\citep{guidi2026fullshapejointanalysistwo},
with the corresponding
full-shape 2PCF$+$3PCF modelling is now being prepared for Euclid
\citep{Guidi2026}.

It has also been used to discriminate modified-gravity models \citep{Sabiu2016,Aviles2023} and interacting dark-energy cosmologies \citep{Moresco2014}, that
two-point statistics alone cannot separate.
Its constraining
power has been quantified extensively\citep{2019MNRAS.483.2078Y}, most recently with the Quijote simulation
suite \citep{VillaescusaNavarro2020}, where the halo bispectrum roughly doubles
the constraint on the summed neutrino mass relative to the power spectrum alone
\citep{Hahn2020, HahnVillaescusa2021}. A broad programme has used Quijote to
forecast many alternative non-Gaussian summaries, among them the marked power
spectrum \citep{Massara2021}, the wavelet scattering transform
\citep{Valogiannis2022}, and nearest-neighbour distributions
\citep{BanerjeeAbel2021}. Much of this effort traces back to HADES
\citep{VillaescusaNavarro2018}, a dedicated suite of massive-neutrino $N$-body
simulations and Quijote's direct predecessor from the same group. HADES is built
around the $\sig$--$\Mnu$ degeneracy: each massive-neutrino cosmology is paired
with a massless one whose $\sig$ is lowered to mimic the neutrino-induced
suppression of clustering, so that any statistic separating the two breaks the
degeneracy on an equal-amplitude footing. Prior work used HADES to show that a
range of higher-order and configuration-space statistics do exactly this,
including the redshift-space halo bispectrum \citep{Hahn2020}, the marked power
spectrum \citep{Massara2021}, and the void size function combined with halo and
matter clustering \citep{Bayer2021}. With few exceptions, all of these analyses are
performed in Fourier space.

Configuration space offers a complementary route to the same information,
with several practical conveniences of its own, athough we emphasize at
the outset that nothing in this work should be read as a claim of superior
constraining power over the Fourier-space analyses. The $N$-point correlation functions
are local in pair separation, so the baryon-acoustic feature and the scales of
interest are cleanly localised. Survey boundaries, masks, and selection effects
are handled transparently through the random catalogue. And although
perturbation theory is developed most naturally in Fourier space, its tree-level
predictions admit closed-form configuration-space counterparts through
separable radial transforms \citep{BarrigaGaztanaga2002,
SlepianEisenstein2015}, so measured $N$-point functions can be confronted
with analytic templates natively in separation space, without an
intermediate window or mapping step, as we do for the 3PCF in
\S\ref{sec:theory}. The main obstacle has
been computational. Naively, an $N$-point count scales as
$\mathcal{O}(N_{\rm g}^{\,n})$ in the number of objects. However, a number of
algorithmic advances have got around this: the edge-corrected estimators of
\citet{SzapudiSzalay1998}, the $\mathcal{O}(N_{\rm g}^{2})$ multipole
decomposition of the 3PCF \citep{SlepianEisenstein2015, SlepianEisenstein2018},
its four-point generalisation \citep{Philcox2022encore}, and graph- and
FFT-based estimators \citep{PhilcoxEisenstein2020, Sabiu2019}. Together these
have made the 3PCF \citep{GaztanagaScoccimarro2005, Slepian2017BAO} and, more
recently, the 4PCF measurable on survey-scale catalogues.

The four-point function itself is not new: it was first measured in the
angular catalogues of the 1970s and 1980s, where the connected term was
detected and its hierarchical amplitude estimated
\citep{Fry1978, Fry1983}, and its perturbative structure was worked out
alongside that of the lower orders \citep{Fry1984}. What kept it from
becoming a standard tool was cost and noise: reliable four-point
measurements demand both large, dense samples and estimators faster than the
naive $\mathcal{O}(N_{\rm g}^{4})$ count, neither of which was available
until recently. In Fourier space its counterpart, the trispectrum, has
meanwhile been developed analytically \citep{Bertolini2016} and shown in
joint power-spectrum--bispectrum--trispectrum forecasts to carry
non-negligible additional information \citep{Gualdi2021}.

Recently, the four-point function has attracted most attention as a probe of parity
violation, for which the parity-odd part of the 4PCF is a uniquely clean
signature \citep{Philcox2022parity, Hou2023parity, CahnSlepianHou2023}. Its role
as a carrier of standard, parity-even cosmological information has been
comparatively little explored, especially in configuration space, and this is
the gap we address here. Fourth order is also the first at which a qualitatively
new feature appears. A connected statistic separates cleanly into two parts. The
{\em disconnected} part is built from products of two-point functions; it is
Gaussian information already contained in the 2PCF. The {\em connected} part
encodes genuinely new four-body correlations [Eq.~\eqref{eq:disc}]. The
three-point function, by contrast, is connected by construction, since the
disconnected three-point contribution vanishes for a zero-mean field. Isolating
the connected 4PCF is therefore essential for a clean accounting of the
{\em incremental} information beyond the two- and three-point functions.

In this work we measure and forecast the cosmological information content of the
configuration-space halo two-, three-, and connected four-point correlation
functions, treated as a ladder,
$2{\rm PCF}\rightarrow{+}3{\rm PCF}\rightarrow{+}\zconn$. We use the Quijote
simulations at a single redshift and a fixed halo number density, matched across
cosmologies so that the response isolates clustering rather than abundance, and we
construct a Fisher forecast for $\{\Om,\Ob,h,n_{\rm s},\sig,\Mnu\}$ in both real
and redshift space. The measurements use a new GPU build of the \gramsci\
graph-database estimator (\citealt{gramsci_gpu}; see also \citealt{Sabiu2019}),
which computes the full-configuration 3PCF and connected 4PCF quickly enough to
process the thousands of simulations that a converged covariance and derivative
ensemble require. We then ground the measured 3PCF in tree-level perturbation
theory, and quantify, parameter by parameter, how much each rung of the ladder
adds. We pay particular attention to $\sig$ and $\Mnu$, whose degeneracy the
higher-order statistics partially break.

The outline of this paper is as follows. In \S\ref{sec:ladder} we define the
$N$-point ladder and the connected/disconnected decomposition. In
\S\ref{sec:data} we describe the simulations and measurements, and in
\S\ref{sec:fisher} the Fisher methodology. In \S\ref{sec:theory} we validate the
measured 3PCF against tree-level theory. We then present the measured statistics
and their cosmology response in \S\ref{sec:response}, and the information ladder
in \S\ref{sec:results}, with robustness tests in \S\ref{sec:robust}. We discuss
the implications in \S\ref{sec:discussion} and conclude in
\S\ref{sec:conclusions}.

\section{The N-point ladder and the connected four-point function}
\label{sec:ladder}
For a continuous overdensity field $\delta(\mathbf{x}) = \rho(\mathbf{x})/\bar\rho - 1$,
the $N$-point correlation functions are the connected moments of $\delta$
evaluated at $N$ distinct points. Statistical homogeneity and isotropy mean that
the two- and three-point functions depend only on the pair separations,
\begin{equation}
\xi(r_{12}) = \langle \delta_1 \delta_2 \rangle, \qquad
\zeta(r_{12}, r_{13}, r_{23}) = \langle \delta_1 \delta_2 \delta_3 \rangle ,
\end{equation}
where $\zeta$ is a function of the full triangle. At fourth order the moment
$\langle\delta_1\delta_2\delta_3\delta_4\rangle$ is not itself connected, and
instead decomposes as
\begin{equation}
\langle\delta_1\delta_2\delta_3\delta_4\rangle
 = \zfour_{\rm conn} + \big[\xi_{12}\xi_{34}+\xi_{13}\xi_{24}+\xi_{14}\xi_{23}\big]
\label{eq:disc}
\end{equation}
into the connected four-point function $\zfour_{\rm conn}$ and a
{\em disconnected} part built from products of two-point functions. The
disconnected term is fixed entirely by $\xi$. It is the contribution that a
Gaussian field would produce, and it carries no information beyond the 2PCF,
whereas $\zfour_{\rm conn}$ encodes genuinely four-body, non-Gaussian
correlations. No such disconnected piece appears at third order: for a zero-mean
field the would-be disconnected three-point terms are each proportional to
$\langle\delta\rangle = 0$, so $\zeta$ is connected by construction. Fourth order
is therefore the lowest order at which a statistic splits explicitly into a
Gaussian part, already contained in the 2PCF, and a genuinely new non-Gaussian
part. This is what makes the connected 4PCF the natural quantity to use when we
count, {\em incrementally}, the information added at each rung of the ladder
$\xi \rightarrow \zeta \rightarrow \zfour_{\rm conn}$.

We estimate each correlation function with the minimum-variance, edge-corrected
estimators of \citet{SzapudiSzalay1998}. These combine the data ($D$) and a
uniform random ($R$) catalogue through the difference field $D-R$: data and
random points are given signed weights ($+1/N_D$ and $-1/N_R$, respectively),
and the $N$-point function is the binned sum of weight products over $N$-tuples,
normalised by the all-random count. For $N=2$ this reduces to the
Landy--Szalay estimator, and the signed-weight construction targets the
connected moment $\langle\delta^N\rangle$ directly \citep{SlepianEisenstein2015}.
We bin each statistic in its independent pair separations: the 2PCF in $r$, the
3PCF in the three triangle sides $(r_{12},r_{13},r_{23})$, and the 4PCF in the
six pairwise separations of the tetrahedron. Configurations that violate the
triangle (or tetrahedron) inequality are discarded. We obtain the connected 4PCF
by subtracting the disconnected expectation of Eq.~\eqref{eq:disc}, which is
evaluated internally from the measured 2PCF in the same binning, so that
$\zfour_{\rm conn}$ is returned directly for every configuration. In the halo
field the disconnected term dominates the raw four-point count in the large
majority of configurations, so this subtraction is essential. The connected
signal is the comparatively small, but genuinely new, residual.

\section{Simulations and measurements}
\label{sec:data}
\subsection{The Quijote halo catalogues}
We use the Quijote suite of $N$-body simulations \citep{VillaescusaNavarro2020},
each of which evolves the matter field in a periodic box of side
$1\,h^{-1}{\rm Gpc}$, and we work with the friends-of-friends halo catalogues at
$z=0$. The fiducial cosmology is set as $\{\Omega_m=0.3175, \Omega_b=0.049, h=0.6711, n_s=0.9624, \sigma_8=0.834, M_{\nu} =0.0~\rm{eV}\}$. The Quijote suite also includes 500 simulations for each variation of the cosmological parameters $\{\Omega_m^{\pm}:\Delta=\pm 0.01, \Omega_b^{\pm}:\Delta=\pm 0.002, h^{\pm}:\Delta=\pm 0.02, n_s^{\pm}:\Delta=\pm 0.02, \sigma_8^{\pm}:\Delta=\pm 0.015\}$. Due to the constraint $M_{\nu}\ge0.0$, the variations in neutrino mass are all positive and include $M_{\nu}=0.1, 0.2, 0.4$ eV.

We estimate the covariance matrix from $\Ncov=5,000$ realizations of the
fiducial cosmology~ in real space and in redshift space. For the derivatives we use the simulations in which a single
parameter is displaced from its fiducial value, $\Nderiv=500$ realizations per
parameter for $\{\Om,\Ob,h,n_{\rm s},\sig\}$, together with the massive-neutrino
simulations described in \S\ref{sec:fisher}.

We want the parameter response to reflect the {\em clustering} of haloes rather
than their {\em abundance}. We therefore impose a fixed comoving number density
$\bar n = 1.5\times10^{-4}\,h^3{\rm Mpc}^{-3}$, selecting the $150,000$ most massive haloes in
every box and applying the same selection across all cosmologies. Fixing the
count removes the halo-abundance (mass-function) response by construction, so
the Fisher information reflects clustering alone. The Quijote bispectrum
analyses reach a comparable fiducial number density with a fixed halo mass cut
($M>3.2\times10^{13}\,h^{-1}M_\odot$), and instead marginalize over the mass
limit to suppress the residual abundance dependence \citep{Hahn2020}. The
fixed-count selection we adopt here achieves the same end directly, as in the
nearest-neighbour analysis of \citet{BanerjeeAbel2021}. The random
catalogues are drawn uniformly within the periodic box.

\subsection{Redshift-space distortions}
\label{sec:rsd}
We construct redshift-space catalogues in the plane-parallel approximation,
displacing each halo along a chosen line of sight $\hat{\mathbf n}$ by its
peculiar velocity,
\begin{equation}
\mathbf{s} = \mathbf{x} + \frac{(1+z)\,v_\parallel}{H(z)}\,\hat{\mathbf n},
\label{eq:rsd}
\end{equation}
where $H(z)$ is evaluated in the cosmology of each simulation and the box is
periodically re-wrapped. We measure the redshift-space monopole of every
statistic. For the two-point function we additionally measure the anisotropic
correlation function $\xi(\sigma,\pi)$, on a grid of $2\Mpch$ cells in the
transverse ($\sigma$) and line-of-sight ($\pi$) separations out to $40\Mpch$,
using the same fixed-density halo sample and an analytic random count in the
periodic box. From this grid we form the quadrupole $\xi_2(s)$ by
bin-integrated Legendre weighting of $\xi(s,\mu)$, in seven radial bins of
width $4\Mpch$ spanning $10$--$38\Mpch$; the multipole is exact only where the
full $\mu$ arc lies inside the measured grid, which restricts the quadrupole to
$s\le40\Mpch$, the regime that dominates its signal-to-noise. This lets us
build the redshift-space Fisher baseline from the monopole and
quadrupole jointly, $\{\xi_0,\xi_2\}$, so that the leading anisotropic
(Kaiser/growth) information enters the two-point baseline rather than being
credited to the higher-order statistics (\S\ref{sec:res-2pcf}). To reduce the
noise in the derivative estimates we repeat the $N$-point measurements
along all three Cartesian axes.

\subsection{$N$-point measurements}
We measure the correlation functions with a new GPU build of the \gramsci\
graph-database estimator \citep{gramsci_gpu}; the underlying algorithm and its
validation are presented in \citet{Sabiu2019}. The GPU
kernels used here are validated at the count level in \citet{gramsci_gpu}:
bin-by-bin agreement with the independent CPU implementation for every
kernel and mode, and regression tests on constructed catalogues with known
analytic counts, including the connected four-point channel. For this
analysis the 2PCF is additionally cross-checked against an independent
pair-count implementation (\S\ref{sec:rsd}) and the 3PCF against tree-level theory (\S\ref{sec:theory}). No
independent cross-code comparison exists for the exact {\em binned}
3PCF/connected-4PCF output. Published estimators work in multipole bases
that are not comparable bin for bin, so these count-level and
cross-implementation tests define the validation scope; the
disconnected-piece subtraction is fixed algebraically by the measured 2PCF
[Eq.~\eqref{eq:disc}].
Once per catalogue, the estimator
builds a graph whose edges join all point pairs within the maximum separation,
and it then evaluates the $N$-point counts by enumerating sub-graphs directly
from this structure. The GPU implementation makes the full-configuration 3PCF and
4PCF fast enough to process the $\sim 38,000$ catalogues that the covariance and
derivative ensembles require. We bin the 2PCF in $20$ linear bins from $10$ to
$150\Mpch$, the 3PCF with each triangle side in $18$ bins out to $100\Mpch$, and
the connected 4PCF with each of the six tetrahedron separations in $5$ bins out
to $65\Mpch$ (all monopole, $n_\mu=1$). After the triangle- and
tetrahedron-inequality cuts, these give data-vector dimensions
$N_d^{\rm 2pcf}=20$, $N_d^{\rm 3pcf}=832$, and $N_d^{\rm 4pcf}=900$. A further $268$ tetrahedron bins are degenerate---their random count
vanishes ($RRRR=0$)---and are excluded from the analysis
(\S\ref{sec:rob-moped}), leaving $N_d^{\rm 4pcf}=632$ and a total ladder data
vector of $N_{\rm d}=20+832+632=1484$. In redshift space the two-point
block is augmented by the quadrupole $\xi_2$ in seven bins
(\S\ref{sec:rsd}), giving $N_d^{\rm 2pcf}=27$ and a total redshift-space
ladder data vector of $N_{\rm d}=1491$.

We carried out the full set of measurements on a single NVIDIA RTX~3090\,Ti.
Measuring the 2-, 3-, and connected 4-point functions of one halo catalogue
along one line of sight takes just $\sim 35$~s of wall-clock time, dominated
by the four-point graph query ($\sim 8.5$~s, against $\sim 2$~s for the
two-point function). The forecast presented here rests on $\sim 38,000]$
such measurements: the $\Ncov$ fiducial realizations and the
parameter-derivative simulations, in real and redshift space. These total
$\sim 390$ GPU-hours, with the four-point function dominating the cost. The
downstream Fisher, compression, and figure analysis is negligible by comparison.

\section{Fisher methodology}
\label{sec:fisher}
\begin{equation}
F_{ij} = \sum_{a,b}\frac{\partial \mu_a}{\partial\theta_i}\,
        \big[C^{-1}\big]_{ab}\,\frac{\partial \mu_b}{\partial\theta_j}.
\label{eq:fisher}
\end{equation}
We forecast the constraining power of each data vector with the Fisher
information matrix of Eq.~\eqref{eq:fisher}, in which $\boldsymbol\mu(\boldsymbol\theta)$
is the mean data vector, $\boldsymbol\theta = \{\Om,\Ob,h,n_{\rm s},\sig,\Mnu\}$,
and $C$ its covariance. The marginalised $1\sigma$ error on $\theta_i$ is then
$[(F^{-1})_{ii}]^{1/2}$. As is standard for this class of forecast, we assume a
Gaussian likelihood with a parameter-independent covariance.

\subsection{Covariance}
We estimate $C$ from the $\Ncov$ fiducial realizations. The inverse of a sample
covariance is biased, so we debias it with the factor of \citet{Hartlap2007},
$(\Ncov - N_d - 2)/(\Ncov - 1)$, which requires $\Ncov > N_d + 2$ for the inverse
to exist at all. We propagate the residual noise in the estimated covariance into
the parameter errors following \citet{DodelsonSchneider2013} and
\citet{Percival2014}. The connected-4PCF data vector is large, so for a given $\Ncov$ it is whether
the covariance can be inverted, not the signal itself, that limits how far up
the ladder we can go. We quote the covariance size used at each rung.

\subsection{Derivatives}
We compute the derivatives $\partial\boldsymbol\mu/\partial\theta_i$ from the
Quijote derivative simulations. For $\{\Om, \Ob, h, n_{\rm s}, \sig\}$ we use a
central finite difference between the catalogues in which the parameter is
stepped above and below its fiducial value, at fixed initial-condition phases to
suppress sample variance. The summed neutrino mass needs special treatment,
because $\Mnu = 0$ is a physical boundary and the massive-neutrino simulations
are generated from Zel'dovich rather than second-order Lagrangian initial
conditions. Following \citet{Hahn2020} we therefore use the one-sided,
higher-order estimate
\begin{equation}
\frac{\partial\boldsymbol\mu}{\partial\Mnu}
 = \frac{-21\,\boldsymbol\mu_{\rm ZA} + 32\,\boldsymbol\mu_{0.1}
   - 12\,\boldsymbol\mu_{0.2} + \boldsymbol\mu_{0.4}}{1.2~{\rm eV}},
\label{eq:mnu}
\end{equation}
where $\boldsymbol\mu_{\rm ZA}$ is measured in the zero-mass Zel'dovich fiducial
and $\boldsymbol\mu_{0.1}$, $\boldsymbol\mu_{0.2}$, $\boldsymbol\mu_{0.4}$ in
simulations with $\Mnu = 0.1, 0.2, 0.4~{\rm eV}$.

\subsection{Finite-sampling corrections}
A Fisher matrix built from a finite number of derivative simulations is itself a
noisy estimate, and this noise biases the forecast errors
\citep{CoultonWandelt2023}. We mitigate it in two ways. First, in redshift space
we measure every derivative simulation along all three Cartesian lines of sight
and average them, which triples the effective number of derivative realizations
entering $\partial\boldsymbol\mu/\partial\theta_i$. Second, we confirm that the
marginalised errors have converged by recomputing them as the number of
derivative simulations is increased (\S\ref{sec:robust}).

\section{Validation of the measured 3PCF against tree-level theory}
\label{sec:theory}
Before we use the higher-order statistics for a forecast, we verify that the
measured halo 3PCF is described by perturbation theory on the scales that enter
the analysis. We compare the fiducial-cosmology measurement, in real and
redshift space, to a tree-level configuration-space model, whose ingredients we summarise here.

\subsection{Tree-level model}
At leading order in standard Eulerian perturbation theory, the biased galaxy
bispectrum is
\begin{equation}
B_g(\mathbf{k}_1,\mathbf{k}_2,\mathbf{k}_3)
 = 2\,K(\mathbf{k}_1,\mathbf{k}_2)\,P(k_1)\,P(k_2) + \text{2 cyc.},
\label{eq:btree}
\end{equation}
with $\mathbf{k}_1+\mathbf{k}_2+\mathbf{k}_3=0$, the linear power spectrum $P$,
and the vertex kernel
\begin{equation}
K(\mathbf{k}_1,\mathbf{k}_2) = b_1^3\,F_2(\mathbf{k}_1,\mathbf{k}_2)
 + \tfrac{1}{2}\,b_1^2 b_2 + \tfrac{1}{2}\,b_1^2 b_{s2}\,S_2(\mu_{12}),
\label{eq:kkernel}
\end{equation}
where $F_2$ is the symmetric second-order density kernel,
$S_2(\mu_{12})=\mu_{12}^2-\tfrac13$ the tidal kernel,
$\mu_{12}=\hat{\mathbf{k}}_1\!\cdot\!\hat{\mathbf{k}}_2$, and $(b_1,b_2,b_{s2})$
the linear, quadratic, and tidal Eulerian galaxy bias
\citep{Fry1984, Goroff1986, Sefusatti2006, McDonaldRoy2009, DesjacquesJeongSchmidt2018}.
The two power-spectrum legs each carry a factor $b_1$, which gives the $b_1^3$
scaling of the $F_2$ term. We slave the tidal bias to its local-Lagrangian
(coevolution) value $b_{s2}=-\tfrac47(b_1-1)$
\citep{Baldauf2015, DesjacquesJeongSchmidt2018}, leaving $(b_1,b_2)$ free.

Since $F_2$ and $S_2$ are at most quadratic in $\mu_{12}$, each cyclic term of
Eq.~\eqref{eq:btree} Fourier-transforms into a separable sum of one-dimensional
radial transforms \citep{SlepianEisenstein2015, SlepianEisenstein2016}. The 
config\-uration-space (pre-cyclic) 3PCF at the vertex where the two legs meet, with
$\mu$ the cosine of the opening angle, then has the closed form
\citep{BarrigaGaztanaga2002, GaztanagaScoccimarro2005}
\begin{align}
\zeta_{\rm pc}(r_1,r_2,\mu) &=
 2\big(\tfrac{17}{21}b_1^3 + \tfrac12 b_1^2 b_2\big)\,\xi(r_1)\,\xi(r_2)
 + b_1^3\big[\xi'(r_1)\Phi'(r_2)+\Phi'(r_1)\xi'(r_2)\big]\,\mu \nonumber\\
 &\quad + 2\big(\tfrac{4}{21}b_1^3 + \tfrac13 b_1^2 b_{s2}\big)
   \big[\bar\xi(r_1)-\xi(r_1)\big]\big[\bar\xi(r_2)-\xi(r_2)\big]\,P_2(\mu),
\label{eq:zpc}
\end{align}
where $\xi$ is the linear correlation function, $\xi'=\mathrm{d}\xi/\mathrm{d}r$,
$\bar\xi(r)=3r^{-3}\!\int_0^r\xi(s)\,s^2\,\mathrm{d}s$, and $\Phi'(r)=r\bar\xi(r)/3$;
the full 3PCF is the sum of Eq.~\eqref{eq:zpc} over the triangle's three vertices.
We compute the linear spectrum with \textsc{camb} \citep{Lewis2000} at the
Quijote fiducial cosmology, and infrared-resum it to capture the damping of the
acoustic feature,
\begin{equation}
P_{\rm IR}(k) = P_{\rm nw}(k) + e^{-k^2\Sigma^2}\,P_{\rm w}(k),
\label{eq:ir}
\end{equation}
with the wiggle/no-wiggle split $P=P_{\rm nw}+P_{\rm w}$ obtained by the
discrete-sine-transform method \citep{Hamann2010, EisensteinHu1998} and the
leading-order displacement dispersion $\Sigma^2$ \citep{EisensteinSeoWhite2007, Blas2016}.
In redshift space we use the tree-level Scoccimarro--Couchman--Frieman kernels,
with linear factor $Z_1(\mathbf{k})=b_1+f\mu_k^2$ ($f$ the growth rate and
$\mu_k$ the line-of-sight cosine) and the associated second-order kernel $Z_2$,
and we take the exact orientation-averaged monopole to match the angle-averaged
estimator \citep{Kaiser1987, ScoccimarroCF1999}.

\subsection{Comparison with the Quijote 3PCF}
We compare Eq.~\eqref{eq:zpc}, and its redshift-space counterpart, with the mean
fiducial-cosmology 3PCF measured from the same catalogues used for the
covariance. To match the estimator, we bin-average the model within each
measured $(r_1,r_2,r_3)$ bin over a sub-grid weighted by the closed-triangle
volume measure $(r_1 r_2 r_3)^2$, which reproduces the triplet-abundance
weighting carried by the random ($\mathrm{RRR}$) normalisation
\citep{SzapudiSzalay1998}. Since $\zeta$ is a polynomial in the bias parameters,
we assemble the bin-averaged template once per bias monomial and evaluate it at
arbitrary $(b_1,b_2)$ at negligible cost. We fit $(b_1,b_2)$ by minimising a
$\chi^2$ against the measured mean, restricting to triangles whose shortest side
exceeds $r_{\rm min}^{\rm fit}=40\Mpch$, beyond which the tree-level model is
expected to hold.

\begin{figure}
\centering
\includegraphics[width=0.7\textwidth]{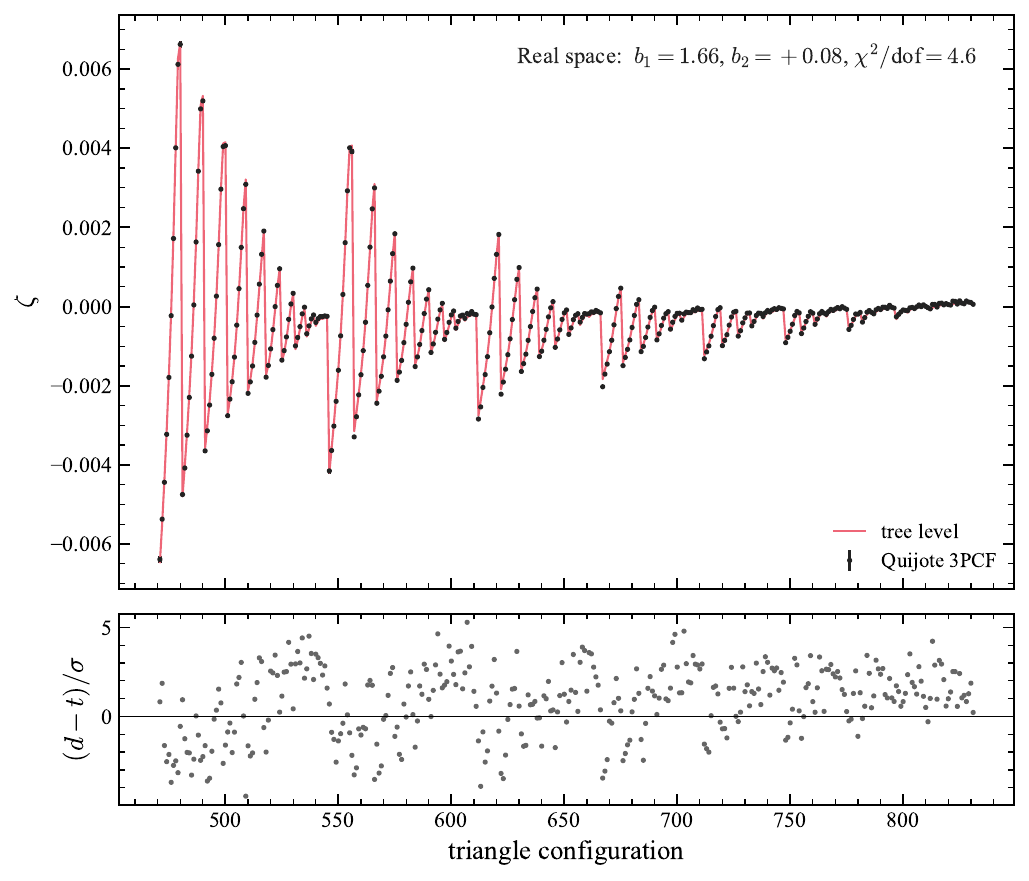}
\includegraphics[width=0.7\textwidth]{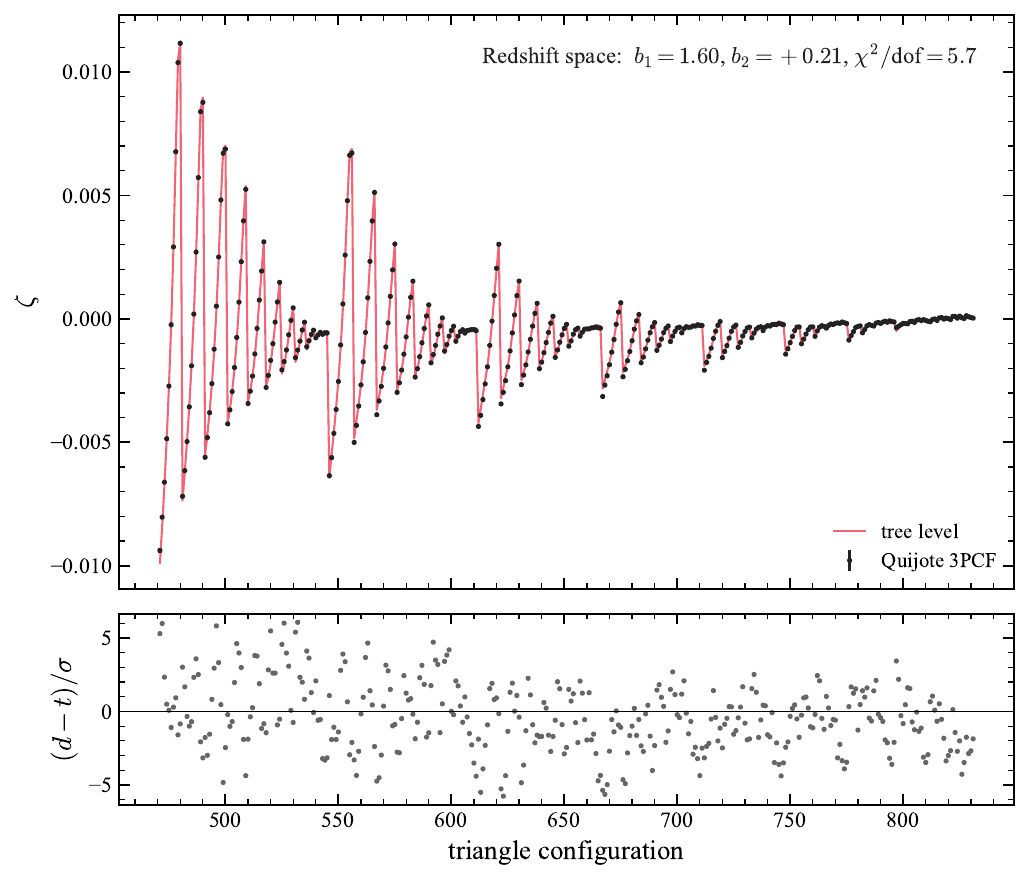}
\caption{Measured Quijote 3PCF (points) versus the tree-level
perturbation-theory model (line), with the $(d-t)/\sigma$ residual panel, in real(top panel)
and redshift(bottom panel) space.}
\label{fig:combined_realrsd}
\end{figure}

The model reproduces the measured 3PCF over the fitted range (Fig.~\ref{fig:combined_realrsd}).
In real space we find $\chi^2/{\rm dof}=4.6$ with $(b_1,b_2)=(1.66,0.08)$, and the
recovered linear bias agrees at the five per cent level with that inferred independently from the
large-scale 2PCF through $\xi_{\rm halo}\simeq b_1^2\,\xi_{\rm lin}$. The
redshift-space monopole is reproduced just as well ($\chi^2/{\rm dof}=5.7$,
$b_1=1.60$, $b_2=0.21$), with the growth rate held at its fiducial value;
its bias sits eight per cent below the two-point reference.

The two-point reference bias is obtained as follows. We fit the mean
fiducial 2PCF to the same linear prediction that enters the
3PCF model, $\xi_{\rm h}(r)=b_1^2\,\bar\xi_{\rm lin}(r)$ in real space and
the linear Kaiser monopole
$\xi_0(s)=(b_1^2+\tfrac{2}{3}b_1 f+\tfrac{1}{5}f^2)\,\bar\xi_{\rm lin}(s)$
in redshift space, growth rate fixed at its fiducial value $f=0.529$, with
the theory bin-averaged over each measured bin and a  covariance matrix built from the fiducial realizations. Restricting to bins lying entirely within
$45$--$101\Mpch$, gives $b_1^{\rm 2PCF}=1.74$ in both real and redshift space. The
formal statistical error of $\pm0.001$ is negligible; the meaningful
uncertainty is the fit-range systematic, with $b_1^{\rm 2PCF}$ varying
between $1.70$ and $1.75$ as the window is moved across
$31$--$150\Mpch$. The 3PCF values, $1.66$ (real space) and $1.60$ (redshift space) 
therefore agree with the two-point reference to within
$5$--$8\%$, which is the level of accuracy expected of a tree-level bias
model on these scales. 
For external context, we compute a standard peak-background-split
expectation: the \citet{Tinker2008} mass function with the
\citet{Tinker2010} bias, number-weighted over a fixed-count sample of
density $\bar n=1.5\times10^{-4}\,h^3{\rm Mpc}^{-3}$ in the fiducial
cosmology. The \citet{Tinker2010} calibration is consistent with the
precision separate-universe measurements of \citet{Lazeyras2016}. This
expectation gives $\bar b_1=1.81$ and implies a threshold mass of
$3.3\times10^{13}\,h^{-1}M_\odot$, consistent with the fixed mass cut of
\citet{Hahn2020} at comparable density. The ${\sim}4\%$ offset from our
measured $1.74$ is within the friends-of-friends versus
spherical-overdensity mass-definition ambiguity.
All three routes thus
bracket $b_1\simeq1.6$--$1.8$ for this sample, and the 3PCF fit lands
inside that bracket.

The reduced $\chi^2$ exceeds unity not because the model fails but because
the fit is to the mean of $5,000$ realizations: its sub-percent statistical error resolves model residuals far below any level that matters for the forecast, so the $\chi^2$ is sensitive to sub-percent departures that a tree-level template cannot be expected to capture \citep{GilMarin2012, Guidi2023}. The physically meaningful validation is therefore the agreement of the recovered bias with the independent 2PCF value, not a reduced $\chi^2$  of order unity.

This agreement shows that the measured 3PCF is physical and can be modelled
analytically on the scales used here, and it motivates the conservative
large-scale cut we adopt when interpreting the information content of the
higher-order statistics (\S\ref{sec:res-cuts}): the same
$r_{\rm min}=40\Mpch$ that bounds the validated regime of this fit defines
the modelling-informed \rev{higher-order selection} under which we re-derive the
information ladder \rev{(the cut restricts the three- and four-point blocks;
the two-point baseline is unchanged)}.

We note in here what this test does and does not establish. It establishes that the measured 3PCF is a physical clustering signal,
described at tree level for triangle sides $\ge40\Mpch$ with a linear bias
consistent with the two-point value. It does not establish that tree-level theory describes
the smaller-scale configurations, down to $10\Mpch$, that enter the
headline signal-to-noise-oriented Fisher vector; those configurations are
included as measurement-defined information, in the standard spirit of a
simulation-based forecast, not because a perturbative model covers them.
The information ladder re-derived under the modelling-informed cut
$r_{\rm min}=40\Mpch$ (\S\ref{sec:res-cuts}) quantifies precisely how much
of each gain survives when the analysis is restricted to the regime this
test validates.

\section{Measured statistics and their cosmology response}
\label{sec:response}
Before we turn to the Fisher forecast, we examine the measured statistics and
their response to the cosmological parameters. Because the configuration-space
correlation functions are local in separation, this also tells us {\em where} the
sensitivity lies, that is, at which scales and triangle or tetrahedron shapes.
This is the physical reason for the information gains of \S\ref{sec:results}.

\begin{figure*}
\centering
\includegraphics[width=0.78\textwidth]{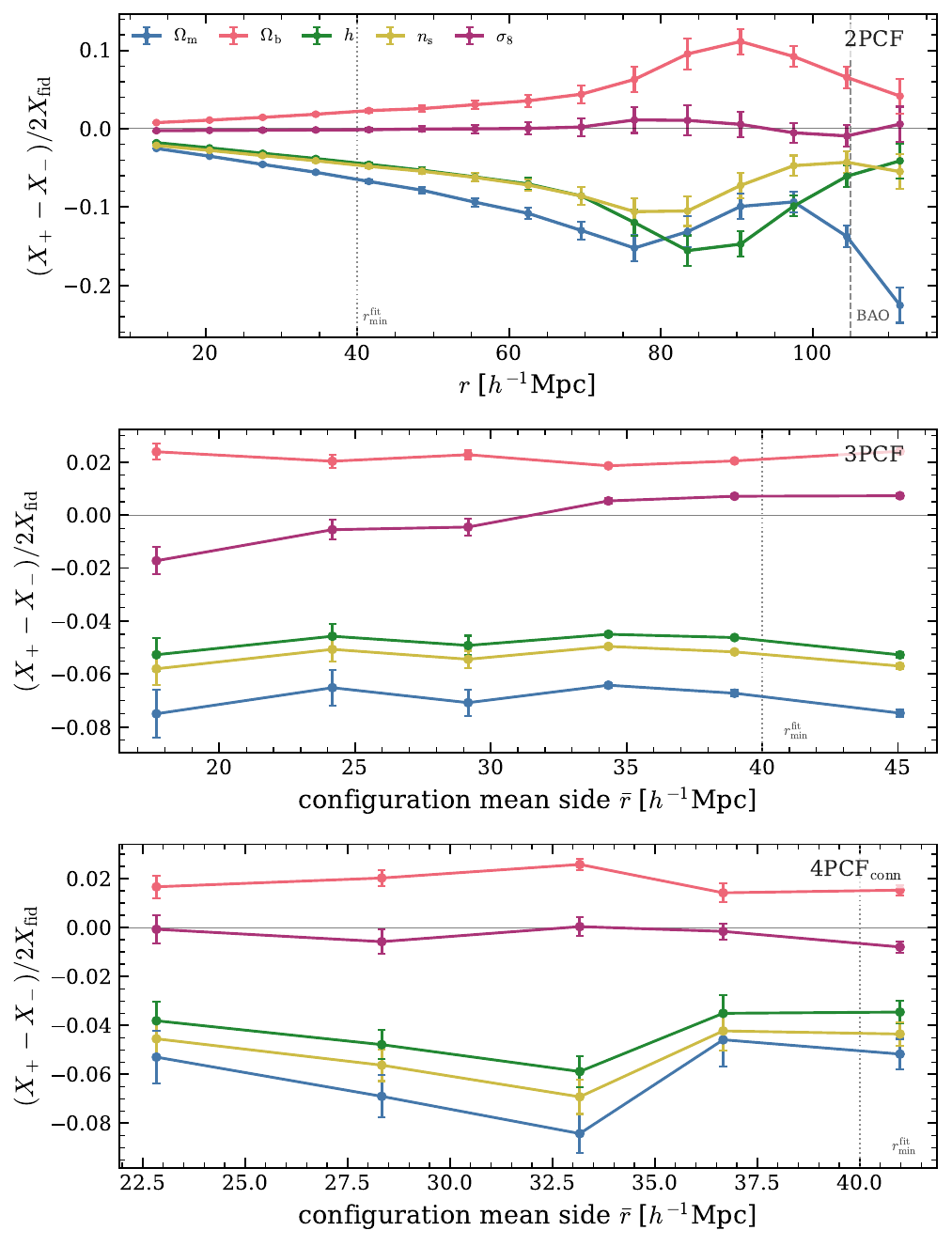}
\caption{Fractional response $(\mu_{+}-\mu_{-})/2\mu_{\rm fid}$ of the 2PCF,
3PCF, and connected 4PCF to the parameters $\{\Om,\Ob,h,n_{\rm s},\sig\}$ in real
space---the quantity the Fisher derivatives are built from. The 2PCF is
shown versus separation; for the 3PCF and connected 4PCF we display the $60$ and
$50$ highest signal-to-noise configurations respectively, ranked by the fiducial
$|\mathrm{mean}|/\sigma$ 
and binned by mean side length, with error bars giving the standard error within
each bin---combining configuration-shape dependence and measurement noise.
Dotted lines mark the tree-level-validated cut
$r^{\rm fit}_{\rm min}=40\Mpch$ (\S\ref{sec:theory},
\S\ref{sec:res-cuts}); the dashed line in the 2PCF panel marks the acoustic
scale. The 2PCF panel stops below the zero crossing of $\xi_{\rm fid}$
($r\simeq122\Mpch$), where the fractional normalization diverges; the
Fisher analysis uses the absolute derivatives, so no such amplification
enters the forecast. The
neutrino-mass response uses the separate four-point derivative scheme and is not
shown here.}
\label{fig:response}
\end{figure*}

Figure~\ref{fig:response} shows the fractional response of each statistic to each
parameter, $(\mu_{+}-\mu_{-})/2\mu_{\rm fid}$, which is the quantity the Fisher
derivatives are built from. Two features stand out. First, the two-point
function responds strongly to the shape parameters $\Om$ and $n_{\rm s}$, which
move the matter--radiation-equality scale and the tilt of the linear spectrum,
but only weakly to $\sig$. This is the configuration-space signature of the
fixed-number-density selection: when $\sig$ is raised, the halo bias of the
fixed-$\bar n$ sample falls, because the same number density then corresponds to
a lower peak height, and the two effects on the clustering amplitude
$b^2\sig^2$ nearly cancel. The 2PCF therefore carries little independent
information on $\sig$, and, through the $\sig$--$\Mnu$ degeneracy, on $\Mnu$.
Second, the three- and four-point responses are not simply rescaled copies of the
two-point response. Their dependence on triangle and tetrahedron {\em shape} is
distinct. This configuration dependence is absent from the 2PCF by construction,
and it restores sensitivity to the amplitude directions and breaks the
bias--amplitude degeneracy.

Three features of Fig.~\ref{fig:response} deserve explicit physical
identification. The localized structure in the 2PCF $\Om$ and $h$ responses
over ${\sim}80$--$110\Mpch$ is acoustic: these parameters shift the sound
horizon, so their derivatives develop the characteristic
differential-of-a-peak shape around the BAO feature (dashed marker); the
growth of the fractional response toward the right-hand edge of the panel
is instead a normalization effect, since $\xi_{\rm fid}$ crosses zero at
$r\simeq122\Mpch$ and the panel is truncated below it. Second, the 3PCF
response to $\sig$ changes sign at a mean side of ${\sim}25$--$30\Mpch$, negative below and positive above. This is the three-point fingerprint of
the fixed-$\bar n$ bias compensation: the vanishing 2PCF response pins
$b_1\sig \simeq$ const on linear scales, and at tree level
$\zeta \propto b_1^3\sig^4 \propto \sig^{+1}$, a weak positive
large-scale response, whereas on
small scales the steeper and opposite-signed response of the quadratic
bias of the fixed-count sample, together with beyond-tree-level
contributions, overcomes it and drives the response negative. The
higher-order statistics thus escape the cancellation that nulls the
two-point amplitude response, which is precisely why they carry the
amplitude information.

\begin{figure}
\centering
\includegraphics[width=\columnwidth]{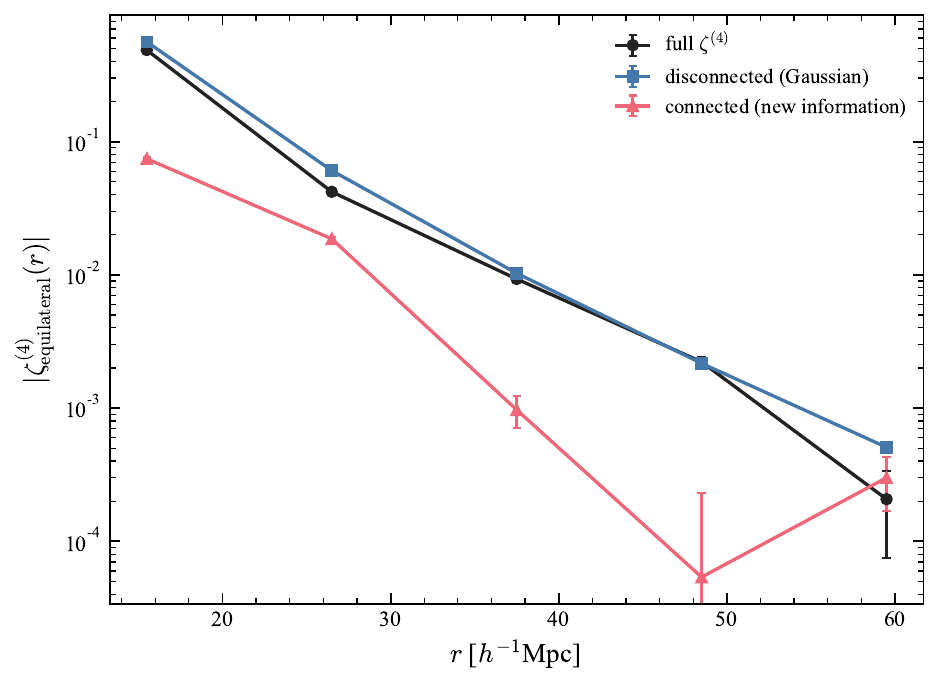}
\caption{Equilateral connected ($\zconn$) and disconnected
($\zeta^{(4)}_{\rm disc}$, Gaussian) four-point functions of the fiducial halo
field versus separation: the disconnected piece dominates the amplitude, while
the connected residual $\zconn$ isolates the comparatively small but genuinely
non-Gaussian four-body information.}
\label{fig:conndisc}
\end{figure}

The connected four-point function is the genuinely non-Gaussian part of the
four-point signal, and Figure~\ref{fig:conndisc} shows why isolating it matters.
Across separations the disconnected piece dominates the four-point amplitude; it
is fixed entirely by the measured 2PCF (Eq.~\ref{eq:disc}). The connected residual
$\zconn$ is comparatively small, but it is the only part that carries four-body
information not already present at second order. Estimating it by subtraction,
configuration by configuration, is what makes the connected rung of the ladder
well defined.

Together these responses show concretely what configuration space buys us. The
higher-order statistics do not just amplify the same information as the 2PCF.
Instead they pick up the shape-dependent sensitivity that two-point clustering of
a fixed-density tracer simply does not have, and this is what breaks the
bias--amplitude degeneracy. We quantify the resulting gain in \S\ref{sec:results}.

\section{The information ladder}
\label{sec:results}
\begin{figure*}
\centering
\includegraphics[width=\textwidth]{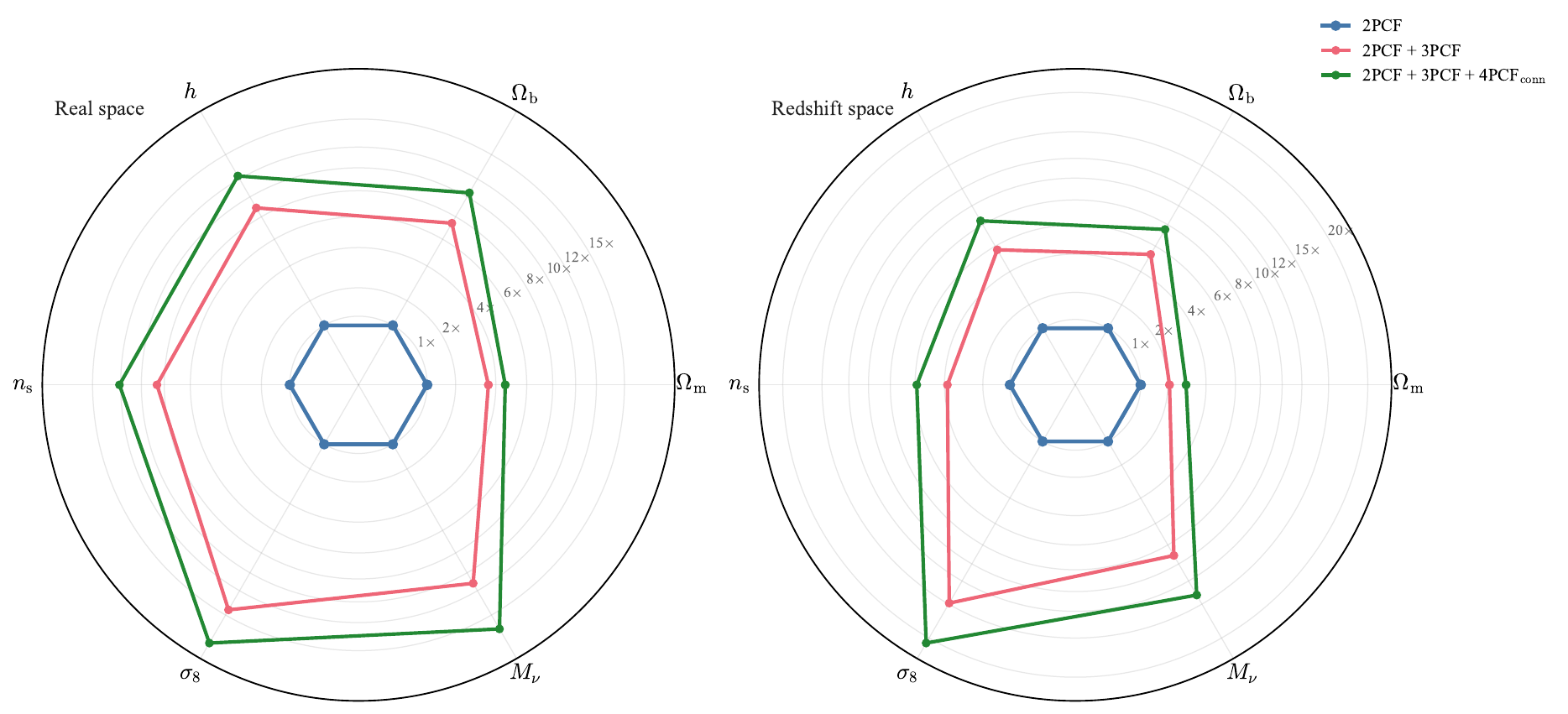}
\caption{Per-parameter information gain $\sigma_{\rm 2PCF}/\sigma$ as the data
vector grows $2\rightarrow2{+}3\rightarrow2{+}3{+}4_{\rm conn}$, in real (left)
and redshift space (right). Rings mark the factor improvement over the 2PCF alone
on a $\sqrt{\cdot}$-spaced radius, so that the enclosed area tracks the
information gained; the bold inner hexagon is the 2PCF baseline. All rungs use the MOPED
score-compressed Fisher (\S\ref{sec:rob-moped}), so the
$2\rightarrow2{+}3\rightarrow2{+}3{+}4_{\rm conn}$ increments are on a consistent
footing; the connected-four-point step adds $\sim1.4$--$1.5$ over $2{+}3$
(\S\ref{sec:res-4pcf}). In redshift space the baseline shown is the
monopole $\xi_0$, so the radial gains are the monopole-referenced factors;
the more conservative gains referenced to the multipole baseline
$\{\xi_0,\xi_2\}$ are given in Table~\ref{tab:fisher} and
\S\ref{sec:res-3pcf}, and the $2{+}3\rightarrow2{+}3{+}4_{\rm conn}$
increment, the spacing between the two outer rings, is
baseline-independent. }
\label{fig:spider}
\end{figure*}

We summarize the cosmological information carried by each rung of the ladder in
Fig.~\ref{fig:spider} and Table~\ref{tab:fisher}. We anchor the two-point
baseline, quantify the gain delivered by the three-point function and the
breaking of the $\sig$--$\Mnu$ degeneracy, compare with the Fourier-space
bispectrum, and set out the status of the connected four-point rung, which the
present covariance does not yet support (\S\ref{sec:robust}). Throughout,
redshift-space results for the three- and four-point functions refer to the monopole of the anisotropic statistic; the
redshift-space two-point baseline is reported both as the monopole alone and
as the joint monopole--quadrupole vector
$\{\xi_0,\xi_2\}$ (\S\ref{sec:rsd}), which we adopt as the reference baseline
for the incremental gains, and
all marginal errors include the Percival covariance-noise correction.

\begin{table}[htbp]
\centering
\caption{Marginalized $1\sigma$ constraints and information gain.\label{tab:fisher}}
\begin{tabular}{lcccccc}
\hline\hline
data vector & $\Om$ & $\Ob$ & $h$ & $n_{\rm s}$ & $\sig$ & $\Mnu$$^{\dagger}$\\
\hline
\multicolumn{7}{c}{\itshape Real space}\\
2PCF      & 0.057 & 0.037 & 0.41 & 0.38 & 0.32 & 0.71\\
$+$3PCF   & 0.016 & 0.0051 & 0.047 & 0.044 & 0.022 & 0.063\\
$+\zconn$ & 0.013 & 0.0036 & 0.034 & 0.031 & 0.017 & 0.042\\
\hline
\multicolumn{7}{c}{\itshape Redshift space (monopole baseline)}\\
2PCF ($\xi_0$)      & 0.041 & 0.036 & 0.36 & 0.21 & 0.41 & 0.81\\
$+$3PCF   & 0.020 & 0.0067 & 0.064 & 0.055 & 0.028 & 0.089\\
$+\zconn$ & 0.014 & 0.0047 & 0.043 & 0.036 & 0.020 & 0.059\\
\hline
\multicolumn{7}{c}{\itshape Redshift space (multipole baseline)}\\
2PCF ($\{\xi_0,\xi_2\}$) & 0.026 & 0.016 & 0.16 & 0.085 & 0.093 & 0.56\\
$+$3PCF   & 0.016 & 0.0066 & 0.062 & 0.046 & 0.027 & 0.089\\
$+\zconn$ & 0.014 & 0.0047 & 0.043 & 0.034 & 0.019 & 0.059\\
\hline
\multicolumn{7}{c}{\rev{\itshape Redshift space (modelling-informed higher-order selection)}}\\
\rev{$+$3PCF ($r_i\ge40\Mpch$)} & \rev{0.019} & \rev{0.0092} & \rev{0.089} & \rev{0.059} & \rev{0.056} & \rev{0.14}\\
\rev{$+\zconn$ ($r_{ij}\ge40\Mpch$)} & \rev{0.019} & \rev{0.0090} & \rev{0.087} & \rev{0.058} & \rev{0.055} & \rev{0.14}\\
\hline\hline
\end{tabular}

\vspace{5pt}
\begin{minipage}{0.95\textwidth}
{\footnotesize
Marginalized $1\sigma$ errors ($\Mnu$ in eV) at $z=0$,
$\Ncov=5000$ in real space and redshift space. The redshift-space entries are computed for monopole ($\xi_0$) and multipole ($\xi_0,\xi_2$) baselines where the multipole-baseline block adds the 2PCF quadrupole to the two-point vector
($N_d^{\rm 2pcf}=27$, full ladder $N_{\rm d}=1491$; \S\ref{sec:rsd}). \rev{The
modelling-informed block repeats the multipole ladder with the higher-order
blocks restricted to the tree-level-validated regime of \S\ref{sec:theory}:
every triangle side $r_i\ge40\Mpch$ (361 of 832 configurations) and every
tetrahedron side $r_{ij}\ge40\Mpch$ (11 of 632). The cut applies to the
higher-order blocks only; the reference $\{\xi_0,\xi_2\}$ row is the
unchanged multipole baseline above ($\xi_0$: $10$--$150\Mpch$, $\xi_2$:
$10$--$38\Mpch$), so the two selections are directly comparable row by row
(\S\ref{sec:res-cuts}).} The $+\zconn$ row is our primary estimator,
the MOPED score-compressed Fisher (\S\ref{sec:rob-moped}): at $N_{\rm d}={1484}$
($1491$ with the quadrupole)
the direct covariance, though invertible, carries large debiasing corrections
(Hartlap ${0.70}$, Percival $m_1={1.42}$) that
do not arise for the compressed score covariance. The direct Fisher serves as a cross-check and
agrees with the compressed result once those corrections are applied
(\S\ref{sec:rob-moped}). For a consistent ladder the $2$ and $+$3PCF rows are
likewise compressed (effectively lossless at their data-vector lengths,
where direct and compressed Fishers coincide). The
connected-four-point increment over the 3PCF is $\approx 1.2$--$1.5$,
under either redshift-space baseline
(\S\ref{sec:res-4pcf}). At the maximal Quijote derivative set
($\Nderiv=500$) the convergence test (\S\ref{sec:rob-mirage}) shows this is
parameter dependent: only $\sig$ is near convergence (a $\sim 15\%$ residual
systematic), while $\Ob$, $h$, and $n_{\rm s}$ are still rising and are
best read as lower bounds ($\Om$ is intermediate; \S\ref{sec:rob-mirage}). $^{\dagger}$The absolute $\Mnu$ column in particular is
scheme-dependent and un-converged---the tabulated value is from the
highest-order (four-point) derivative scheme, the noisiest of the three, and
is spuriously tightened by the finite-sample mirage; the
scheme-independent lower bound is $\sigma(\Mnu)\gtrsim0.1$~eV (real), so we do
not treat these $\Mnu$ entries as a forecast but as non-converged
diagnostic values, marked $^{\dagger}$ throughout; the per-parameter
convergence status of every rung is summarized in
Table~\ref{tab:conv} (\S\ref{sec:rob-mirage}).
}
\end{minipage}
\end{table}

\subsection{The two-point baseline}
\label{sec:res-2pcf}
The first block of Table~\ref{tab:fisher} gives the constraints from the 2PCF
alone. For a single $(1\,h^{-1}{\rm Gpc})^3$ volume the amplitude-like
directions are poorly constrained: the 2PCF determines $\sig$ only to
$\simeq0.32$ and $\Mnu$ to $\simeq0.71$~eV in real space. These directions are
limited by the degeneracy between the clustering amplitude, the linear halo
bias, and the small-scale power suppression from massive neutrinos, which
two-point clustering cannot separate. In redshift space the $\sig$ and $\Mnu$
constraints from the monopole alone do not improve. Instead they weaken to $\simeq0.41$ and
$\simeq0.81$~eV, because the monopole folds the Kaiser boost into the same
amplitude direction without the anisotropy needed to pin down the growth rate.
Including the quadrupole recovers exactly this anisotropic information,
and it strengthens the two-point baseline substantially: relative to the
monopole alone, the joint $\{\xi_0,\xi_2\}$ vector tightens $\sig$ by a factor
of $4.4$ (to $\simeq0.093$), the shape parameters $\Ob$, $h$, and $n_{\rm s}$
by $2.2$--$2.5$, $\Om$ by $1.6$, and $\Mnu$ by $1.5$ (to $\simeq0.56$~eV;
Table~\ref{tab:fisher}). The quadrupole separates the Kaiser anisotropy from
the isotropic amplitude, so the growth-rate direction that the monopole folds
away becomes accessible already at two-point order. All incremental
higher-order gains quoted below are therefore referenced to this stronger
$\{\xi_0,\xi_2\}$ baseline; the quadrupole is measured over
$10$--$38\Mpch$, where its signal-to-noise is concentrated (\S\ref{sec:rsd}).
Extending the anisotropic treatment to the higher multipoles of the three- and
four-point functions is left to future work, for which the necessary
machinery already exists: self-consistent frameworks for the joint
anisotropic two- and three-point analysis \citep{Sugiyama2021}, optimized
line-of-sight choices for the anisotropic 3PCF \citep{Garcia2022}, and, most
recently, coordinated modelling and measurement of the anisotropic halo 3PCF
\citep{Farina2026}.

\subsection{Information gain from the three-point function}
\label{sec:res-3pcf}
Adding the 3PCF tightens every parameter, and it helps most where it matters
(Fig.~\ref{fig:spider}). In real space, where the two-point function is
isotropic and the monopole is the complete baseline, the shape parameters $\Om$, $\Ob$, $h$, and $n_{\rm s}$ gain by factors of $\sim$4--9. The amplitude directions gain far
more: the marginal error on $\sig$ improves by a factor $\simeq14$ and that on
$\Mnu$ by $\sim$11  in real space. In redshift space, referenced to the
multipole baseline $\{\xi_0,\xi_2\}$, the gains are more moderate but remain
substantial: $1.6$--$2.6$ on the shape parameters, $3.5$ on $\sig$, and
$6.3$ on $\Mnu$. The neutrino mass is the direction the quadrupole helps
least and the 3PCF helps most. Either way, the $2{+}3$ rung
reaches $\sigma(\sig)\simeq0.02$--$0.03$ and $\sigma(\Mnu)\simeq0.06$--$0.09$~eV from
this single volume. These factors are referenced to the configuration-space
2PCF, whose six-parameter marginal errors are large in both the real-space
and monopole cases and are themselves
mirage-sensitive (\S\ref{sec:rob-mirage}); the largest ratios are therefore best read as indicative. The quantity that does not depend on this weak baseline is the relative rung-to-rung gain (\S\ref{sec:res-4pcf}).

The 3PCF carries this leverage because its dependence on
triangle shape is not just a global amplitude rescaling; the configuration
information absent from the 2PCF is precisely what separates bias from
amplitude. However, these two highest-gain directions are {\em preliminary} at
the current covariance size. The Fisher's-mirage diagnostic of
\S\ref{sec:robust} shows that $\sig$ and $\Mnu$ have not yet converged, and we
expect their gains to moderate as $\Ncov$ and $\Nderiv$ grow. The two are not equally recoverable, however: $\sig$ is nearly converged and its constraint
($\sigma(\sig)\simeq0.02$--$0.03$) is robust to within a $\sim10\%$ convergence
systematic, whereas the $\sigma(\Mnu)\simeq0.06$--$0.09$~eV quoted above is the
value of the highest-order (four-point) derivative scheme---the noisiest of the
three (\S\ref{sec:rob-mirage})---and is spuriously tightened by the finite-sample
mirage; it is scheme-dependent and un-converged, so we treat it as a lower bound ($\sigma(\Mnu)\gtrsim0.1$~eV) rather than a forecast (\S\ref{sec:rob-mirage}).

\subsection{Breaking the \texorpdfstring{$\sig$--$\Mnu$}{sigma8-Mnu} degeneracy}
\label{sec:res-deg}
\begin{figure*}
\centering
\includegraphics[width=\textwidth]{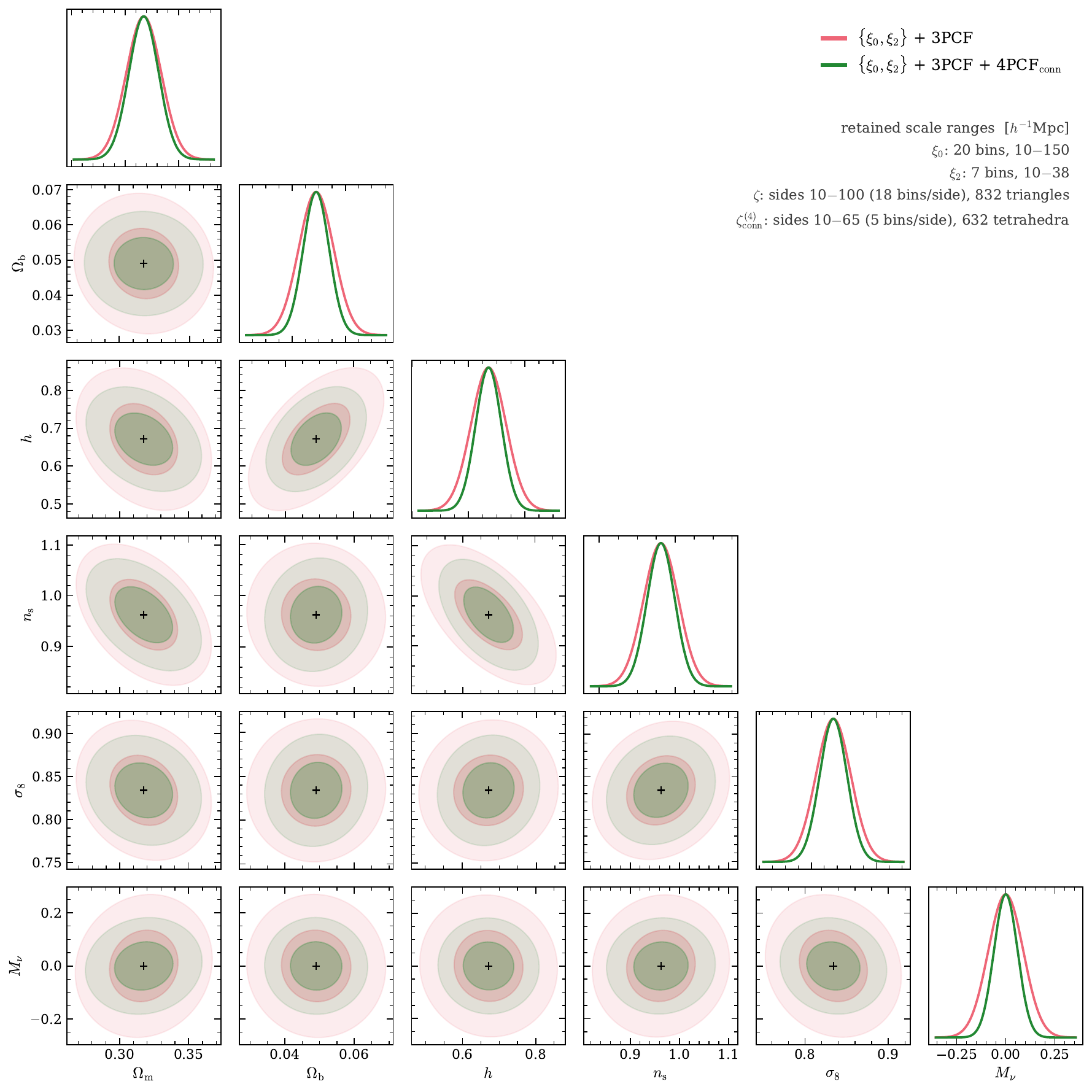}
\caption{Marginalized $68\%$ and $95\%$ constraints on the six parameters in
redshift space, for the two higher-order rungs $\{\xi_0,\xi_2\}{+}3$PCF (pink) and
$\{\xi_0,\xi_2\}{+}3{\rm PCF}{+}\zconn$ (green, the tightest), on the
multipole two-point baseline of Table~\ref{tab:fisher}. The retained
scale ranges of every block are annotated on the figure: $\xi_0$ in $20$
bins over $10$--$150\Mpch$; $\xi_2$ in $7$ bins over $10$--$38\Mpch$
(\S\ref{sec:rsd}); the 3PCF with each triangle side in $18$ bins over
$10$--$100\Mpch$ with $832$ configurations after the triangle-inequality cut;
and the connected 4PCF with each tetrahedron separation in $5$ bins over
$10$--$65\Mpch$ with $632$ configurations after the tetrahedron-inequality and
$RRRR>0$ cuts; \S\ref{sec:data}. We omit the 2PCF here: it is several times
looser, so on a common scale it sets the axis limits and compresses the
$2{+}3\rightarrow2{+}3{+}\zconn$ comparison this figure exists to resolve. The
2PCF constraint appears as the spider baseline (Fig.~\ref{fig:spider}), in
Table~\ref{tab:fisher}, and per statistic in Fig.~\ref{fig:s8mnu}; here we show
instead the connected-four-point increment over the 3PCF.}
\label{fig:corner}
\end{figure*}

\begin{figure*}
\centering
\includegraphics[width=0.92\textwidth]{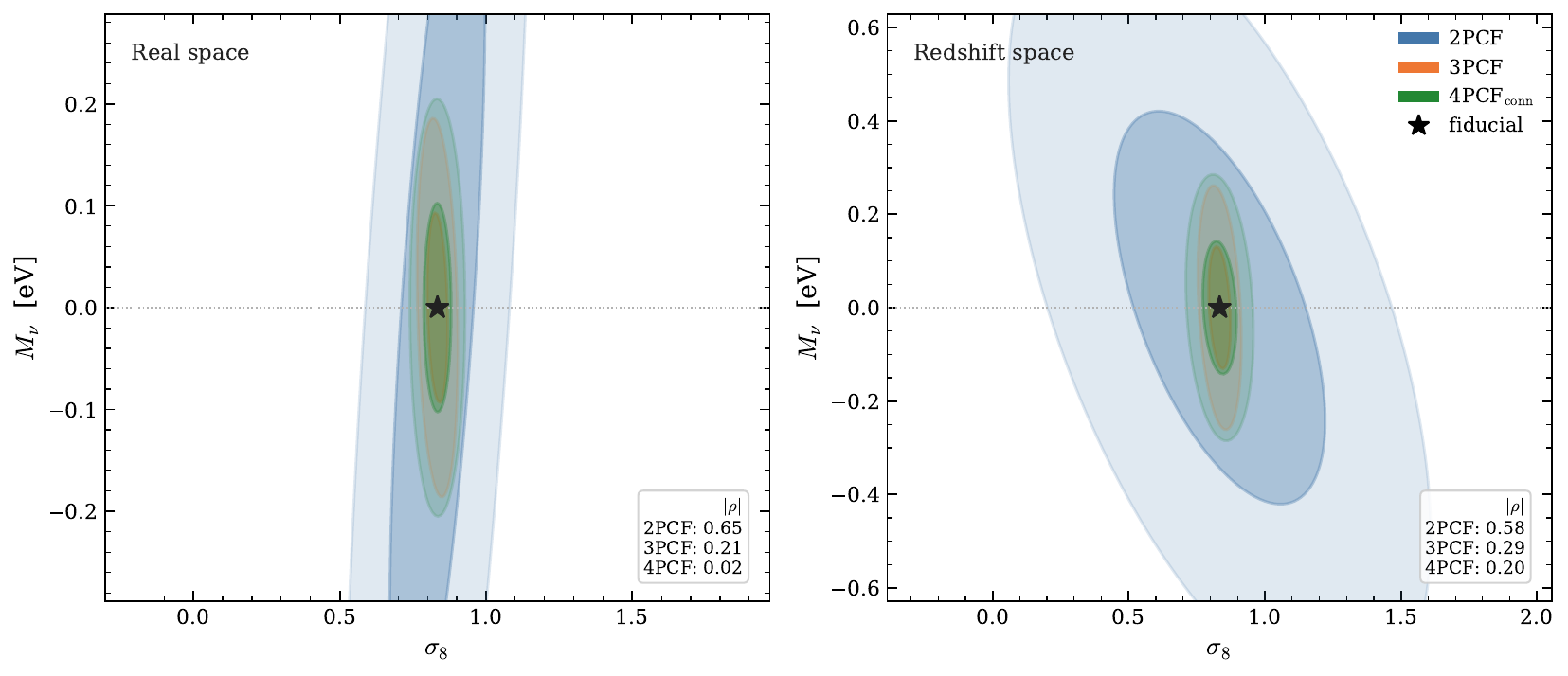}
\caption{
The intrinsic $\sig$--$\Mnu$ constraint from each statistic on
its own (2PCF, 3PCF, and $\zconn$; $68\%$ and $95\%$ ellipses), with the other
four parameters held fixed, in real (left) and redshift space (right); the
star marks the fiducial model. The 2PCF carries a marked $\sig$--$\Mnu$
degeneracy ($|\rho|\simeq0.65$ in real and $0.58$ in redshift space). The 3PCF
alone reduces it to $|\rho|\simeq0.21$ ($0.29$), the configuration-space
counterpart of the Fourier-space bispectrum degeneracy break of
\citet{Hahn2020}, and in real space the connected 4PCF alone is nearly degeneracy-free
($|\rho|\simeq0.02$), its constraint almost orthogonal to the
2PCF's, while in redshift space it reaches $|\rho|\simeq0.20$, modestly below the 3PCF's. 
The ellipses use the Percival-corrected parameter covariance at
$\Ncov=5000$; the fully marginalized contours of the combined rungs are shown
in Fig.~\ref{fig:corner}
}
\label{fig:s8mnu}
\end{figure*}

Figs.~\ref{fig:corner} and \ref{fig:s8mnu} make the origin of these gains
explicit. Fig.~\ref{fig:s8mnu} shows the intrinsic $\sig$--$\Mnu$ constraint
from {\em each statistic on its own}, the other four parameters held fixed. The
2PCF carries a pronounced degeneracy ($|\rho|\simeq0.65$ in real and $0.58$ in
redshift space), but the higher orders are not merely tighter---their degeneracy directions are also different. The 3PCF alone reduces the
correlation to $|\rho|\simeq0.21$ and $0.29$, and in real-space the connected 4PCF alone is
nearly degeneracy-{\em free} ($|\rho|\simeq0.02$), its constraint almost orthogonal to the 2PCF's; in redshift space the connected 4PCF reaches $|\rho|\simeq0.20$, modestly below the 3PCF's, so both higher rungs markedly reduce the 2PCF degeneracy, although neither approaches in redshift space the near-orthogonality that the connected 4PCF attains in real space. This degeneracy reduction
is the configuration-space analogue of the bispectrum degeneracy break of
\citet{Hahn2020}: it is why the combined ladder collapses the joint constraint
(Fig.~\ref{fig:corner}) so sharply---the statistics pin $\sig$ and $\Mnu$ along
complementary axes rather than repeating the same measurement. This degeneracy {\em breaking}---the reduction of the $\sig$--$\Mnu$
correlation and the attendant sharpening of $\sig$---is a {\em relative}
statement and is the robust content of this plane; it does not yet support a
converged {\em absolute} $\Mnu$ error, whose finite-sample and
derivative-scheme sensitivities are discussed in \S\ref{sec:res-4pcf} and
\S\ref{sec:rob-mirage}.

One subtlety bears on the interpretation. In the fully marginalized six-parameter
analysis the dominant partner of $\Mnu$ is not $\sig$ but $h$, and the strongest
residual degeneracies are $\Ob$--$h$ and $\Om$--$n_{\rm s}$. The 3PCF tightens
these as well, so the marginal $\sig$ and $\Mnu$ improvements reflect a global
degeneracy-breaking rather than the rotation of a single direction.

\subsection{Comparison with the Fourier-space bispectrum}
\label{sec:res-hahn}
It is useful to set these gains beside the Fourier-space halo bispectrum
forecast of \citet{Hahn2020} on the same simulation suite. Their
power-spectrum baseline includes the redshift-space monopole and quadrupole,
so our $\{\xi_0,\xi_2\}$ baseline is the appropriate point of
comparison. Referenced to it, the redshift-space 3PCF-over-2PCF factors are
$1.6$--$2.6$ for $\{\Om,\Ob,h,n_{\rm s}\}$, $3.5$ for $\sig$, and $6.3$ for
$\Mnu$, against their bispectrum-over-power-spectrum factors of
$1.9$--$3.6$, $2.6$, and $\sim5$. The two ladders now agree, not only in fact that the three-point statistic helps $\sig$ and $\Mnu$ the most, but
to within factors of order unity in every direction, with our $\Mnu$ factor
moderately larger. The residual differences have identifiable sources. Our
fine configuration-space binning down to $10\Mpch$, the different large-scale
reach of the two baselines, $k_{\max}=0.5\,h\,{\rm Mpc}^{-1}$ versus our
$10$--$150\Mpch$ monopole with the quadrupole over $10$--$38\Mpch$, and the
finite-sample effects of \S\ref{sec:rob-mirage}. Referenced to the
monopole-only baseline the factors are
$\sim2$--$6$, $\simeq15$, and $\simeq9$; most of the excess over
\citet{Hahn2020} disappears once the anisotropic two-point information is
restored to the baseline, confirming that it reflected the weak denominator
rather than the higher-order statistics. Even so, the two analyses remain
unmatched in several respects. Firstly, regarding tracer selection, we use a fixed number density while they used a fixed mass cut with the mass limit marginalized. Secondly the scale cuts of $k_{\max}=0.5\,h\,{\rm Mpc}^{-1}$ versus our configuration ranges. We therefore read the comparison as a consistency check
of the degeneracy-breaking pattern, and present the configuration-space ladder as an {\em independent and
complementary} route to higher-order information, not as an improvement on the
Fourier bispectrum.

\subsection{The connected four-point function}
\label{sec:res-4pcf}
The genuinely non-Gaussian four-point information resides in the connected 4PCF,
$\zconn$, which we separate on the fly from its disconnected (Gaussian) part.
Including it enlarges the data vector to $N_{\rm d}=1484$ configurations
(degenerate tetrahedra excluded; \S\ref{sec:data}). At the current covariance
size the direct Fisher for this rung carries large covariance-debiasing
corrections (\S\ref{sec:rob-cov}), so we adopt the better-conditioned MOPED
score-compressed Fisher as our primary estimator and retain the direct Fisher as
a cross-check (\S\ref{sec:rob-moped}). The connected 4PCF tightens the
marginalized constraints by a further factor of $1.2$--$1.5$ over the
$2{+}3$ combination (largest for $\Mnu$), an increment that is
robust to the compression regularization and that the direct Fisher confirms up
to the covariance-debiasing factor. This increment is also independent of
the two-point baseline: over the monopole-based rung the connected 4PCF adds
factors of $1.4$--$1.5$ ($1.52$ for $\Mnu$), and over the multipole-based rung
$\{\xi_0,\xi_2\}{+}\zeta$ it adds $1.2$--$1.5$ with a near identical $1.51$ for
$\Mnu$ (Table~\ref{tab:fisher}). The connected four-point gain is therefore
not fed by two-point anisotropy missing from the baseline; it is genuine
higher-order information. This {\em relative} increment is the
robust result of this rung: as the Fisher's-mirage test of
\S\ref{sec:rob-mirage} shows (Fig.~\ref{fig:conv}, bottom row), the
finite-sample tightening common to the $2{+}3$ and $2{+}3{+}\zconn$ rungs
largely cancels in their ratio, so the connected-four-point {\em information
gain} survives the convergence test even where the absolute errors do not.
This is a modest but real addition. Most of the accessible
higher-order information is already captured by the 3PCF, with the four-point
function contributing a further $\sim$20--50\% tightening. The {\em
absolute} $\Mnu$ error of this ladder must, by contrast, be read with care. It
is strongly scheme-dependent: repeating the forecast with the forward, three-,
and four-point neutrino-derivative schemes gives $\sigma(\Mnu)\simeq0.10$,
$0.055$, and $0.042$~eV in real space (and $0.14$, $0.077$, and $0.059$~eV in
redshift space), a spread of $\sim2.3\times$. Tellingly, the {\em
noisier} the scheme, the {\em tighter} the apparent constraint---the signature
of a finite-sample mirage, whose spurious tightening grows with the derivative
noise. None of the schemes has converged at the maximal
Quijote derivative set (\S\ref{sec:rob-mirage}), and a $1/\Nderiv$ extrapolation
places the true $\sigma(\Mnu)\gtrsim0.1$~eV in real space, so the apparently
tight four-point value is largely a mirage artifact. We therefore do {\em not}
report a single absolute $\Mnu$ forecast from this single
$(1\,h^{-1}{\rm Gpc})^3$ volume at $z=0$; what the connected four-point function
robustly delivers is the {\em relative} gain in neutrino information over the
$2{+}3$ ladder, which is real and survives the convergence test. This
higher-order route to $\Mnu$ information is a
complementary and methodologically independent counterpart to the Fourier-space
bispectrum.

\subsection{Scale cuts and physically motivated configurations}
\label{sec:res-cuts}
\begin{figure*}
\centering
\includegraphics[width=\textwidth]{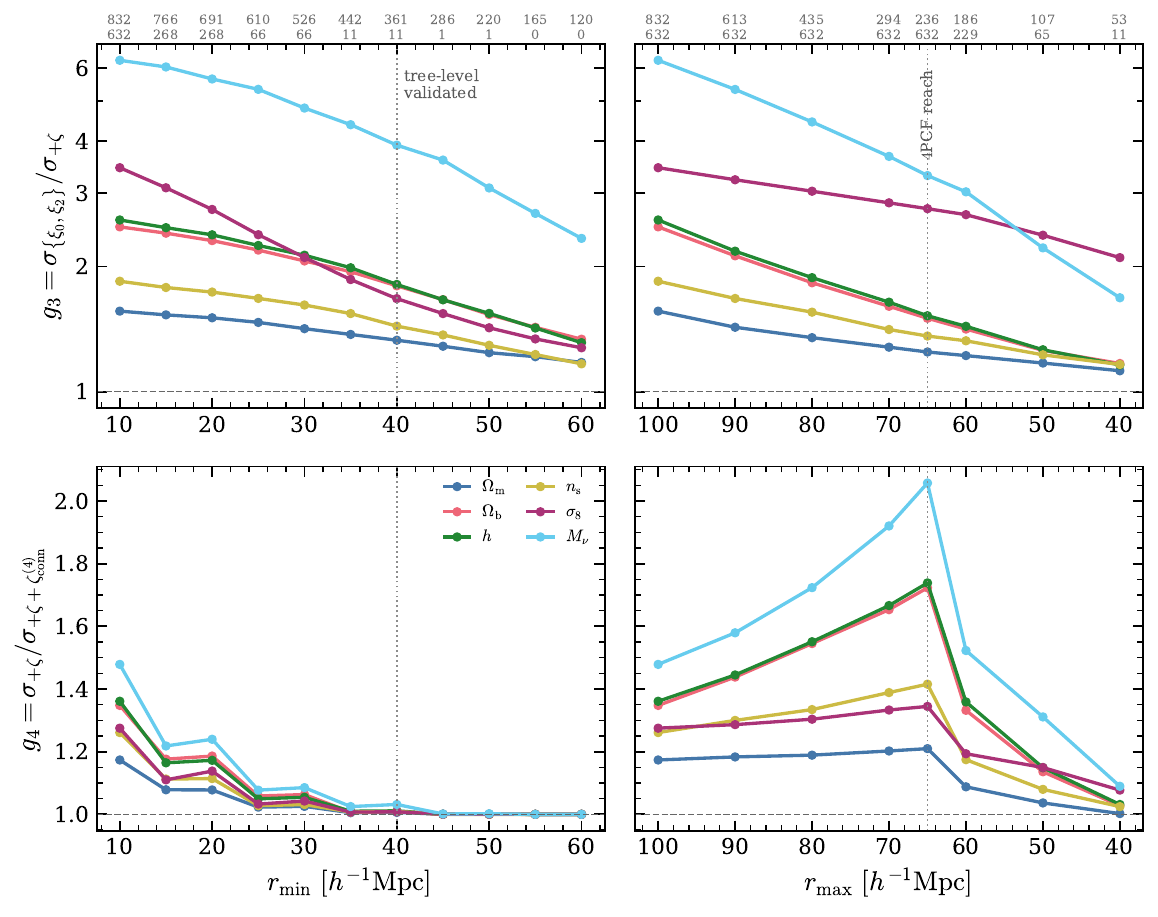}
\caption{Incremental information gains under scale cuts, at the fixed
redshift-space $\{\xi_0,\xi_2\}$ baseline. {\em Left column:} cumulative
gains versus the minimum-side cut $r_{\rm min}$ (all side bins of a retained
configuration lie entirely above the cut, applied to both the 3PCF and the
connected 4PCF); {\em right column:} versus the maximum-side cut
$r_{\rm max}$. {\em Top row:} the 3PCF gain
$g_3=\sigma_{\{\xi_0,\xi_2\}}/\sigma_{+\zeta}$ (log scale; grey annotations
give the surviving 3PCF/4PCF configuration counts); {\em bottom row:} the
connected-four-point gain $g_4=\sigma_{+\zeta}/\sigma_{+\zeta+\zconn}$. The
dotted lines mark the tree-level-validated cut ($r_{\rm min}=40\Mpch$,
\S\ref{sec:theory}) and the four-point reach ($65\Mpch$).}
\label{fig:cuts}
\end{figure*}

The configuration vectors used above extend to the smallest measured
separations, a selection oriented toward signal-to-noise. To establish which
conclusions persist under physically motivated cuts, we repeat the
redshift-space ladder, holding the $\{\xi_0,\xi_2\}$ baseline
fixed, while sweeping a minimum-side cut $r_{\rm min}$ (every side bin of a
configuration must lie entirely above the cut) and, separately, a
maximum-side cut $r_{\rm max}$, applied to the 3PCF and connected-4PCF
blocks. Figure~\ref{fig:cuts} shows the cumulative incremental gains
$g_3=\sigma_{\{\xi_0,\xi_2\}}/\sigma_{+\zeta}$ and
$g_4=\sigma_{+\zeta}/\sigma_{+\zeta+\zconn}$ along both sweeps.

Three conclusions can be drawn from this test. First, the three-point gain survives conservative
cuts. In the \rev{modelling-informed higher-order selection}
$r_{\rm min}=40\Mpch$, the
regime in which the tree-level model of \S\ref{sec:theory} is validated
\rev{(the cut defines the selection for the three- and four-point blocks
only, while the $\{\xi_0,\xi_2\}$ baseline is unchanged, with $\xi_0$ over
$10$--$150\Mpch$ and $\xi_2$ over $10$--$38\Mpch$)}, the
3PCF with 361 of 832 configurations still tightens the shape parameters by
$1.3$--$1.8$, $\sig$ by $1.7$, and $\Mnu$ by $3.9$, reaching
$\sigma(\sig)\simeq0.056$ and $\sigma(\Mnu)\simeq0.14$~eV, where the latter is a non-converged diagnostic value; \S\ref{sec:rob-mirage}. The
neutrino-mass direction, the least aided by the quadrupole, retains the
largest higher-order gain even in the fully theory-tractable regime.
\rev{This selection is tabulated side by side with the full-range,
signal-to-noise-oriented vectors as the final block of
Table~\ref{tab:fisher}, so the two cases can be compared directly, row by
row: the full-range rows quantify what the simulations measure at high
signal-to-noise, the modelling-informed rows what survives in the regime our
own tree-level model validates. The near-coincidence of its $+$3PCF and
$+\zconn$ rows (only 11 tetrahedra survive the cut) displays at a glance the
contrast established below: the three-point gain is robust to the selection,
the connected-four-point increment is not.}

Second, the connected-four-point increment is a small-scale statement. Its
tetrahedra reach only $65\Mpch$, so minimum-side cuts deplete it rapidly, with 632, 268, 66, and 11 configurations at $r_{\rm min}=10$, $15$, $25$, and
$40\Mpch$. Excluding every configuration that touches the shortest side bin of $10$--$21\Mpch$ retains roughly half of the $\Mnu$ increment
$g_4=1.25$ versus $1.48$ by $r_{\rm min}\simeq40\Mpch$ the surviving
configurations add essentially nothing ($g_4\le1.05$). The
connected-four-point gain therefore originates in configurations with at
least one side below ${\sim}30\Mpch$, which is where a connected four-point
signal is expected to peak, but also where analytic modelling is only now emerging \citep{Ortola2025}; we flag this explicitly in \S\ref{sec:disc-caveats}.

Third, the maximum-side sweep separates the large-scale contributions.
Restricting the 3PCF to $r_{\rm max}\le65\Mpch$ (the four-point
reach) roughly halves its $\Mnu$ gain from $6.3$ to $3.3$ while leaving the $\sig$ gain largely intact from $3.5$ to $2.8$: the three-point neutrino information draws substantially on the $65$--$100\Mpch$ domain, whereas the amplitude information is predominantly sub-$65\Mpch$. With the
3PCF so restricted the four-point increment grows ($g_4$ on $\Mnu$ rises
from $1.51$ to $2.2$), showing that the two higher-order statistics
partially substitute for one another on their common scales. Finally, we
note that neither higher-order vector reaches the acoustic scale: the 3PCF
sides extend to $100\Mpch$ and the 4PCF to $65\Mpch$, while the BAO peak
($\simeq105\Mpch$) is carried only by the 2PCF baseline
($10$--$150\Mpch$). The ladder gains reported here are therefore sub-BAO
configuration information, and the higher-order BAO domain remains
unexplored territory for this estimator, though BAO features have now been detected in survey 3PCFs \citep{2021Moresco, Kamalinejad2026}.

\rev{Finally, because imposing $r_{\rm min}=40\Mpch$ changes the dimension
and composition of the data vector, the derivative-convergence diagnostics
of \S\ref{sec:rob-mirage} for the full-range vectors do not automatically
carry over, and we have therefore subjected the modelling-informed selection
to the same $\Nderiv$ sweep at fixed covariance. The absolute errors behave
like their full-range counterparts: still rising at $\Nderiv=500$, with
last-step drifts of $3$--$13\%$ and $1/\Nderiv$-extrapolated residuals of
$4$--$18\%$ ($\Mnu$ worst), so the cut errors quoted above inherit the same
reading as the full-range entries of Table~\ref{tab:fisher}: lower bounds,
with $\Mnu$ a non-converged diagnostic. The gain itself is much more
stable. The key modelling-informed endpoint,
$g_3(\Mnu;r_{\rm min}=40\Mpch)$, drifts from $3.93$ to $3.90$ over the last
derivative step (below $1\%$), and is scheme-independent at the per-cent
level: $3.91$, $3.95$, and $3.90$ across the forward, three-point, and
four-point neutrino schemes, whose absolute $\sigma(\Mnu)$ under the cut
still differ by a factor of $2.4$. Like the four-point ratio of
\S\ref{sec:rob-mirage}, this gain is bounded below by unity by construction
and the mirage inflates it from above, so derivative noise can only
overstate it, never manufacture it; a linear $1/\Nderiv$ extrapolation of
its slow decline suggests an asymptote of $g_3(\Mnu)\simeq2.9$ (and
$\simeq1.6$ for $\sig$, versus $1.68$ at $\Nderiv=500$), with every
parameter's asymptote remaining above $1.2$. We accordingly quote the
modelling-informed gains at the Quijote derivative ceiling ($3.9$ on
$\Mnu$), subject to a residual derivative-noise systematic of up to
${\sim}25\%$ in the downward direction. The four-point conclusion is
unaffected: $g_4$ under the cut stays within $1.01$--$1.06$ across the entire
$\Nderiv$ range.}

\section{Robustness and convergence}
\label{sec:robust}
Because the forecast is built from a finite number of simulations, both the
covariance and the derivatives carry estimation noise, and these have opposite
effects on the inferred errors. We correct the covariance noise, mitigate and
diagnose the derivative noise, and verify that the gains do not hinge on a single
binning choice.

\subsection{Covariance-matrix corrections}
\label{sec:rob-cov}
The finite number $\Ncov$ of fiducial realizations biases the inverse covariance
and propagates noise into the parameter errors, and we correct for both. We
debias the inverse sample covariance with the Hartlap factor
$(\Ncov-N_{\rm d}-2)/(\Ncov-1)$ \citep{Hartlap2007}, and we correct the residual
propagation of covariance-estimation noise into the {\em parameter} covariance
with the factor $m_1$ of \citet{Percival2014}, applied as
$F^{-1}\!\rightarrow m_1\,F^{-1}$; the same noise underlies the
parameter-variance inflation quantified by \citet{DodelsonSchneider2013}. Both
corrections grow with the data-vector length $N_{\rm d}$, and so they bear
hardest on the higher rungs, whose covariance is the most expensive to estimate.

At $\Ncov=5000$ the 2PCF rung is essentially uncorrected
($m_1=1.00$), and the $2{+}3$ rung is well-conditioned (Hartlap $0.83$,
$m_1=1.20$, a $20\%$ inflation of the marginal errors that is already included in
Table~\ref{tab:fisher}). The full $2{+}3{+}4_{\rm conn}$ rung
($N_{\rm d}=1484$) is by contrast poorly conditioned (Hartlap $0.70$,
$m_1=1.42$): a $42\%$ inflation of the parameter covariance, applied to a
$\sim$1500-dimensional inverse estimated from $5000$ realizations. The
redshift-space analysis with the quadrupole included ($N_{\rm d}=1491$) sits at the same conditioning: Hartlap $0.70$ and $m_1=1.42$ for the full
rung, $0.83$ and $1.20$ for the $\{\xi_0,\xi_2\}{+}3$PCF rung. This is
the quantitative basis for adopting the score-compressed Fisher as our primary
estimator for the connected four-point rung (\S\ref{sec:rob-moped}), with the
direct, fully-corrected Fisher retained as a cross-check. The $2$ and $2{+}3$
rungs are already in their well-corrected regime, where direct and
compressed Fishers coincide and the compression is verified lossless in
practice.

\subsection{The Fisher's-mirage test}
\label{sec:rob-mirage}
\begin{figure}
\centering
\includegraphics[width=\columnwidth]{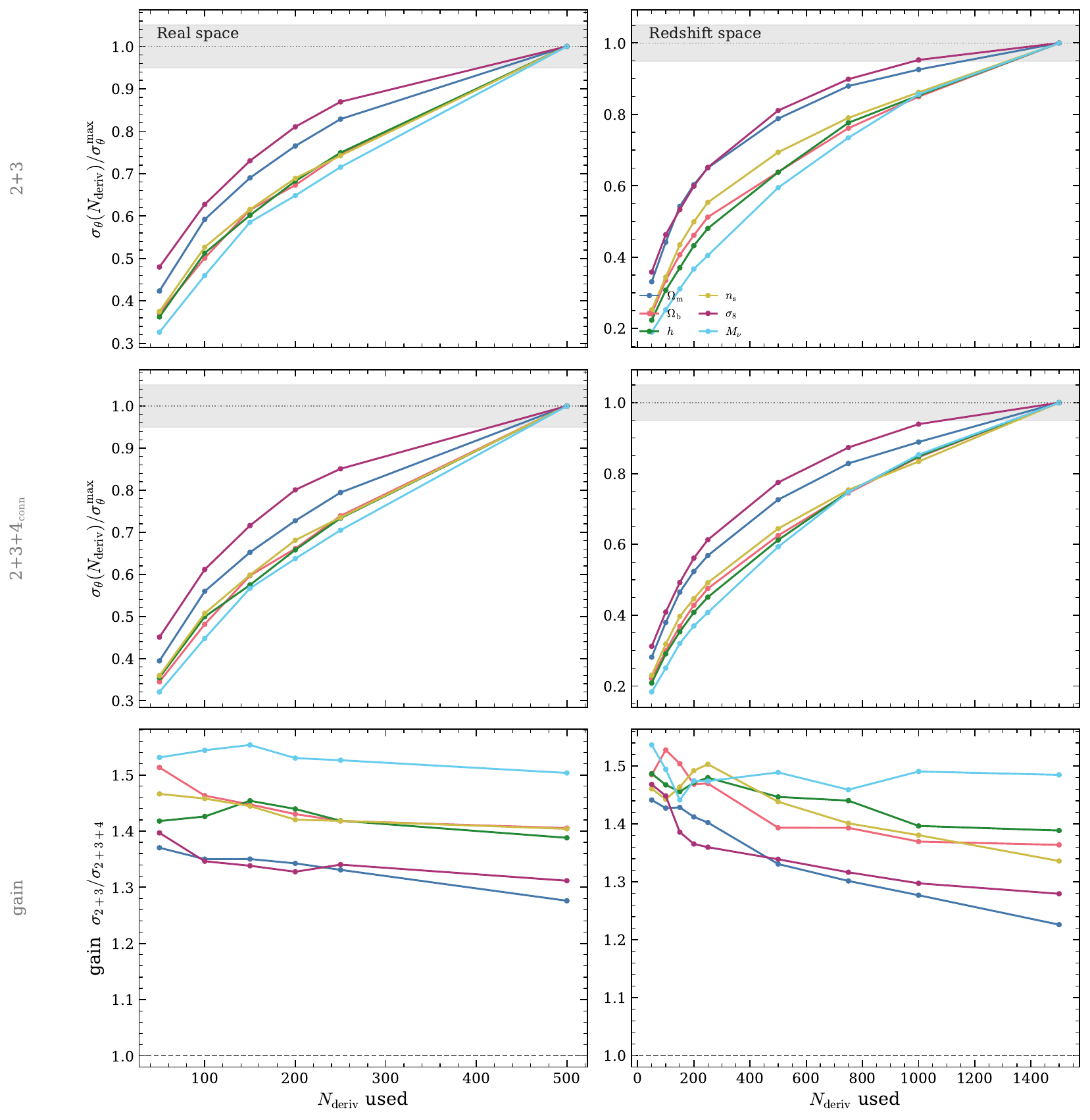}
\caption{
Fisher's-mirage convergence test: the marginal error
$\sigma_\theta(\Nderiv)/\sigma_\theta(\Nderiv^{\rm max})$ versus the number of
derivative realizations retained, at fixed covariance. Columns are real (left)
and redshift space (right); rows are the well-conditioned $2{+}3$ rung (top) and
the compressed $2{+}3{+}4_{\rm conn}$ rung (middle, the reported result),
so the test covers the headline data vector including the noisier four-point
derivatives. A plateau indicates a constraint no longer limited by derivative
noise \citep{WilsonBean2025}. The curves are still rising at the current
$\Nderiv$ for most parameters---only $\sig$ has nearly plateaued
(\S\ref{sec:rob-mirage})---so the absolute constraints, and the $\Mnu$ gains
of Fig.~\ref{fig:spider} in particular, remain preliminary.
{\em Bottom row:} the information-gain ratio
$\sigma_{2+3}/\sigma_{2+3+4}$ versus $\Nderiv$. A gain manufactured purely by the
mirage of the noisier four-point derivatives would collapse toward unity (dashed
line) as $\Nderiv$ grows; instead the ratio holds at $\sim1.2$--$1.5$, declining
by at most $\sim15\%$ over the full $30\times$ span in $\Nderiv$ (least of
all for $\Mnu$, $\lesssim4\%$).
The connected-four-point information is therefore {\em real}, not a finite-sample
artifact: the common mirage largely cancels in the ratio, so the relative gain is
robust even while the absolute constraints firm up.
}
\label{fig:conv}
\end{figure}

A distinct finite-sample effect arises from the {\em derivative} simulations. The
noise in derivatives estimated from a finite number $\Nderiv$ of realizations
does not loosen but spuriously {\em tightens} the marginalized Fisher
constraints, through the marginalization \citep{WilsonBean2025}. This has the
opposite sign to the covariance corrections above, and it is most dangerous for
the highest-gain directions. We guard against it in two ways. First, we suppress the
derivative noise by averaging each redshift-space derivative over the three
line-of-sight axes of the periodic box. Because the three axes sample the same
underlying density field they are correlated, so the effective $\Nderiv$ grows by
a factor that is appreciable but smaller than the nominal three; in direct
single- versus three-axis tests the averaging reduces the spurious tightening by
$\sim25\%$. Second, we diagnose convergence directly: Fig.~\ref{fig:conv}
shows $\sigma_\theta(\Nderiv)/\sigma_\theta(\Nderiv^{\rm max})$ at fixed
covariance, so that a plateau signals a constraint no longer limited by
derivative noise \citep[cf.][]{CoultonWandelt2023}. We now run this diagnostic out to the full Quijote derivative set,
$\Nderiv=500$ realizations per step---the maximum available---and find that
convergence is strongly {\em parameter dependent} rather than uniform. Only
$\sig$ is close to converged: its curve has all but flattened, with a last-step
drift of only $\sim3$--$4\%$ and a $1/\Nderiv$-extrapolated residual of
$\sim15\%$ (real) / $\sim12\%$ (redshift space). $\Om$ is intermediate
($\sim 16$--$20\%$ residual. The remaining directions have not converged even at
the Quijote ceiling: $\Ob$, $h$, and $n_{\rm s}$ are still rising at $\Nderiv=500$
(last-step drift $\sim 8$--$13\%$, extrapolated residual $\sim 22$--$25\%$), and
$\Mnu$ is the worst case, with no sign of a plateau. The physical distinction is
that these are the weak, mutually degenerate directions: the derivative signal is
small and the derivative noise is a large fraction of it, so the marginalization
never drains the mirage. We therefore report $\sig$ as the one reportable
higher-order constraint, $\sigma(\sig)\simeq0.017$ (real) / $0.020$ (redshift
space) at $\Nderiv=500$, carrying a stated $\sim15\%$ convergence systematic;
the absolute errors on $\Om$, $\Ob$, $h$, and $n_{\rm s}$ we present as {\em
lower bounds} rather than forecasts, since their curves are still climbing at the
maximal derivative set.

We attempted an explicit derivative-noise debiasing
\citep[the analytic and split-sample estimators of][]{CoultonWandelt2023} but
found it unstable here: the connected 4PCF and the four-point $\Mnu$ combination
have per-realization derivatives noisy enough that the estimated bias exceeds the
signal and the debiased Fisher loses positive-definiteness. We therefore do not
debias analytically, relying instead on the line-of-sight averaging, this
diagnostic, and a still-growing derivative suite; the connected-four-point
constraints will be finalized as it grows.

The neutrino mass demands a separate and more stringent test, because it is
both the highest-gain and the least-converged direction. We recompute
$\sigma(\Mnu)$ using three finite-difference schemes for the neutrino response of
contrasting noise---forward, three-point, and four-point (our default)---which
carry roughly $14\times$, $25\times$, and $33\times$ the per-realization
derivative noise, respectively. If the constraint were converged the three
schemes would agree; instead all three are still rising at $\Nderiv=500$ and they
disagree by a factor of $\sim2.3$. The marginal errors are $\sigma(\Mnu)=0.098$
(forward), $0.055$ (three-point derivative), and $0.042$~eV (four-point derivative) in real space, and
$0.138$, $0.077$, and $0.059$~eV in redshift space. The ordering is the
tell-tale signature of the mirage: the {\em noisier} the scheme, the {\em
tighter} the apparent constraint, i.e.\ the apparent tightness scales with the
derivative noise rather than with any real information. The four-point scheme's $\sim0.04$~eV is thus
largely a finite-sample artifact, not a physical constraint. Extrapolating along
the noise ordering, the true marginal error satisfies
$\sigma(\Mnu)\gtrsim0.1$~eV (real) and $\gtrsim0.14$~eV (redshift space), and
even these are lower bounds, since every scheme is still climbing at the Quijote
maximum. We accordingly do {\em not} report a single $\Mnu$ forecast: the
absolute neutrino-mass constraint is not convergeable with the present derivative
suite. Table~\ref{tab:conv} collects these convergence diagnostics
compactly, per parameter and rung, for the multipole-baseline
redshift-space ladder; its drifts and extrapolated residuals are consistent
with the monopole-based curves of Fig.~\ref{fig:conv}, as expected from the
baseline-insensitivity of the higher-order blocks.

\begin{table}[htbp]
\centering
\caption{Convergence summary for the multipole-baseline
redshift-space rungs.\label{tab:conv}}

\begin{tabular}{lcccccccl}
\hline\hline
 & \multicolumn{3}{c}{$\{\xi_0,\xi_2\}{+}\zeta$} &
   \multicolumn{3}{c}{$\{\xi_0,\xi_2\}{+}\zeta{+}\zconn$} & & \\
\cline{2-4}\cline{5-7}
 & $\sigma$ & drift & resid. & $\sigma$ & drift & resid. & $g_4$ & status\\
\hline
$\Om$              & 0.016  & 4\%  & 7\%  & 0.014  & 6\%  & 11\% & 1.20 & intermediate\\
$\Ob$              & 0.0066 & 13\% & 15\% & 0.0047 & 13\% & 16\% & 1.39 & rising\\
$h$                & 0.062  & 12\% & 15\% & 0.043  & 13\% & 16\% & 1.44 & rising\\
$n_{\rm s}$        & 0.046  & 9\%  & 12\% & 0.034  & 10\% & 13\% & 1.34 & rising\\
$\sig$             & 0.027  & 4\%  & 9\%  & 0.019  & 5\%  & 9\% & 1.38 & near-converged\\
$\Mnu$   & 0.089  & 12\% & 18\% & 0.059  & 14\% & 17\% & 1.51 & un-converged\\
\hline\hline
\end{tabular}

\vspace{5pt}
\begin{minipage}{0.95\textwidth}
{\footnotesize
Ingredients common to every entry: covariance from $\Ncov=5{,}000$
realization-aligned fiducial boxes (LOS $z$); derivative ensembles of
$\Nderiv=500$ realizations per cosmology ($N$-point blocks averaged over 3
lines of sight, $1{,}500$ samples; quadrupole 500); central finite
differences at fixed initial phases for $\{\Om,\Ob,h,n_{\rm s},\sig\}$ and
the ZA-anchored one-sided schemes of Eq.~\eqref{eq:mnu} for $\Mnu$
(four-point default); MOPED score compression with shrinkage
$\lambda=0.1$ (as Table~\ref{tab:fisher}). $\sigma$: marginalized error at
the maximal derivative set. Drift: last-step rise of the convergence curve,
$1-\sigma(350)/\sigma(500)$, at fixed covariance. Resid.: remaining rise
implied by a linear $1/\Nderiv$ extrapolation,
$1-\sigma(500)/\sigma(\infty)$; both are lower limits to the residual
mirage since the curves are still climbing. $g_4$: connected-four-point
gain ratio at $\Nderiv=500$ (baseline-independent; \S\ref{sec:res-4pcf}).
}
\end{minipage}

\end{table}

What survives all of this is the {\em
relative} information gain, which is robust to the mirage precisely because the
mirage is {\em common} to the rungs being compared and cancels in their ratio.
The bottom row of Fig.~\ref{fig:conv} tracks the information-gain ratio
$\sigma_{2+3}/\sigma_{2+3+4}$ as a function of $\Nderiv$. A gain manufactured
purely by the mirage of the noisier four-point derivatives would fall toward
unity as $\Nderiv$ grows; instead the measured ratio holds at $\sim1.5$
for the neutrino direction and stays above $1.2$ in every direction across the full derivative set---to
$\Nderiv=500$ realizations in real space and, exploiting the three line-of-sight
axes, to $1500$ samples in redshift space, a $30\times$ span. Over that range it
declines by at most $\sim15\%$, and its step-to-step drift shrinks to a few
percent at the largest $\Nderiv$: the ratio {\em plateaus} well above the no-gain
limit rather than collapsing toward it. Crucially this cancellation holds even
for $\Mnu$: its gain ratio is nearly flat (declining by $\lesssim4\%$ over the
same span), so although the absolute error is scheme-dependent and un-converged,
the connected-four-point {\em gain} on $\Mnu$ is stable, and the four-point
function genuinely adds neutrino information even though the absolute
$\sigma(\Mnu)$ cannot yet be pinned down. The gain ratio is likewise
insensitive to the two-point baseline: computed over the multipole-based
$\{\xi_0,\xi_2\}{+}\zeta$ rung instead of the monopole-based one, it is
$1.2$--$1.5$ with the $\Mnu$ value essentially unchanged at $1.51$ versus $1.52$
(\S\ref{sec:res-4pcf}), a further indication that it measures four-point
information rather than any deficiency of the baseline. \rev{The same
diagnostic applied to the modelling-informed higher-order selection of
\S\ref{sec:res-cuts} shows the same behaviour: its three-point gain
$g_3(r_{\rm min}=40\Mpch)$ is flat over the last derivative step
($3.93\to3.90$ for $\Mnu$, a drift below $1\%$) and scheme-independent at
the per-cent level, with a linear $1/\Nderiv$ asymptote of $\simeq2.9$
(\S\ref{sec:res-cuts}).} The downward drift of the
weakest directions ($\Om$, $n_{\rm s}$) does not signal a collapse toward unity,
for two reasons. First, the ratio is
bounded below {\em by construction}: the $2{+}3{+}\zconn$ data vector contains
the $2{+}3$ vector, so adding the connected four-point block cannot reduce the
Fisher information and $g\ge1$ necessarily. The mirage inflates $g$ at finite
$\Nderiv$, so the ratio declines toward the truth {\em from above} and the trend
cannot invert; the worst case is that the four-point function adds nothing, never
that its apparent gain is spurious. Second, extrapolating each ratio to infinite
$\Nderiv$---with both a linear and a curvature-aware quadratic $1/\Nderiv$ fit,
since the weak directions do bend---and bootstrapping the seed-matched derivative
steps jointly, places {\em every} parameter's asymptote above unity even at the
conservative (quadratic) end: at $\ge11\sigma$ in redshift space and $\ge5\sigma$
in real space. The weak directions are the least certain in magnitude ($\Om$:
$g_\infty\simeq1.14$--$1.22$), but the neutrino gain that carries the physics is
flat and robust ($g_\infty\simeq1.5$). The connected-four-point information content---the central claim of this
work---is therefore real and not a finite-sample artifact, even while the
absolute errors on the weak directions remain lower bounds and $\Mnu$ itself is
not a reportable forecast.

\subsection{Binning}
\label{sec:rob-binning}
The absolute gains depend on the binning of each statistic, so we attach no
significance to any single gain factor. The constraints reported here use the
fixed measurement binning: $20$, $18$, and $5$ bins in pair separation out to
$150$, $100$, and $65\Mpch$ for the 2-, 3-, and 4-point functions respectively.
As a cross-check, the pipeline can thin the three-point configurations onto a
coarser sub-grid. Because this discards configurations it is conservative by
construction, {\em under}-stating the gain, so stability under coarsening bounds
the gain from below rather than pinning it. We defer a full binning-convergence
sweep to future work. 

\subsection{Compression of the connected four-point rung}
\label{sec:rob-moped}
We adopt score (MOPED) compression \citep{Heavens2000, AlsingWandelt2018} as our
primary estimator for the connected-four-point rung, and use the direct Fisher as
a cross-check. The motivation is conditioning, not invertibility: once degenerate
tetrahedra are excluded from the data vector (zero random count, $RRRR=0$;
\S\ref{sec:data}), the direct $2{+}3{+}\zconn$ covariance ($N_{\rm d}=1484$)
{\em does} invert, but at the current covariance size it carries large debiasing
corrections (Hartlap $0.70$, Percival $m_1=1.42$; \S\ref{sec:rob-cov})
that both inflate the parameter errors and inject covariance-estimation noise.
Compression re-locates the inversion: each realization is reduced to one score per
parameter, $t_a=\sum_{ij}D_{a,i}\,\mathsf{W}_{ij}\,(x_j-\mu_j)$, and only the
resulting $6\times6$ compressed covariance is estimated from the simulations and
inverted. Its Hartlap factor is $\approx1$ and no appreciable Percival inflation
is needed, so the compressed Fisher is well-conditioned where the full
$N_{\rm d}\times N_{\rm d}$ inverse is noisy. It is worth noting what this
does and does not buy. It removes the debiasing corrections attached to
inverting a $\sim$1500-dimensional sample covariance, because that
inversion is no longer performed. However, this does not remove the sampling
noise of the weight matrix $\mathsf{W}$, which is estimated from the same
finite, shrinkage-regularized fiducial ensemble, nor the noise of the
numerical derivatives that define the score directions; a noisy
$\mathsf{W}$ or $D$ yields a suboptimal, though still unbiased, set of
compression vectors, and the derivative noise propagates into the mirage
diagnosed in \S\ref{sec:rob-mirage}. Losslessness holds only in the limit
of well-estimated weights and derivatives and cannot be inferred from the
dimension of the score covariance alone; we therefore test the stability of
the compression directly, below. We take the weighting matrix to be a
shrinkage-regularized inverse covariance
\citep{LedoitWolf2004, PopeSzapudi2008},
$\mathsf{W}=[(1-\lambda)\hat{C}+\lambda\,\mathrm{diag}\,\hat{C}]^{-1}$. This
regularization is a separate ingredient from the compression itself, and it is
essential, because the unregularized sample precision $\hat{C}^{-1}$ reintroduces
the very covariance noise we are trying to avoid. The compression is lossless only in the limit of
a well-estimated covariance and well-estimated derivatives, and we verify that it reproduces the direct 2PCF and
$2{+}3$ constraints; the recovered connected-four-point information gain is stable
against the regularization, varying by less than $10\%$ as $\lambda$ ranges over
$[0.02,0.5]$, and we adopt $\lambda=0.1$. 

A useful consistency check comes from the direct
$2{+}3{+}\zconn$ Fisher, whose covariance is non-singular once the geometric
mask is applied. The marginalized errors of it agree with the compressed ones up
to a single overall factor of $\sim1.4$, the same for all six
parameters, just the Hartlap--Percival penalty attached to inverting the
full covariance, which does not arise for the $6\times6$ score covariance
and which itself relaxes toward unity as the fiducial ensemble grows. We
stress that this is a consistency check, not a proof of losslessness:
correlated covariance and derivative errors enter both pipelines, and the
shrinkage can impose common structure on the two estimates.
That the offset is a single global scaling, not a
parameter-dependent distortion, is consistent with compression discarding no parameter
information.

The direct test of the residual weight and derivative noise is a
split-half stability analysis, which we run on the multipole-baseline
redshift-space rungs of Table~\ref{tab:fisher}. (i) On the covariance side, we build $\mathsf{W}$ from one
interleaved half of the fiducial ensemble and apply the resulting
compression to the held-out half, estimating the score covariance only from
realizations never used for $\mathsf{W}$. The two split assignments agree
with each other to within ${\sim}3\%$ in every marginalized error and every
gain ratio, below $1\%$ for most entries. The compressed estimate is reproducible against the specific
realizations used to build the weights. At half the ensemble the errors are
${\sim}1.4\times$ larger than the full-ensemble result and the
connected-four-point gain is {\em smaller} ($g_4$ on $\Mnu$: $1.34$ versus
$1.51$). Weight noise degrades the compression and erodes the four-point
gain rather than manufacturing it, so the reported gain is conservative
with respect to the quality of $\mathsf{W}$. (ii) On the derivative side: 
rebuilding the score directions from each independent half of the
derivative ensembles shifts the absolute errors downward by $12$--$28\%$,
which is just the finite-$\Nderiv$ mirage expected at half the derivative
set (Table~\ref{tab:conv}) and is why the absolute errors carry the labels
of \S\ref{sec:rob-mirage}; the two halves agree with each other to
${\sim}5\%$, and the gain ratio is stable at the few-per-cent level
($g_4$ on $\Mnu$: $1.47$--$1.57$). (iii) {\em Shrinkage:} across
$\lambda\in[0.02,0.5]$ the errors vary by at most ${\sim}17\%$ and the gain
ratios by less than $7\%$, less than $3\%$ over $[0.02,0.2]$. The
constraints, and especially the rung-to-rung gains, are therefore stable
against the estimated ingredients of the compression. What is not stable, is the absolute-error dependence on $\Nderiv$, which is precisely what the
mirage analysis quantifies and what the non-converged labels flag.

\section{Discussion}
\label{sec:discussion}

\subsection{An independent route to higher-order information}
\label{sec:disc-route}
The configuration-space ladder reaches the same physics as the Fourier-space
bispectrum by an independent route: the breaking of the bias--amplitude and
$\sig$--$\Mnu$ degeneracies. To within the expected differences, its gains track
those of \citet{Hahn2020} (\S\ref{sec:res-hahn}). Working in
separation space brings three practical advantages. First, the baryon-acoustic
feature and the scales that carry the signal are {\em local} in pair separation,
so the information is not spread across a window-convolved range of wavenumbers.
Second, survey boundaries, masks, and selection functions are absorbed
transparently through the random catalogue, rather than through a window-matrix
deconvolution \citep{SzapudiSzalay1998, PhilcoxEisenstein2020, Sabiu2019}. Third, the
large-scale signal connects to the same analytic perturbation theory
that underlies Fourier-space modelling, through closed-form
configuration-space templates, so theory comparison proceeds natively in
separation space. Our
tree-level validation of the 3PCF demonstrates this (\S\ref{sec:theory}).
Methods that approach the field-level information content, such as the wavelet
scattering transform \citep{Valogiannis2022}, can extract more in total, but they
are harder to interpret analytically. The $N$-point ladder sits in between: it is
easy to interpret, it connects to theory, and it is ready for real surveys.

\subsection{Diminishing returns and the value of fourth order}
\label{sec:disc-returns}
A clear feature of the ladder is that the largest step is the first: the 3PCF
delivers most of the accessible higher-order information, tightening $\sig$ and
$\Mnu$ several-fold over the 2PCF. The {\em incremental} value of the connected
4PCF over the $2{+}3$ combination is the natural next question. Although the
{\em absolute} constraints at fourth order have not converged, the incremental
gain is a {\em ratio} of rungs, and the convergence test of
\S\ref{sec:rob-mirage} shows that this ratio is stable: the common finite-sample
mirage cancels between the numerator and denominator, so the connected four-point
function adds a real $\sim20$--$50\%$ tightening rather than a manufactured one.
That this increment is modest is itself informative: it shows that the
three-point function already captures most of the non-Gaussian information
available at these scales and this number density, and it sets a quantitative
ceiling on the cost/benefit of measuring a 632-dimensional four-point
data vector. Whether the increment grows into a more decisive gain at lower
redshift, higher number density, or fourth-order theory that we do not yet model
remains an open question.

\subsection{Comparison with previous studies}
\label{sec:compare-refs}

Several earlier studies have quantified the cosmological information carried by
higher-order statistics, and it is useful to place our configuration-space
results in that context. Using the Quijote $N$-body suite and Fisher forecasts,
\citet{Hahn2020} and \citet{HahnVillaescusa2021} established the constraining
power of the Fourier-space bispectrum, with particular attention to the summed
neutrino mass, and both found that adding the bispectrum to the power spectrum
tightens the constraints on the cosmological parameters in general and on $\Mnu$
in particular. \citet{Hahn2020} showed that in redshift space the halo
bispectrum responds differently to $\Mnu$ and $\sig$, so that varying the two
leaves distinct imprints and the bispectrum breaks the degeneracy that limits
the power spectrum. Quantitatively, at $k_{\max}=0.5\,h\,{\rm Mpc}^{-1}$ they
found the bispectrum to tighten the constraints on
$\{\Om,\Ob,h,n_{\rm s},\sig\}$ by factors of $1.9$, $2.6$, $3.1$, $3.6$, and
$2.6$ relative to the power spectrum, and that on $\Mnu$ by a factor of $5$.
\citet{HahnVillaescusa2021} extended this analysis to a halo-occupation
galaxy-bias model, computing the galaxy bispectrum into the nonlinear regime in
redshift space; combined with the power spectrum it tightened the same
parameters by $2.8$, $3.1$, $3.8$, $4.2$, and $4.2$, and $\Mnu$ by $4.6$. Even
under {\em Planck} priors the galaxy bispectrum delivered on average a factor of
${\sim}2$ improvement over the power spectrum, with the gain concentrated in the
nonlinear regime, $k_{\max}>0.2\,h\,{\rm Mpc}^{-1}$.

Whereas those analyses were performed on simulations,
\citet{DESIDR12_2026} applied the same approach to data, combining the DESI DR1
LRG bispectrum with DR2 BAO measurements. Following \citet{Hahn2020} and
\citet{HahnVillaescusa2021}, they worked with the redshift-space bispectrum
monopole, and by consistently modelling the cross-covariance between the two
datasets they showed that the DESI DR1 bispectrum sharpens the sensitivity to
the neutrino mass. In combination with CMB data the constraint tightens
substantially, reaching $\Sigma m_\nu < 0.059\,{\rm eV}$ at $95\%$ confidence.
They further reported that the bispectrum helps break the degeneracies among the
cosmological parameters, the growth rate $f$, and $\sig$, and that it shifts the
posteriors toward $\Lambda$CDM, weakening the evidence for time-varying dark
energy relative to the power-spectrum-only result. Although their parameter set
differs from ours, this illustrates the practical value of higher-order
clustering for survey data and motivates extending our configuration-space
ladder to higher redshift and to extended cosmological models.

Closest in spirit to the present work is \citet{Labate2026}, who
studied the neutrino-mass sensitivity of the three-point function directly in
configuration space. Using $2000$ Quijote $N$-body realizations at three
redshifts and four neutrino masses, they measured the connected and reduced 3PCF
with the estimator of \citet{SlepianEisenstein2015}, binning the 2PCF out to
$150\Mpch$ and the 3PCF out to $145\Mpch$. With a methodology and simulation set
close to ours, they examined how the neutrino imprint varies with triangle shape
and scale, finding that elongated triangles---which trace the filamentary
structures of the cosmic web---carry most of the neutrino sensitivity, with
complementary information from right-angled configurations in the reduced 3PCF.
On this basis they concluded that the configuration-space 3PCF, and not only the
bispectrum studied previously, can break the $\Mnu$--$\sig$ degeneracy,
consistent with what we find here.

\subsection{Caveats}
\label{sec:disc-caveats}
Several limitations should be kept in view. The forecast is a Fisher forecast, so
it assumes a Gaussian parameter likelihood and locally-linear derivatives, which
can be optimistic for a high-dimensional, manifestly non-Gaussian data vector; a
simulation-based-inference treatment would be needed to test that assumption. The
covariance and derivative estimates carry residual noise, which we quantify and
correct as far as possible in \S\ref{sec:robust}. The $\sig$ amplitude
constraint is nearly converged (a residual systematic of order $15\%$), but the
absolute constraints on the weak directions ($\Ob$, $h$, $n_{\rm s}$) and on
$\Mnu$ are still rising at the largest derivative set and should be read as lower
bounds rather than forecasts. The neutrino-mass response is the most sensitive of
all: it is obtained by finite differencing the massive-neutrino simulations, and
the resulting $\sigma(\Mnu)$ both fails to converge at the Quijote ceiling and
varies by $\sim2.3\times$ with the finite-difference scheme, with the noisier scheme
yielding the {\em tighter} error---the signature of a finite-sample mirage rather
than of genuine constraining power. We therefore treat the absolute $\Mnu$ error
as a scheme-dependent lower bound; the {\em relative} $\Mnu$ gains along the
ladder, which cancel this mirage, are robust. The analysis is of dark-matter {\em haloes} at a
single redshift ($z=0$) and a fixed number density. Holding $\bar n$ fixed
isolates clustering from abundance (\S\ref{sec:data}), but it treats the halo bias
as effectively known. The {\em absolute} constraints are therefore optimistic
relative to a galaxy survey that must marginalize the halo--galaxy connection.
The {\em relative} gains along the ladder are far more robust to this.
Finally, the measurements are made in a periodic box, free of the survey window,
mask, and fibre-collision effects of a real catalogue, and so far only the
3PCF, not the connected 4PCF, has been validated against an analytic model;
the scale-cut analysis of \S\ref{sec:res-cuts} shows the connected-four-point
increment to originate in configurations with at least one side below
${\sim}30\Mpch$, precisely where such a model is farthest out of reach, so
this rung should be read as a measurement-driven information statement
rather than a theory-backed forecast. These
are the natural targets of the extensions set out in
\S\ref{sec:conclusions}.

\section{Conclusions}
\label{sec:conclusions}
We have measured the configuration-space two-, three-, and connected four-point
correlation functions of the Quijote dark-matter halo field, in both real and
redshift space at $z=0$ and a fixed comoving number density, and we have used
them to forecast the cosmological information available on
$\{\Om,\Ob,h,n_{\rm s},\sig,\Mnu\}$. We separate the connected four-point
function from its disconnected Gaussian part on the fly, so the genuinely
non-Gaussian four-body signal is measured directly rather than by subtraction.
A tree-level perturbation-theory model reproduces the measured 3PCF over the
fitted range, with a galaxy bias consistent with that inferred from the 2PCF
(\S\ref{sec:theory}), which establishes analytic contact for the lowest
higher-order rung. Adding the 3PCF tightens every parameter, and most strongly
$\sig$ and $\Mnu$, through a configuration-space breaking of the degeneracy
between clustering amplitude, halo bias, and neutrino free-streaming. The
pattern of per-parameter gains tracks the Fourier-space halo bispectrum
\citep{Hahn2020}, with the three-point statistic helping $\sig$ and $\Mnu$
most, which confirms that configuration space offers an
independent and complementary route to the same higher-order information.
In redshift space the two-point baseline includes
the quadrupole, $\{\xi_0,\xi_2\}$, so the leading anisotropic (Kaiser)
information is counted at two-point order: the quadrupole strengthens the
two-point baseline by up to $4.4\times$ ($\sig$), the 3PCF gains referenced
to it are a more conservative $1.6$--$6.3$ (largest for $\Mnu$), and on this
more nearly matched footing the gains agree with the bispectrum-over-power-spectrum
factors of \citet{Hahn2020} to within factors of order unity, a consistency
check of the two routes.
Two results survive our full convergence analysis and carry the paper. The first is a
reportable amplitude constraint: the $2{+}3$ rung reaches
$\sigma(\sig)\simeq0.02$--$0.03$, nearly converged, with a residual
convergence systematic of order $15\%$; the full ladder tightens this to
$\sigma(\sig)\simeq0.017$ (real) / $0.019$ (redshift space,
quadrupole included). The second, and central, result is the
{\em information-content ladder} itself---the ordered, several-fold gains from
$2\rightarrow2{+}3\rightarrow2{+}3{+}\zconn$. These {\em relative} gains are robust:
the finite-sample tightening (the Fisher's mirage) is common to the rungs it
connects and cancels in the ratio (\S\ref{sec:rob-mirage}), so the connected
four-point function genuinely {\em adds} information---including on $\Mnu$---even
where the absolute errors have not converged. The neutrino mass is the honest
exception at the level of an absolute forecast. Its {\em relative} gains along the
ladder are real, but its {\em absolute} $\sigma(\Mnu)$ does not converge even at
the maximal Quijote derivative set and is strongly scheme-dependent, varying by
$\sim2.3$--$2.4$$\times$ between finite-difference schemes; the apparent tightness of the
highest-order scheme is largely a finite-sample mirage that grows with derivative
noise, so we quote $\sigma(\Mnu)$ as a {\em lower bound}
($\gtrsim0.1$~eV) rather than a forecast. We therefore report the $2$ and $2{+}3$
amplitude constraints and the full information-gain ladder as our robust findings,
and present every absolute $\Mnu$ constraint, together with the weak, still-rising
directions ($\Ob$, $h$, $n_{\rm s}$), as un-converged and hence conservative.

This analysis opens several extensions. The most immediate is to move from dark matter halos to mock galaxies. Populating the haloes with a
halo-occupation model and marginalizing its parameters would convert the present
halo forecast into a survey-realistic galaxy forecast, and moving from periodic
boxes to lightcones with realistic windows, redshift distributions, masks, and fibre collisions would
exercise the configuration-space estimators under survey conditions. An analytic
or emulated model for the connected 4PCF (so far only the 3PCF is
theory-validated) would extend analytic contact to fourth order, and a joint
configuration-plus-Fourier analysis would combine
the complementary strengths of the two spaces in a single inference. Stage-IV
surveys are now mapping the non-Gaussian regime, so a configuration-space ladder
of this kind, which is easy to interpret and connects to theory, is a useful
addition to the higher-order toolkit.

\begin{acknowledgments}
We thank the anonymous referee for their detailed suggestions and comments, which significantly improved this work.

CGS and IP  acknowledge support from the Basic Science Research Program \linebreak
(2018\-R1A6A1A06024977) through Korea's NRF funded by the Ministry of Education.

This work was also supported by the UBAI computing resources at the University of Seoul.

This manuscript was written with the aid of AI, specifically Claude Opus 4.8 and Fable. Generated text was approved by all co-authors, who take full responsibility for the contents of this work.

\end{acknowledgments}

\bibliographystyle{JHEP}
\bibliography{refs}

@ARTICLE{SzapudiSzalay1998,
       author = {{Szapudi}, Istv{\'a}n and {Szalay}, Alexander S.},
        title = "{A New Class of Estimators for the N-Point Correlations}",
      journal = {\apjl},
         year = 1998,
        month = feb,
       volume = {494},
       number = {1},
        pages = {L41-L44},
          doi = {10.1086/311146},
       eprint = {astro-ph/9704241},
 archivePrefix = {arXiv},
   primaryClass = {astro-ph},
       adsurl = {https://ui.adsabs.harvard.edu/abs/1998ApJ...494L..41S}
}

@ARTICLE{SlepianEisenstein2015,
       author = {{Slepian}, Zachary and {Eisenstein}, Daniel J.},
        title = "{Computing the three-point correlation function of galaxies in O(N\^2) time}",
      journal = {\mnras},
         year = 2015,
        month = dec,
       volume = {454},
       number = {4},
        pages = {4142-4158},
          doi = {10.1093/mnras/stv2119},
archivePrefix = {arXiv},
       eprint = {1506.02040},
 primaryClass = {astro-ph.CO},
       adsurl = {https://ui.adsabs.harvard.edu/abs/2015MNRAS.454.4142S}
}

@ARTICLE{SlepianEisenstein2018,
       author = {{Slepian}, Zachary and {Eisenstein}, Daniel J.},
        title = "{A practical computational method for the anisotropic redshift-space three-point correlation function}",
      journal = {\mnras},
         year = 2018,
        month = aug,
       volume = {478},
       number = {2},
        pages = {1468-1483},
          doi = {10.1093/mnras/sty1063},
archivePrefix = {arXiv},
       eprint = {1709.10150},
 primaryClass = {astro-ph.CO},
       adsurl = {https://ui.adsabs.harvard.edu/abs/2018MNRAS.478.1468S}
}

@ARTICLE{Slepian2017BAO,
       author = {{Slepian}, Zachary and {Eisenstein}, Daniel J. and {Brownstein}, Joel R. and {Chuang}, Chia-Hsun and {Gil-Mar{\'\i}n}, H{\'e}ctor and {Ho}, Shirley and {Kitaura}, Francisco-Shu and {Percival}, Will J. and {Ross}, Ashley J. and {Rossi}, Graziano and {Seo}, Hee-Jong and {Slosar}, An{\v{z}}e and {Vargas-Maga{\~n}a}, Mariana},
        title = "{Detection of baryon acoustic oscillation features in the large-scale three-point correlation function of SDSS BOSS DR12 CMASS galaxies}",
      journal = {\mnras},
         year = 2017,
        month = aug,
       volume = {469},
       number = {2},
        pages = {1738-1751},
          doi = {10.1093/mnras/stx488},
archivePrefix = {arXiv},
       eprint = {1607.06097},
 primaryClass = {astro-ph.CO},
       adsurl = {https://ui.adsabs.harvard.edu/abs/2017MNRAS.469.1738S}
}

@ARTICLE{PhilcoxEisenstein2020,
       author = {{Philcox}, Oliver H.~E. and {Eisenstein}, Daniel J.},
        title = "{Computing the small-scale galaxy power spectrum and bispectrum in configuration space}",
      journal = {\mnras},
         year = 2020,
        month = feb,
       volume = {492},
       number = {1},
        pages = {1214-1242},
          doi = {10.1093/mnras/stz3335},
archivePrefix = {arXiv},
       eprint = {1912.01010},
 primaryClass = {astro-ph.CO},
       adsurl = {https://ui.adsabs.harvard.edu/abs/2020MNRAS.492.1214P}
}

@ARTICLE{GaztanagaScoccimarro2005,
       author = {{Gazta{\~n}aga}, E. and {Scoccimarro}, R.},
        title = "{The three-point function in large-scale structure: redshift distortions and galaxy bias}",
      journal = {\mnras},
         year = 2005,
        month = aug,
       volume = {361},
       number = {3},
        pages = {824-836},
          doi = {10.1111/j.1365-2966.2005.09234.x},
archivePrefix = {arXiv},
       eprint = {astro-ph/0501637},
 primaryClass = {astro-ph},
       adsurl = {https://ui.adsabs.harvard.edu/abs/2005MNRAS.361..824G}
}

@BOOK{Peebles1980,
       author = {{Peebles}, P.~J.~E.},
        title = "{The large-scale structure of the universe}",
         year = 1980,
         publisher = "{Princeton University Press}",
       adsurl = {https://ui.adsabs.harvard.edu/abs/1980lssu.book.....P}
}

@ARTICLE{Sabiu2019,
       author = {{Sabiu}, Cristiano G. and {Hoyle}, Ben and {Kim}, Juhan and {Li}, Xiao-Dong},
        title = "{Graph Database Solution for Higher-order Spatial Statistics in the Era of Big Data}",
      journal = {\apjs},
         year = 2019,
        month = jun,
       volume = {242},
       number = {2},
          eid = {29},
        pages = {29},
          doi = {10.3847/1538-4365/ab22b5},
archivePrefix = {arXiv},
       eprint = {1901.00296},
 primaryClass = {astro-ph.CO},
       adsurl = {https://ui.adsabs.harvard.edu/abs/2019ApJS..242...29S}
}

@ARTICLE{Philcox2022encore,
       author = {{Philcox}, Oliver H.~E. and {Slepian}, Zachary and {Hou}, Jiamin and {Warner}, Craig and {Cahn}, Robert N. and {Eisenstein}, Daniel J.},
        title = "{ENCORE: an O (N$_{g}$$^{2}$) estimator for galaxy N-point correlation functions}",
      journal = {\mnras},
         year = 2022,
        month = jan,
       volume = {509},
       number = {2},
        pages = {2457-2481},
          doi = {10.1093/mnras/stab3025},
archivePrefix = {arXiv},
       eprint = {2105.08722},
 primaryClass = {astro-ph.IM},
       adsurl = {https://ui.adsabs.harvard.edu/abs/2022MNRAS.509.2457P}
}

@ARTICLE{Philcox2022parity,
       author = {{Philcox}, Oliver H.~E.},
        title = "{Probing parity violation with the four-point correlation function of BOSS galaxies}",
      journal = {\prd},
         year = 2022,
        month = sep,
       volume = {106},
       number = {6},
          eid = {063501},
        pages = {063501},
          doi = {10.1103/PhysRevD.106.063501},
archivePrefix = {arXiv},
       eprint = {2206.04227},
 primaryClass = {astro-ph.CO},
       adsurl = {https://ui.adsabs.harvard.edu/abs/2022PhRvD.106f3501P}
}

@ARTICLE{Hou2023parity,
       author = {{Hou}, Jiamin and {Slepian}, Zachary and {Cahn}, Robert N.},
        title = "{Measurement of parity-odd modes in the large-scale 4-point correlation function of Sloan Digital Sky Survey Baryon Oscillation Spectroscopic Survey twelfth data release CMASS and LOWZ galaxies}",
      journal = {\mnras},
         year = 2023,
        month = may,
       volume = {522},
       number = {4},
        pages = {5701-5739},
          doi = {10.1093/mnras/stad1062},
archivePrefix = {arXiv},
       eprint = {2206.03625},
 primaryClass = {astro-ph.CO},
       adsurl = {https://ui.adsabs.harvard.edu/abs/2023MNRAS.522.5701H}
}

@ARTICLE{CahnSlepianHou2023,
       author = {{Cahn}, Robert N. and {Slepian}, Zachary and {Hou}, Jiamin},
        title = "{Test for Cosmological Parity Violation Using the 3D Distribution of Galaxies}",
      journal = {\prl},
         year = 2023,
        month = may,
       volume = {130},
       number = {20},
          eid = {201002},
        pages = {201002},
          doi = {10.1103/PhysRevLett.130.201002},
archivePrefix = {arXiv},
       eprint = {2110.12004},
 primaryClass = {astro-ph.CO},
       adsurl = {https://ui.adsabs.harvard.edu/abs/2023PhRvL.130t1002C}
}

@ARTICLE{Bertolini2016,
       author = {{Bertolini}, Daniele and {Schutz}, Katelin and {Solon}, Mikhail P. and {Zurek}, Kathryn M.},
        title = "{The trispectrum in the Effective Field Theory of Large Scale Structure}",
      journal = {\jcap},
         year = 2016,
        month = jun,
       volume = {2016},
       number = {6},
          eid = {052},
        pages = {052},
          doi = {10.1088/1475-7516/2016/06/052},
archivePrefix = {arXiv},
       eprint = {1604.01770},
 primaryClass = {astro-ph.CO},
       adsurl = {https://ui.adsabs.harvard.edu/abs/2016JCAP...06..052B}
}

@ARTICLE{Gualdi2021,
       author = {{Gualdi}, Davide and {Gil-Mar{\'\i}n}, H{\'e}ctor and {Verde}, Licia},
        title = "{Joint analysis of anisotropic power spectrum, bispectrum and trispectrum: application to N-body simulations}",
      journal = {\jcap},
         year = 2021,
        month = jul,
       volume = {2021},
       number = {7},
          eid = {008},
        pages = {008},
          doi = {10.1088/1475-7516/2021/07/008},
archivePrefix = {arXiv},
       eprint = {2104.03976},
 primaryClass = {astro-ph.CO},
       adsurl = {https://ui.adsabs.harvard.edu/abs/2021JCAP...07..008G}
}

@ARTICLE{Fry1984,
       author = {{Fry}, J.~N.},
        title = "{The Galaxy correlation hierarchy in perturbation theory}",
      journal = {\apj},
         year = 1984,
        month = apr,
       volume = {279},
        pages = {499-510},
          doi = {10.1086/161913},
       adsurl = {https://ui.adsabs.harvard.edu/abs/1984ApJ...279..499F}
}

@ARTICLE{Sefusatti2006,
       author = {{Sefusatti}, Emiliano and {Crocce}, Mart{\'\i}n and {Pueblas}, Sebasti{\'a}n and {Scoccimarro}, Rom{\'a}n},
        title = "{Cosmology and the bispectrum}",
      journal = {\prd},
         year = 2006,
        month = jul,
       volume = {74},
       number = {2},
          eid = {023522},
        pages = {023522},
          doi = {10.1103/PhysRevD.74.023522},
archivePrefix = {arXiv},
       eprint = {astro-ph/0604505},
 primaryClass = {astro-ph},
       adsurl = {https://ui.adsabs.harvard.edu/abs/2006PhRvD..74b3522S}
}

@ARTICLE{GilMarin2017,
       author = {{Gil-Mar{\'\i}n}, H{\'e}ctor and {Percival}, Will J. and {Verde}, Licia and {Brownstein}, Joel R. and {Chuang}, Chia-Hsun and {Kitaura}, Francisco-Shu and {Rodr{\'\i}guez-Torres}, Sergio A. and {Olmstead}, Matthew D.},
        title = "{The clustering of galaxies in the SDSS-III Baryon Oscillation Spectroscopic Survey: RSD measurement from the power spectrum and bispectrum of the DR12 BOSS galaxies}",
      journal = {\mnras},
         year = 2017,
        month = feb,
       volume = {465},
       number = {2},
        pages = {1757-1788},
          doi = {10.1093/mnras/stw2679},
archivePrefix = {arXiv},
       eprint = {1606.00439},
 primaryClass = {astro-ph.CO},
       adsurl = {https://ui.adsabs.harvard.edu/abs/2017MNRAS.465.1757G}
}

@ARTICLE{Hahn2020,
       author = {{Hahn}, ChangHoon and {Villaescusa-Navarro}, Francisco and {Castorina}, Emanuele and {Scoccimarro}, Roman},
        title = "{Constraining M$_{{\ensuremath{\nu}}}$ with the bispectrum. Part I. Breaking parameter degeneracies}",
      journal = {\jcap},
         year = 2020,
        month = mar,
       volume = {2020},
       number = {3},
          eid = {040},
        pages = {040},
          doi = {10.1088/1475-7516/2020/03/040},
archivePrefix = {arXiv},
       eprint = {1909.11107},
 primaryClass = {astro-ph.CO},
       adsurl = {https://ui.adsabs.harvard.edu/abs/2020JCAP...03..040H}
}

@ARTICLE{HahnVillaescusa2021,
       author = {{Hahn}, ChangHoon and {Villaescusa-Navarro}, Francisco},
        title = "{Constraining M$_{{\ensuremath{\nu}}}$ with the bispectrum. Part II. The information content of the galaxy bispectrum monopole}",
      journal = {\jcap},
         year = 2021,
        month = apr,
       volume = {2021},
       number = {4},
          eid = {029},
        pages = {029},
          doi = {10.1088/1475-7516/2021/04/029},
archivePrefix = {arXiv},
       eprint = {2012.02200},
 primaryClass = {astro-ph.CO},
       adsurl = {https://ui.adsabs.harvard.edu/abs/2021JCAP...04..029H}
}

@ARTICLE{Massara2021,
       author = {{Massara}, Elena and {Villaescusa-Navarro}, Francisco and {Ho}, Shirley and {Dalal}, Neal and {Spergel}, David N.},
        title = "{Using the Marked Power Spectrum to Detect the Signature of Neutrinos in Large-Scale Structure}",
      journal = {\prl},
         year = 2021,
        month = jan,
       volume = {126},
       number = {1},
          eid = {011301},
        pages = {011301},
          doi = {10.1103/PhysRevLett.126.011301},
archivePrefix = {arXiv},
       eprint = {2001.11024},
 primaryClass = {astro-ph.CO},
       adsurl = {https://ui.adsabs.harvard.edu/abs/2021PhRvL.126a1301M}
}

@ARTICLE{Valogiannis2022,
       author = {{Valogiannis}, Georgios and {Dvorkin}, Cora},
        title = "{Towards an optimal estimation of cosmological parameters with the wavelet scattering transform}",
      journal = {\prd},
         year = 2022,
        month = may,
       volume = {105},
       number = {10},
          eid = {103534},
        pages = {103534},
          doi = {10.1103/PhysRevD.105.103534},
archivePrefix = {arXiv},
       eprint = {2108.07821},
 primaryClass = {astro-ph.CO},
       adsurl = {https://ui.adsabs.harvard.edu/abs/2022PhRvD.105j3534V}
}

@ARTICLE{BanerjeeAbel2021,
       author = {{Banerjee}, Arka and {Abel}, Tom},
        title = "{Nearest neighbour distributions: New statistical measures for cosmological clustering}",
      journal = {\mnras},
         year = 2021,
        month = jan,
       volume = {500},
       number = {4},
        pages = {5479-5499},
          doi = {10.1093/mnras/staa3604},
archivePrefix = {arXiv},
       eprint = {2007.13342},
 primaryClass = {astro-ph.CO},
       adsurl = {https://ui.adsabs.harvard.edu/abs/2021MNRAS.500.5479B}
}

@ARTICLE{GilMarin2012,
       author = {{Gil-Mar{\'\i}n}, H{\'e}ctor and {Wagner}, Christian and {Fragkoudi}, Frantzeska and {Jimenez}, Raul and {Verde}, Licia},
        title = "{An improved fitting formula for the dark matter bispectrum}",
      journal = {\jcap},
         year = 2012,
        month = feb,
       volume = {2012},
       number = {2},
          eid = {047},
        pages = {047},
          doi = {10.1088/1475-7516/2012/02/047},
archivePrefix = {arXiv},
       eprint = {1111.4477},
 primaryClass = {astro-ph.CO},
       adsurl = {https://ui.adsabs.harvard.edu/abs/2012JCAP...02..047G}
}

@ARTICLE{VillaescusaNavarro2018,
       author = {{Villaescusa-Navarro}, Francisco and {Banerjee}, Arka and {Dalal}, Neal and {Castorina}, Emanuele and {Scoccimarro}, Roman and {Angulo}, Raul and {Spergel}, David N.},
        title = "{The Imprint of Neutrinos on Clustering in Redshift Space}",
      journal = {\apj},
         year = 2018,
        month = jul,
       volume = {861},
       number = {1},
          eid = {53},
        pages = {53},
          doi = {10.3847/1538-4357/aac6bf},
archivePrefix = {arXiv},
       eprint = {1708.01154},
 primaryClass = {astro-ph.CO},
       adsurl = {https://ui.adsabs.harvard.edu/abs/2018ApJ...861...53V}
}

@ARTICLE{Bayer2021,
       author = {{Bayer}, Adrian E. and {Villaescusa-Navarro}, Francisco and {Massara}, Elena and {Liu}, Jia and {Spergel}, David N. and {Verde}, Licia and {Wandelt}, Benjamin D. and {Viel}, Matteo and {Ho}, Shirley},
        title = "{Detecting Neutrino Mass by Combining Matter Clustering, Halos, and Voids}",
      journal = {\apj},
         year = 2021,
        month = sep,
       volume = {919},
       number = {1},
          eid = {24},
        pages = {24},
          doi = {10.3847/1538-4357/ac0e91},
archivePrefix = {arXiv},
       eprint = {2102.05049},
 primaryClass = {astro-ph.CO},
       adsurl = {https://ui.adsabs.harvard.edu/abs/2021ApJ...919...24B}
}

@ARTICLE{VillaescusaNavarro2020,
       author = {{Villaescusa-Navarro}, Francisco and {Hahn}, ChangHoon and {Massara}, Elena and {Banerjee}, Arka and {Delgado}, Ana Maria and {Ramanah}, Doogesh Kodi and {Charnock}, Tom and {Giusarma}, Elena and {Li}, Yin and {Allys}, Erwan and {Brochard}, Antoine and {Uhlemann}, Cora and {Chiang}, Chi-Ting and {He}, Siyu and {Pisani}, Alice and {Obuljen}, Andrej and {Feng}, Yu and {Castorina}, Emanuele and {Contardo}, Gabriella and {Kreisch}, Christina D. and {Nicola}, Andrina and {Alsing}, Justin and {Scoccimarro}, Roman and {Verde}, Licia and {Viel}, Matteo and {Ho}, Shirley and {Mallat}, Stephane and {Wandelt}, Benjamin and {Spergel}, David N.},
        title = "{The Quijote Simulations}",
      journal = {\apjs},
         year = 2020,
        month = sep,
       volume = {250},
       number = {1},
          eid = {2},
        pages = {2},
          doi = {10.3847/1538-4365/ab9d82},
archivePrefix = {arXiv},
       eprint = {1909.05273},
 primaryClass = {astro-ph.CO},
       adsurl = {https://ui.adsabs.harvard.edu/abs/2020ApJS..250....2V}
}

@ARTICLE{LesgourguesPastor2006,
       author = {{Lesgourgues}, Julien and {Pastor}, Sergio},
        title = "{Massive neutrinos and cosmology}",
      journal = {\physrep},
         year = 2006,
        month = jul,
       volume = {429},
       number = {6},
        pages = {307-379},
          doi = {10.1016/j.physrep.2006.04.001},
archivePrefix = {arXiv},
       eprint = {astro-ph/0603494},
 primaryClass = {astro-ph},
       adsurl = {https://ui.adsabs.harvard.edu/abs/2006PhR...429..307L}
}

@ARTICLE{DESI2016,
       author = {{DESI Collaboration} and {Aghamousa}, Amir and {Aguilar}, Jessica and {Ahlen}, Steve and {Alam}, Shadab and {Allen}, Lori E. and {Allende Prieto}, Carlos and {Annis}, James and {Bailey}, Stephen and {Balland}, Christophe and {Ballester}, Otger and {Baltay}, Charles and {Beaufore}, Lucas and {Bebek}, Chris and {Beers}, Timothy C. and {Bell}, Eric F. and {Bernal}, Jos{\'e} Luis and {Besuner}, Robert and {Beutler}, Florian and {Blake}, Chris and {Bleuler}, Hannes and {Blomqvist}, Michael and {Blum}, Robert and {Bolton}, Adam S. and {Briceno}, Cesar and {Brooks}, David and {Brownstein}, Joel R. and {Buckley-Geer}, Elizabeth and {Burden}, Angela and {Burtin}, Etienne and {Busca}, Nicolas G. and {Cahn}, Robert N. and {Cai}, Yan-Chuan and {Cardiel-Sas}, Laia and {Carlberg}, Raymond G. and {Carton}, Pierre-Henri and {Casas}, Ricard and {Castander}, Francisco J. and {Cervantes-Cota}, Jorge L. and {Claybaugh}, Todd M. and {Close}, Madeline and {Coker}, Carl T. and {Cole}, Shaun and {Comparat}, Johan and {Cooper}, Andrew P. and {Cousinou}, M.-C. and {Crocce}, Martin and {Cuby}, Jean-Gabriel and {Cunningham}, Daniel P. and {Davis}, Tamara M. and {Dawson}, Kyle S. and {de la Macorra}, Axel and {De Vicente}, Juan and {Delubac}, Timoth{\'e}e and {Derwent}, Mark and {Dey}, Arjun and {Dhungana}, Govinda and {Ding}, Zhejie and {Doel}, Peter and {Duan}, Yutong T. and {Ealet}, Anne and {Edelstein}, Jerry and {Eftekharzadeh}, Sarah and {Eisenstein}, Daniel J. and {Elliott}, Ann and {Escoffier}, St{\'e}phanie and {Evatt}, Matthew and {Fagrelius}, Parker and {Fan}, Xiaohui and {Fanning}, Kevin and {Farahi}, Arya and {Farihi}, Jay and {Favole}, Ginevra and {Feng}, Yu and {Fernandez}, Enrique and {Findlay}, Joseph R. and {Finkbeiner}, Douglas P. and {Fitzpatrick}, Michael J. and {Flaugher}, Brenna and {Flender}, Samuel and {Font-Ribera}, Andreu and {Forero-Romero}, Jaime E. and {Fosalba}, Pablo and {Frenk}, Carlos S. and {Fumagalli}, Michele and {Gaensicke}, Boris T. and {Gallo}, Giuseppe and {Garcia-Bellido}, Juan and {Gaztanaga}, Enrique and {Pietro Gentile Fusillo}, Nicola and {Gerard}, Terry and {Gershkovich}, Irena and {Giannantonio}, Tommaso and {Gillet}, Denis and {Gonzalez-de-Rivera}, Guillermo and {Gonzalez-Perez}, Violeta and {Gott}, Shelby and {Graur}, Or and {Gutierrez}, Gaston and {Guy}, Julien and {Habib}, Salman and {Heetderks}, Henry and {Heetderks}, Ian and {Heitmann}, Katrin and {Hellwing}, Wojciech A. and {Herrera}, David A. and {Ho}, Shirley and {Holland}, Stephen and {Honscheid}, Klaus and {Huff}, Eric and {Hutchinson}, Timothy A. and {Huterer}, Dragan and {Hwang}, Ho Seong and {Illa Laguna}, Joseph Maria and {Ishikawa}, Yuzo and {Jacobs}, Dianna and {Jeffrey}, Niall and {Jelinsky}, Patrick and {Jennings}, Elise and {Jiang}, Linhua and {Jimenez}, Jorge and {Johnson}, Jennifer and {Joyce}, Richard and {Jullo}, Eric and {Juneau}, St{\'e}phanie and {Kama}, Sami and {Karcher}, Armin and {Karkar}, Sonia and {Kehoe}, Robert and {Kennamer}, Noble and {Kent}, Stephen and {Kilbinger}, Martin and {Kim}, Alex G. and {Kirkby}, David and {Kisner}, Theodore and {Kitanidis}, Ellie and {Kneib}, Jean-Paul and {Koposov}, Sergey and {Kovacs}, Eve and {Koyama}, Kazuya and {Kremin}, Anthony and {Kron}, Richard and {Kronig}, Luzius and {Kueter-Young}, Andrea and {Lacey}, Cedric G. and {Lafever}, Robin and {Lahav}, Ofer and {Lambert}, Andrew and {Lampton}, Michael and {Landriau}, Martin and {Lang}, Dustin and {Lauer}, Tod R. and {Le Goff}, Jean-Marc and {Le Guillou}, Laurent and {Le Van Suu}, Auguste and {Lee}, Jae Hyeon and {Lee}, Su-Jeong and {Leitner}, Daniela and {Lesser}, Michael and {Levi}, Michael E. and {L'Huillier}, Benjamin and {Li}, Baojiu and {Liang}, Ming and {Lin}, Huan and {Linder}, Eric and {Loebman}, Sarah R. and {Luki{\'c}}, Zarija and {Ma}, Jun and {MacCrann}, Niall and {Magneville}, Christophe and {Makarem}, Laleh and {Manera}, Marc and {Manser}, Christopher J. and {Marshall}, Robert and {Martini}, Paul and {Massey}, Richard and {Matheson}, Thomas and {McCauley}, Jeremy and {McDonald}, Patrick and {McGreer}, Ian D. and {Meisner}, Aaron and {Metcalfe}, Nigel and {Miller}, Timothy N. and {Miquel}, Ramon and {Moustakas}, John and {Myers}, Adam and {Naik}, Milind and {Newman}, Jeffrey A. and {Nichol}, Robert C. and {Nicola}, Andrina and {Nicolati da Costa}, Luiz and {Nie}, Jundan and {Niz}, Gustavo and {Norberg}, Peder and {Nord}, Brian and {Norman}, Dara and {Nugent}, Peter and {O'Brien}, Thomas and {Oh}, Minji and {Olsen}, Knut A.~G.},
        title = "{The DESI Experiment Part I: Science,Targeting, and Survey Design}",
      journal = {arXiv e-prints},
         year = 2016,
        month = oct,
          eid = {arXiv:1611.00036},
        pages = {arXiv:1611.00036},
          doi = {10.48550/arXiv.1611.00036},
archivePrefix = {arXiv},
       eprint = {1611.00036},
 primaryClass = {astro-ph.IM},
       adsurl = {https://ui.adsabs.harvard.edu/abs/2016arXiv161100036D}
}

@ARTICLE{Laureijs2011,
       author = {{Laureijs}, R. and {Amiaux}, J. and {Arduini}, S. and {Augu{\`e}res}, J.  -L. and {Brinchmann}, J. and {Cole}, R. and {Cropper}, M. and {Dabin}, C. and {Duvet}, L. and {Ealet}, A. and {Garilli}, B. and {Gondoin}, P. and {Guzzo}, L. and {Hoar}, J. and {Hoekstra}, H. and {Holmes}, R. and {Kitching}, T. and {Maciaszek}, T. and {Mellier}, Y. and {Pasian}, F. and {Percival}, W. and {Rhodes}, J. and {Saavedra Criado}, G. and {Sauvage}, M. and {Scaramella}, R. and {Valenziano}, L. and {Warren}, S. and {Bender}, R. and {Castander}, F. and {Cimatti}, A. and {Le F{\`e}vre}, O. and {Kurki-Suonio}, H. and {Levi}, M. and {Lilje}, P. and {Meylan}, G. and {Nichol}, R. and {Pedersen}, K. and {Popa}, V. and {Rebolo Lopez}, R. and {Rix}, H.  -W. and {Rottgering}, H. and {Zeilinger}, W. and {Grupp}, F. and {Hudelot}, P. and {Massey}, R. and {Meneghetti}, M. and {Miller}, L. and {Paltani}, S. and {Paulin-Henriksson}, S. and {Pires}, S. and {Saxton}, C. and {Schrabback}, T. and {Seidel}, G. and {Walsh}, J. and {Aghanim}, N. and {Amendola}, L. and {Bartlett}, J. and {Baccigalupi}, C. and {Beaulieu}, J.  -P. and {Benabed}, K. and {Cuby}, J.  -G. and {Elbaz}, D. and {Fosalba}, P. and {Gavazzi}, G. and {Helmi}, A. and {Hook}, I. and {Irwin}, M. and {Kneib}, J.  -P. and {Kunz}, M. and {Mannucci}, F. and {Moscardini}, L. and {Tao}, C. and {Teyssier}, R. and {Weller}, J. and {Zamorani}, G. and {Zapatero Osorio}, M.~R. and {Boulade}, O. and {Foumond}, J.~J. and {Di Giorgio}, A. and {Guttridge}, P. and {James}, A. and {Kemp}, M. and {Martignac}, J. and {Spencer}, A. and {Walton}, D. and {Bl{\"u}mchen}, T. and {Bonoli}, C. and {Bortoletto}, F. and {Cerna}, C. and {Corcione}, L. and {Fabron}, C. and {Jahnke}, K. and {Ligori}, S. and {Madrid}, F. and {Martin}, L. and {Morgante}, G. and {Pamplona}, T. and {Prieto}, E. and {Riva}, M. and {Toledo}, R. and {Trifoglio}, M. and {Zerbi}, F. and {Abdalla}, F. and {Douspis}, M. and {Grenet}, C. and {Borgani}, S. and {Bouwens}, R. and {Courbin}, F. and {Delouis}, J.  -M. and {Dubath}, P. and {Fontana}, A. and {Frailis}, M. and {Grazian}, A. and {Koppenh{\"o}fer}, J. and {Mansutti}, O. and {Melchior}, M. and {Mignoli}, M. and {Mohr}, J. and {Neissner}, C. and {Noddle}, K. and {Poncet}, M. and {Scodeggio}, M. and {Serrano}, S. and {Shane}, N. and {Starck}, J.  -L. and {Surace}, C. and {Taylor}, A. and {Verdoes-Kleijn}, G. and {Vuerli}, C. and {Williams}, O.~R. and {Zacchei}, A. and {Altieri}, B. and {Escudero Sanz}, I. and {Kohley}, R. and {Oosterbroek}, T. and {Astier}, P. and {Bacon}, D. and {Bardelli}, S. and {Baugh}, C. and {Bellagamba}, F. and {Benoist}, C. and {Bianchi}, D. and {Biviano}, A. and {Branchini}, E. and {Carbone}, C. and {Cardone}, V. and {Clements}, D. and {Colombi}, S. and {Conselice}, C. and {Cresci}, G. and {Deacon}, N. and {Dunlop}, J. and {Fedeli}, C. and {Fontanot}, F. and {Franzetti}, P. and {Giocoli}, C. and {Garcia-Bellido}, J. and {Gow}, J. and {Heavens}, A. and {Hewett}, P. and {Heymans}, C. and {Holland}, A. and {Huang}, Z. and {Ilbert}, O. and {Joachimi}, B. and {Jennins}, E. and {Kerins}, E. and {Kiessling}, A. and {Kirk}, D. and {Kotak}, R. and {Krause}, O. and {Lahav}, O. and {van Leeuwen}, F. and {Lesgourgues}, J. and {Lombardi}, M. and {Magliocchetti}, M. and {Maguire}, K. and {Majerotto}, E. and {Maoli}, R. and {Marulli}, F. and {Maurogordato}, S. and {McCracken}, H. and {McLure}, R. and {Melchiorri}, A. and {Merson}, A. and {Moresco}, M. and {Nonino}, M. and {Norberg}, P. and {Peacock}, J. and {Pello}, R. and {Penny}, M. and {Pettorino}, V. and {Di Porto}, C. and {Pozzetti}, L. and {Quercellini}, C. and {Radovich}, M. and {Rassat}, A. and {Roche}, N. and {Ronayette}, S. and {Rossetti}, E.},
        title = "{Euclid Definition Study Report}",
      journal = {arXiv e-prints},
         year = 2011,
        month = oct,
          eid = {arXiv:1110.3193},
        pages = {arXiv:1110.3193},
          doi = {10.48550/arXiv.1110.3193},
archivePrefix = {arXiv},
       eprint = {1110.3193},
 primaryClass = {astro-ph.CO},
       adsurl = {https://ui.adsabs.harvard.edu/abs/2011arXiv1110.3193L}
}

@ARTICLE{Dore2014,
       author = {{Dor{\'e}}, Olivier and {Bock}, Jamie and {Ashby}, Matthew and {Capak}, Peter and {Cooray}, Asantha and {de Putter}, Roland and {Eifler}, Tim and {Flagey}, Nicolas and {Gong}, Yan and {Habib}, Salman and {Heitmann}, Katrin and {Hirata}, Chris and {Jeong}, Woong-Seob and {Katti}, Raj and {Korngut}, Phil and {Krause}, Elisabeth and {Lee}, Dae-Hee and {Masters}, Daniel and {Mauskopf}, Phil and {Melnick}, Gary and {Mennesson}, Bertrand and {Nguyen}, Hien and {{\"O}berg}, Karin and {Pullen}, Anthony and {Raccanelli}, Alvise and {Smith}, Roger and {Song}, Yong-Seon and {Tolls}, Volker and {Unwin}, Steve and {Venumadhav}, Tejaswi and {Viero}, Marco and {Werner}, Mike and {Zemcov}, Mike},
        title = "{Cosmology with the SPHEREX All-Sky Spectral Survey}",
      journal = {arXiv e-prints},
         year = 2014,
        month = dec,
          eid = {arXiv:1412.4872},
        pages = {arXiv:1412.4872},
          doi = {10.48550/arXiv.1412.4872},
archivePrefix = {arXiv},
       eprint = {1412.4872},
 primaryClass = {astro-ph.CO},
       adsurl = {https://ui.adsabs.harvard.edu/abs/2014arXiv1412.4872D}
}

@ARTICLE{Hartlap2007,
       author = {{Hartlap}, J. and {Simon}, P. and {Schneider}, P.},
        title = "{Why your model parameter confidences might be too optimistic. Unbiased estimation of the inverse covariance matrix}",
      journal = {\aap},
         year = 2007,
        month = mar,
       volume = {464},
       number = {1},
        pages = {399-404},
          doi = {10.1051/0004-6361:20066170},
archivePrefix = {arXiv},
       eprint = {astro-ph/0608064},
 primaryClass = {astro-ph},
       adsurl = {https://ui.adsabs.harvard.edu/abs/2007A&A...464..399H}
}

@ARTICLE{Percival2014,
       author = {{Percival}, Will J. and {Ross}, Ashley J. and {S{\'a}nchez}, Ariel G. and {Samushia}, Lado and {Burden}, Angela and {Crittenden}, Robert and {Cuesta}, Antonio J. and {Magana}, Mariana Vargas and {Manera}, Marc and {Beutler}, Florian and {Chuang}, Chia-Hsun and {Eisenstein}, Daniel J. and {Ho}, Shirley and {McBride}, Cameron K. and {Montesano}, Francesco and {Padmanabhan}, Nikhil and {Reid}, Beth and {Saito}, Shun and {Schneider}, Donald P. and {Seo}, Hee-Jong and {Tojeiro}, Rita and {Weaver}, Benjamin A.},
        title = "{The clustering of Galaxies in the SDSS-III Baryon Oscillation Spectroscopic Survey: including covariance matrix errors}",
      journal = {\mnras},
         year = 2014,
        month = apr,
       volume = {439},
       number = {3},
        pages = {2531-2541},
          doi = {10.1093/mnras/stu112},
archivePrefix = {arXiv},
       eprint = {1312.4841},
 primaryClass = {astro-ph.CO},
       adsurl = {https://ui.adsabs.harvard.edu/abs/2014MNRAS.439.2531P}
}

@ARTICLE{DodelsonSchneider2013,
       author = {{Dodelson}, Scott and {Schneider}, Michael D.},
        title = "{The effect of covariance estimator error on cosmological parameter constraints}",
      journal = {\prd},
         year = 2013,
        month = sep,
       volume = {88},
       number = {6},
          eid = {063537},
        pages = {063537},
          doi = {10.1103/PhysRevD.88.063537},
archivePrefix = {arXiv},
       eprint = {1304.2593},
 primaryClass = {astro-ph.CO},
       adsurl = {https://ui.adsabs.harvard.edu/abs/2013PhRvD..88f3537D}
}

@ARTICLE{CoultonWandelt2023,
       author = {{Coulton}, William R. and {Wandelt}, Benjamin D.},
        title = "{How to estimate Fisher information matrices from simulations}",
      journal = {arXiv e-prints},
         year = 2023,
        month = may,
          eid = {arXiv:2305.08994},
        pages = {arXiv:2305.08994},
          doi = {10.48550/arXiv.2305.08994},
archivePrefix = {arXiv},
       eprint = {2305.08994},
 primaryClass = {stat.ME},
       adsurl = {https://ui.adsabs.harvard.edu/abs/2023arXiv230508994C}
}

@ARTICLE{ScoccimarroCF1999,
       author = {{Scoccimarro}, Rom{\'a}n and {Couchman}, H.~M.~P. and {Frieman}, Joshua A.},
        title = "{The Bispectrum as a Signature of Gravitational Instability in Redshift Space}",
      journal = {\apj},
         year = 1999,
        month = jun,
       volume = {517},
       number = {2},
        pages = {531-540},
          doi = {10.1086/307220},
archivePrefix = {arXiv},
       eprint = {astro-ph/9808305},
 primaryClass = {astro-ph},
       adsurl = {https://ui.adsabs.harvard.edu/abs/1999ApJ...517..531S}
}

@ARTICLE{Bernardeau2002,
       author = {{Bernardeau}, F. and {Colombi}, S. and {Gazta{\~n}aga}, E. and {Scoccimarro}, R.},
        title = "{Large-scale structure of the Universe and cosmological perturbation theory}",
      journal = {\physrep},
         year = 2002,
        month = sep,
       volume = {367},
       number = {1-3},
        pages = {1-248},
          doi = {10.1016/S0370-1573(02)00135-7},
archivePrefix = {arXiv},
       eprint = {astro-ph/0112551},
 primaryClass = {astro-ph},
       adsurl = {https://ui.adsabs.harvard.edu/abs/2002PhR...367....1B}
}

@ARTICLE{Blas2016,
       author = {{Blas}, Diego and {Garny}, Mathias and {Ivanov}, Mikhail M. and {Sibiryakov}, Sergey},
        title = "{Time-sliced perturbation theory II: baryon acoustic oscillations and infrared resummation}",
      journal = {\jcap},
         year = 2016,
        month = jul,
       volume = {2016},
       number = {7},
          eid = {028},
        pages = {028},
          doi = {10.1088/1475-7516/2016/07/028},
archivePrefix = {arXiv},
       eprint = {1605.02149},
 primaryClass = {astro-ph.CO},
       adsurl = {https://ui.adsabs.harvard.edu/abs/2016JCAP...07..028B}
}

@ARTICLE{Lewis2000,
       author = {{Lewis}, Antony and {Challinor}, Anthony and {Lasenby}, Anthony},
        title = "{Efficient Computation of Cosmic Microwave Background Anisotropies in Closed Friedmann-Robertson-Walker Models}",
      journal = {\apj},
         year = 2000,
        month = aug,
       volume = {538},
       number = {2},
        pages = {473-476},
          doi = {10.1086/309179},
archivePrefix = {arXiv},
       eprint = {astro-ph/9911177},
 primaryClass = {astro-ph},
       adsurl = {https://ui.adsabs.harvard.edu/abs/2000ApJ...538..473L}
}

@article{gramsci_gpu,
      title={Fast Graph-based Higher-Order Clustering Statistics on the GPU}, 
      author={Cristiano G. Sabiu},
      year={2026},
      journal={\apjs~submitted},
      eprint={2607.06604},
      archivePrefix={arXiv},
      primaryClass={astro-ph.IM},
      url={https://arxiv.org/abs/2607.06604}, 
}

@article{SlepianEisenstein2016,
       author = {{Slepian}, Zachary and {Eisenstein}, Daniel J.},
        title = "{Accelerating the two-point and three-point galaxy correlation functions using Fourier transforms}",
      journal = {\mnras},
         year = 2016,
        month = jan,
       volume = {455},
       number = {1},
        pages = {L31-L35},
          doi = {10.1093/mnrasl/slv133},
archivePrefix = {arXiv},
       eprint = {1506.04746},
 primaryClass = {astro-ph.CO},
       adsurl = {https://ui.adsabs.harvard.edu/abs/2016MNRAS.455L..31S}
}

@article{BarrigaGaztanaga2002,
       author = {{Barriga}, J. and {Gazta{\~n}aga}, E.},
        title = "{The three-point function in large-scale structure - I. The weakly non-linear regime in N-body simulations}",
      journal = {\mnras},
         year = 2002,
        month = jun,
       volume = {333},
       number = {2},
        pages = {443-453},
          doi = {10.1046/j.1365-8711.2002.05431.x},
archivePrefix = {arXiv},
       eprint = {astro-ph/0112278},
 primaryClass = {astro-ph},
       adsurl = {https://ui.adsabs.harvard.edu/abs/2002MNRAS.333..443B}
}

@article{EisensteinHu1998,
       author = {{Eisenstein}, Daniel J. and {Hu}, Wayne},
        title = "{Baryonic Features in the Matter Transfer Function}",
      journal = {\apj},
         year = 1998,
        month = mar,
       volume = {496},
       number = {2},
        pages = {605-614},
          doi = {10.1086/305424},
archivePrefix = {arXiv},
       eprint = {astro-ph/9709112},
 primaryClass = {astro-ph},
       adsurl = {https://ui.adsabs.harvard.edu/abs/1998ApJ...496..605E}
}

@article{Hamann2010,
       author = {{Hamann}, Jan and {Hannestad}, Steen and {Lesgourgues}, Julien and {Rampf}, Cornelius and {Wong}, Yvonne Y.~Y.},
        title = "{Cosmological parameters from large scale structure - geometric versus shape information}",
      journal = {\jcap},
         year = 2010,
        month = jul,
       volume = {2010},
       number = {7},
          eid = {022},
        pages = {022},
          doi = {10.1088/1475-7516/2010/07/022},
archivePrefix = {arXiv},
       eprint = {1003.3999},
 primaryClass = {astro-ph.CO},
       adsurl = {https://ui.adsabs.harvard.edu/abs/2010JCAP...07..022H}
}

@article{Lazeyras2016,
       author = {{Lazeyras}, Titouan and {Wagner}, Christian and {Baldauf}, Tobias and {Schmidt}, Fabian},
        title = "{Precision measurement of the local bias of dark matter halos}",
      journal = {\jcap},
         year = 2016,
        month = feb,
       volume = {2016},
       number = {2},
        pages = {018-018},
          doi = {10.1088/1475-7516/2016/02/018},
archivePrefix = {arXiv},
       eprint = {1511.01096},
 primaryClass = {astro-ph.CO},
       adsurl = {https://ui.adsabs.harvard.edu/abs/2016JCAP...02..018L}
}

@article{McDonaldRoy2009,
       author = {{McDonald}, Patrick and {Roy}, Arabindo},
        title = "{Clustering of dark matter tracers: generalizing bias for the coming era of precision LSS}",
      journal = {\jcap},
         year = 2009,
        month = aug,
       volume = {2009},
       number = {8},
          eid = {020},
        pages = {020},
          doi = {10.1088/1475-7516/2009/08/020},
archivePrefix = {arXiv},
       eprint = {0902.0991},
 primaryClass = {astro-ph.CO},
       adsurl = {https://ui.adsabs.harvard.edu/abs/2009JCAP...08..020M}
}

@article{DesjacquesJeongSchmidt2018,
       author = {{Desjacques}, Vincent and {Jeong}, Donghui and {Schmidt}, Fabian},
        title = "{Large-scale galaxy bias}",
      journal = {\physrep},
         year = 2018,
        month = feb,
       volume = {733},
        pages = {1-193},
          doi = {10.1016/j.physrep.2017.12.002},
archivePrefix = {arXiv},
       eprint = {1611.09787},
 primaryClass = {astro-ph.CO},
       adsurl = {https://ui.adsabs.harvard.edu/abs/2018PhR...733....1D}
}

@article{Baldauf2015,
       author = {{Baldauf}, Tobias and {Mirbabayi}, Mehrdad and {Simonovi{\'c}}, Marko and {Zaldarriaga}, Matias},
        title = "{Equivalence principle and the baryon acoustic peak}",
      journal = {\prd},
         year = 2015,
        month = aug,
       volume = {92},
       number = {4},
          eid = {043514},
        pages = {043514},
          doi = {10.1103/PhysRevD.92.043514},
archivePrefix = {arXiv},
       eprint = {1504.04366},
 primaryClass = {astro-ph.CO},
       adsurl = {https://ui.adsabs.harvard.edu/abs/2015PhRvD..92d3514B}
}

@article{Goroff1986,
       author = {{Goroff}, M.~H. and {Grinstein}, B. and {Rey}, S.-J. and {Wise}, M.~B.},
        title = "{Coupling of modes of cosmological mass density fluctuations}",
      journal = {\apj},
         year = 1986,
        month = dec,
       volume = {311},
        pages = {6-14},
          doi = {10.1086/164749},
       adsurl = {https://ui.adsabs.harvard.edu/abs/1986ApJ...311....6G}
}

@article{EisensteinSeoWhite2007,
       author = {{Eisenstein}, Daniel J. and {Seo}, Hee-Jong and {White}, Martin},
        title = "{On the Robustness of the Acoustic Scale in the Low-Redshift Clustering of Matter}",
      journal = {\apj},
         year = 2007,
        month = aug,
       volume = {664},
       number = {2},
        pages = {660-674},
          doi = {10.1086/518755},
archivePrefix = {arXiv},
       eprint = {astro-ph/0604361},
 primaryClass = {astro-ph},
       adsurl = {https://ui.adsabs.harvard.edu/abs/2007ApJ...664..660E}
}

@article{Kaiser1987,
       author = {{Kaiser}, Nick},
        title = "{Clustering in real space and in redshift space}",
      journal = {\mnras},
         year = 1987,
        month = jul,
       volume = {227},
        pages = {1-21},
          doi = {10.1093/mnras/227.1.1},
       adsurl = {https://ui.adsabs.harvard.edu/abs/1987MNRAS.227....1K}
}

@ARTICLE{WilsonBean2025,
       author = {{Wilson}, Christopher and {Bean}, Rachel},
        title = "{Implications of noisy numerical derivatives for simulation-based cosmological inference and Fisher forecasts}",
      journal = {\prd},
         year = 2025,
        month = may,
       volume = {111},
       number = {10},
          eid = {103532},
        pages = {103532},
          doi = {10.1103/PhysRevD.111.103532},
archivePrefix = {arXiv},
       eprint = {2406.06067},
 primaryClass = {astro-ph.CO},
       adsurl = {https://ui.adsabs.harvard.edu/abs/2025PhRvD.111j3532W}
}

@ARTICLE{Sabiu2016,
       author = {{Sabiu}, Cristiano G. and {Mota}, David F. and {Llinares}, Claudio and {Park}, Changbom},
        title = "{Probing scalar tensor theories for gravity in redshift space}",
      journal = {\aap},
         year = 2016,
        month = jul,
       volume = {592},
          eid = {A38},
        pages = {A38},
          doi = {10.1051/0004-6361/201527776},
archivePrefix = {arXiv},
       eprint = {1603.05750},
 primaryClass = {astro-ph.CO},
       adsurl = {https://ui.adsabs.harvard.edu/abs/2016A&A...592A..38S}
}

@ARTICLE{Heavens2000,
       author = {{Heavens}, Alan F. and {Jimenez}, Raul and {Lahav}, Ofer},
        title = "{Massive lossless data compression and multiple parameter estimation from galaxy spectra}",
      journal = {\mnras},
         year = 2000,
        month = oct,
       volume = {317},
       number = {4},
        pages = {965-972},
          doi = {10.1046/j.1365-8711.2000.03692.x},
archivePrefix = {arXiv},
       eprint = {astro-ph/9911102},
 primaryClass = {astro-ph},
       adsurl = {https://ui.adsabs.harvard.edu/abs/2000MNRAS.317..965H}
}

@ARTICLE{AlsingWandelt2018,
       author = {{Alsing}, Justin and {Wandelt}, Benjamin},
        title = "{Generalized massive optimal data compression}",
      journal = {\mnras},
         year = 2018,
        month = may,
       volume = {476},
       number = {1},
        pages = {L60-L64},
          doi = {10.1093/mnrasl/sly029},
archivePrefix = {arXiv},
       eprint = {1712.00012},
 primaryClass = {astro-ph.CO},
       adsurl = {https://ui.adsabs.harvard.edu/abs/2018MNRAS.476L..60A}
}

@ARTICLE{LedoitWolf2004,
title = {A well-conditioned estimator for large-dimensional covariance matrices},
journal = {Journal of Multivariate Analysis},
volume = {88},
number = {2},
pages = {365-411},
year = {2004},
issn = {0047-259X},
doi = {https://doi.org/10.1016/S0047-259X(03)00096-4},
url = {https://www.sciencedirect.com/science/article/pii/S0047259X03000964},
author = {Olivier Ledoit and Michael Wolf},
}

@ARTICLE{PopeSzapudi2008,
       author = {{Pope}, Adrian C. and {Szapudi}, Istv{\'a}n},
        title = "{Shrinkage estimation of the power spectrum covariance matrix}",
      journal = {\mnras},
         year = 2008,
        month = sep,
       volume = {389},
       number = {2},
        pages = {766-774},
          doi = {10.1111/j.1365-2966.2008.13561.x},
archivePrefix = {arXiv},
       eprint = {0711.2509},
 primaryClass = {astro-ph},
       adsurl = {https://ui.adsabs.harvard.edu/abs/2008MNRAS.389..766P}
}

@ARTICLE{Guidi2026,
       author = {{Euclid Collaboration} and {Guidi}, M. and {Veropalumbo}, A. and {Pugno}, A. and {Moresco}, M. and {Sefusatti}, E. and {Porciani}, C. and {Branchini}, E. and {Breton}, M.-A. and {Camacho Quevedo}, B. and {Crocce}, M. and {de la Torre}, S. and {Desjacques}, V. and {Eggemeier}, A. and {Farina}, A. and {K{\"a}rcher}, M. and {Linde}, D. and {Marinucci}, M. and {Moradinezhad Dizgah}, A. and {Moretti}, C. and {Pardede}, K. and {Pezzotta}, A. and {Sarpa}, E. and {Amara}, A. and {Andreon}, S. and {Auricchio}, N. and {Baccigalupi}, C. and {Bagot}, D. and {Baldi}, M. and {Bardelli}, S. and {Battaglia}, P. and {Biviano}, A. and {Brescia}, M. and {Camera}, S. and {Ca{\~n}as-Herrera}, G. and {Capobianco}, V. and {Carbone}, C. and {Cardone}, V.~F. and {Carretero}, J. and {Castellano}, M. and {Castignani}, G. and {Cavuoti}, S. and {Chambers}, K.~C. and {Cimatti}, A. and {Colodro-Conde}, C. and {Congedo}, G. and {Conversi}, L. and {Copin}, Y. and {Courbin}, F. and {Courtois}, H.~M. and {Da Silva}, A. and {Degaudenzi}, H. and {De Lucia}, G. and {Dole}, H. and {Douspis}, M. and {Dubath}, F. and {Dupac}, X. and {Dusini}, S. and {Escoffier}, S. and {Farina}, M. and {Farinelli}, R. and {Faustini}, F. and {Ferriol}, S. and {Finelli}, F. and {Fosalba}, P. and {Fotopoulou}, S. and {Frailis}, M. and {Franceschi}, E. and {Fumana}, M. and {Galeotta}, S. and {Gillis}, B. and {Giocoli}, C. and {Gracia-Carpio}, J. and {Grazian}, A. and {Grupp}, F. and {Guzzo}, L. and {Haugan}, S.~V.~H. and {Holmes}, W. and {Hormuth}, F. and {Hornstrup}, A. and {Jahnke}, K. and {Jhabvala}, M. and {Joachimi}, B. and {Keih{\"a}nen}, E. and {Kermiche}, S. and {Kiessling}, A. and {Kubik}, B. and {K{\"u}mmel}, M. and {Kunz}, M. and {Kurki-Suonio}, H. and {Le Brun}, A.~M.~C. and {Ligori}, S. and {Lilje}, P.~B. and {Lindholm}, V. and {Lloro}, I. and {Mainetti}, G. and {Maino}, D. and {Maiorano}, E. and {Mansutti}, O. and {Marcin}, S. and {Marggraf}, O. and {Markovic}, K. and {Martinelli}, M. and {Martinet}, N. and {Marulli}, F. and {Massey}, R. and {Medinaceli}, E. and {Mei}, S. and {Melchior}, M. and {Mellier}, Y. and {Meneghetti}, M. and {Merlin}, E. and {Meylan}, G. and {Mora}, A. and {Morin}, B. and {Moscardini}, L. and {Munari}, E. and {Neissner}, C. and {Niemi}, S.-M. and {Padilla}, C. and {Paltani}, S. and {Pasian}, F. and {Pedersen}, K. and {Percival}, W.~J. and {Pettorino}, V. and {Pires}, S. and {Polenta}, G. and {Poncet}, M. and {Popa}, L.~A. and {Raison}, F. and {Rebolo}, R. and {Renzi}, A. and {Rhodes}, J. and {Riccio}, G. and {Romelli}, E. and {Roncarelli}, M. and {Saglia}, R. and {Sakr}, Z. and {S{\'a}nchez}, A.~G. and {Sapone}, D. and {Sartoris}, B. and {Schewtschenko}, J.~A. and {Schneider}, P. and {Schrabback}, T. and {Scodeggio}, M. and {Secroun}, A. and {Seidel}, G. and {Seiffert}, M. and {Serrano}, S. and {Simon}, P. and {Sirignano}, C. and {Sirri}, G. and {Spurio Mancini}, A. and {Stanco}, L. and {Steinwagner}, J. and {Tallada-Cresp{\'\i}}, P. and {Tavagnacco}, D. and {Taylor}, A.~N. and {Tereno}, I. and {Tessore}, N. and {Toft}, S. and {Toledo-Moreo}, R. and {Torradeflot}, F. and {Tsyganov}, A. and {Tutusaus}, I. and {Valenziano}, L. and {Valiviita}, J. and {Vassallo}, T. and {Verdoes Kleijn}, G. and {Wang}, Y. and {Weller}, J. and {Zamorani}, G. and {Zerbi}, F.~M. and {Zucca}, E. and {Allevato}, V. and {Ballardini}, M. and {Bolzonella}, M. and {Bozzo}, E. and {Burigana}, C. and {Cabanac}, R. and {Calabrese}, M. and {Cappi}, A. and {Di Ferdinando}, D. and {Escartin Vigo}, J.~A. and {Gabarra}, L. and {Mart{\'\i}n-Fleitas}, J. and {Matthew}, S. and {Maturi}, M. and {Mauri}, N. and {Metcalf}, R.~B. and {Nucita}, A.~A. and {P{\"o}ntinen}, M. and {Risso}, I. and {Scottez}, V. and {Sereno}, M. and {Tenti}, M. and {Viel}, M. and {Wiesmann}, M. and {Akrami}, Y. and {Andika}, I.~T.},
        title = "{Euclid preparation: LXXVIII. Full-shape modelling of two-point and three-point correlation functions in real space}",
      journal = {\aap},
         year = 2026,
        month = mar,
       volume = {707},
          eid = {A228},
        pages = {A228},
          doi = {10.1051/0004-6361/202556177},
archivePrefix = {arXiv},
       eprint = {2506.22257},
 primaryClass = {astro-ph.CO},
       adsurl = {https://ui.adsabs.harvard.edu/abs/2026A&A...707A.228E}
}

@ARTICLE{DESIDR12_2026,
       author = {{Forero-S{\'a}nchez}, D. and {Novell Masot}, S. and {Gil-Mar{\'\i}n}, H. and {Verde}, L. and {Aguilar}, J. and {Ahlen}, S. and {Bianchi}, D. and {Brodzeller}, A. and {Brooks}, D. and {Castander}, F.~J. and {Cole}, S. and {de la Macorra}, A. and {Della Costa}, J. and {Dey}, Biprateep and {Doel}, P. and {Ferraro}, S. and {Font-Ribera}, A. and {Forero-Romero}, J.~E. and {Gontcho}, Satya Gontcho A and {Gutierrez}, G. and {Hahn}, C. and {Herrera-Alcantar}, H.~K. and {Honscheid}, K. and {Huterer}, D. and {Ishak}, M. and {Kirkby}, D. and {Kremin}, A. and {Lahav}, O. and {Lamman}, C. and {Landriau}, M. and {Le Guillou}, L. and {Levi}, M.~E. and {Manera}, M. and {Meisner}, A. and {Miquel}, R. and {Moustakas}, J. and {Nadathur}, S. and {Newman}, J.~A. and {Niz}, G. and {Palanque-Delabrouille}, N. and {Percival}, W.~J. and {Prada}, F. and {P{\'e}rez-R{\`a}fols}, I. and {Rossi}, G. and {Samushia}, L. and {Sanchez}, E. and {Schlegel}, D. and {Schubnell}, M. and {Silber}, J. and {Tarl{\'e}}, G. and {Weaver}, B.~A.},
        title = "{Cosmological constraints from the DESI DR1 Bispectrum Full-Shape and DR2 BAO}",
      journal = {arXiv e-prints},
         year = 2026,
        month = jun,
          eid = {arXiv:2606.23936},
        pages = {arXiv:2606.23936},
archivePrefix = {arXiv},
       eprint = {2606.23936},
 primaryClass = {astro-ph.CO},
       adsurl = {https://ui.adsabs.harvard.edu/abs/2026arXiv260623936F}
}

@misc{guidi2026fullshapejointanalysistwo,
      title={First full-shape joint analysis of the two- and three-point correlation functions on real data: $\Lambda$CDM cosmological constraints from BOSS DR12}, 
      author={Massimo Guidi and Michele Moresco and Alfonso Veropalumbo and Lorenzo Cavazzini and Antonio Farina and Andrea Labate},
      year={2026},
      eprint={2606.26237},
      archivePrefix={arXiv},
      primaryClass={astro-ph.CO},
      url={https://arxiv.org/abs/2606.26237}, 
}

@ARTICLE{2019MNRAS.483.2078Y,
       author = {{Yankelevich}, Victoria and {Porciani}, Cristiano},
        title = "{Cosmological information in the redshift-space bispectrum}",
      journal = {\mnras},
         year = 2019,
        month = feb,
       volume = {483},
       number = {2},
        pages = {2078-2099},
          doi = {10.1093/mnras/sty3143},
archivePrefix = {arXiv},
       eprint = {1807.07076},
 primaryClass = {astro-ph.CO},
       adsurl = {https://ui.adsabs.harvard.edu/abs/2019MNRAS.483.2078Y}
}

@ARTICLE{Labate2026,
       author = {{Labate}, Andrea and {Guidi}, Massimo and {Moresco}, Michele and {Veropalumbo}, Alfonso},
        title = "{The imprints of massive neutrinos on the three-point correlation function of large-scale structures}",
      journal = {\aap},
         year = 2026,
        month = apr,
       volume = {708},
          eid = {A210},
        pages = {A210},
          doi = {10.1051/0004-6361/202558719},
archivePrefix = {arXiv},
       eprint = {2512.16992},
 primaryClass = {astro-ph.CO},
       adsurl = {https://ui.adsabs.harvard.edu/abs/2026A&A...708A.210L}
}

@ARTICLE{Fry1993,
       author = {{Fry}, J.~N. and {Gaztanaga}, Enrique},
        title = "{Biasing and Hierarchical Statistics in Large-Scale Structure}",
      journal = {\apj},
         year = 1993,
        month = aug,
       volume = {413},
        pages = {447},
          doi = {10.1086/173015},
archivePrefix = {arXiv},
       eprint = {astro-ph/9302009},
 primaryClass = {astro-ph},
       adsurl = {https://ui.adsabs.harvard.edu/abs/1993ApJ...413..447F}
}

@ARTICLE{Frieman1994,
       author = {{Frieman}, Joshua A. and {Gaztanaga}, Enrique},
        title = "{The Three-Point Function as a Probe of Models for Large-Scale Structure}",
      journal = {\apj},
         year = 1994,
        month = apr,
       volume = {425},
        pages = {392},
          doi = {10.1086/173995},
archivePrefix = {arXiv},
       eprint = {astro-ph/9306018},
 primaryClass = {astro-ph},
       adsurl = {https://ui.adsabs.harvard.edu/abs/1994ApJ...425..392F}
}

@ARTICLE{Gaztanaga1994,
       author = {{Gaztanaga}, E.},
        title = "{High-Order Galaxy Correlation Functions in the APM Galaxy Survey}",
      journal = {\mnras},
         year = 1994,
        month = jun,
       volume = {268},
        pages = {913},
          doi = {10.1093/mnras/268.4.913},
archivePrefix = {arXiv},
       eprint = {astro-ph/9309019},
 primaryClass = {astro-ph},
       adsurl = {https://ui.adsabs.harvard.edu/abs/1994MNRAS.268..913G}
}

@ARTICLE{Jing1998,
       author = {{Jing}, Y.~P. and {B{\"o}rner}, G.},
        title = "{The Three-Point Correlation Function of Galaxies Determined from the Las Campanas Redshift Survey}",
      journal = {\apj},
         year = 1998,
        month = aug,
       volume = {503},
       number = {1},
        pages = {37-47},
          doi = {10.1086/305997},
archivePrefix = {arXiv},
       eprint = {astro-ph/9802011},
 primaryClass = {astro-ph},
       adsurl = {https://ui.adsabs.harvard.edu/abs/1998ApJ...503...37J}
}

@ARTICLE{Jing2004,
       author = {{Jing}, Y.~P. and {B{\"o}rner}, G.},
        title = "{The Three-Point Correlation Function of Galaxies Determined from the Two-Degree Field Galaxy Redshift Survey}",
      journal = {\apj},
         year = 2004,
        month = may,
       volume = {607},
       number = {1},
        pages = {140-163},
          doi = {10.1086/383343},
archivePrefix = {arXiv},
       eprint = {astro-ph/0311585},
 primaryClass = {astro-ph},
       adsurl = {https://ui.adsabs.harvard.edu/abs/2004ApJ...607..140J}
}

@ARTICLE{Kayo2004,
       author = {{Kayo}, Issha and {Suto}, Yasushi and {Nichol}, Robert C. and {Pan}, Jun and {Szapudi}, Istv{\'a}n and {Connolly}, Andrew J. and {Gardner}, Jeff and {Jain}, Bhuvnesh and {Kulkarni}, Gauri and {Matsubara}, Takahiko and {Sheth}, Ravi and {Szalay}, Alexander S. and {Brinkmann}, Jon},
        title = "{Three-Point Correlation Functions of SDSS Galaxies in Redshift Space: Morphology, Color, and Luminosity Dependence}",
      journal = {\pasj},
         year = 2004,
        month = jun,
       volume = {56},
       number = {3},
        pages = {415-423},
          doi = {10.1093/pasj/56.3.415},
archivePrefix = {arXiv},
       eprint = {astro-ph/0403638},
 primaryClass = {astro-ph},
       adsurl = {https://ui.adsabs.harvard.edu/abs/2004PASJ...56..415K}
}

@ARTICLE{Marin2008,
       author = {{Mar{\'\i}n}, Felipe A. and {Wechsler}, Risa H. and {Frieman}, Joshua A. and {Nichol}, Robert C.},
        title = "{Modeling the Galaxy Three-Point Correlation Function}",
      journal = {\apj},
         year = 2008,
        month = jan,
       volume = {672},
       number = {2},
        pages = {849-860},
          doi = {10.1086/523628},
archivePrefix = {arXiv},
       eprint = {0704.0255},
 primaryClass = {astro-ph},
       adsurl = {https://ui.adsabs.harvard.edu/abs/2008ApJ...672..849M}
}

@ARTICLE{Gaztanaga2009,
       author = {{Gazta{\~n}aga}, Enrique and {Cabr{\'e}}, Anna and {Castander}, Francisco and {Crocce}, Martin and {Fosalba}, Pablo},
        title = "{Clustering of luminous red galaxies - III. Baryon acoustic peak in the three-point correlation}",
      journal = {\mnras},
         year = 2009,
        month = oct,
       volume = {399},
       number = {2},
        pages = {801-811},
          doi = {10.1111/j.1365-2966.2009.15313.x},
archivePrefix = {arXiv},
       eprint = {0807.2448},
 primaryClass = {astro-ph},
       adsurl = {https://ui.adsabs.harvard.edu/abs/2009MNRAS.399..801G}
}

@ARTICLE{McBride2011,
       author = {{McBride}, Cameron K. and {Connolly}, Andrew J. and {Gardner}, Jeffrey P. and {Scranton}, Ryan and {Newman}, Jeffrey A. and {Scoccimarro}, Rom{\'a}n and {Zehavi}, Idit and {Schneider}, Donald P.},
        title = "{Three-point Correlation Functions of SDSS Galaxies: Luminosity and Color Dependence in Redshift and Projected Space}",
      journal = {\apj},
         year = 2011,
        month = jan,
       volume = {726},
       number = {1},
          eid = {13},
        pages = {13},
          doi = {10.1088/0004-637X/726/1/13},
archivePrefix = {arXiv},
       eprint = {1007.2414},
 primaryClass = {astro-ph.CO},
       adsurl = {https://ui.adsabs.harvard.edu/abs/2011ApJ...726...13M}
}

@ARTICLE{Moresco2014,
       author = {{Moresco}, Michele and {Marulli}, Federico and {Baldi}, Marco and {Moscardini}, Lauro and {Cimatti}, Andrea},
        title = "{Disentangling interacting dark energy cosmologies with the three-point correlation function}",
      journal = {\mnras},
         year = 2014,
        month = oct,
       volume = {443},
       number = {4},
        pages = {2874-2886},
          doi = {10.1093/mnras/stu1359},
archivePrefix = {arXiv},
       eprint = {1312.4530},
 primaryClass = {astro-ph.CO},
       adsurl = {https://ui.adsabs.harvard.edu/abs/2014MNRAS.443.2874M}
}

@ARTICLE{Moresco2017,
       author = {{Moresco}, M. and {Marulli}, F. and {Moscardini}, L. and {Branchini}, E. and {Cappi}, A. and {Davidzon}, I. and {Granett}, B.~R. and {de la Torre}, S. and {Guzzo}, L. and {Abbas}, U. and {Adami}, C. and {Arnouts}, S. and {Bel}, J. and {Bolzonella}, M. and {Bottini}, D. and {Carbone}, C. and {Coupon}, J. and {Cucciati}, O. and {De Lucia}, G. and {Franzetti}, P. and {Fritz}, A. and {Fumana}, M. and {Garilli}, B. and {Ilbert}, O. and {Iovino}, A. and {Krywult}, J. and {Le Brun}, V. and {Le F{\`e}vre}, O. and {Ma{\l}ek}, K. and {McCracken}, H.~J. and {Polletta}, M. and {Pollo}, A. and {Scodeggio}, M. and {Tasca}, L.~A.~M. and {Tojeiro}, R. and {Vergani}, D. and {Zanichelli}, A.},
        title = "{The VIMOS Public Extragalactic Redshift Survey (VIPERS) . Exploring the dependence of the three-point correlation function on stellar mass and luminosity at 0.5 $<$ z $<$ 1.1}",
      journal = {\aap},
         year = 2017,
        month = aug,
       volume = {604},
          eid = {A133},
        pages = {A133},
          doi = {10.1051/0004-6361/201628589},
archivePrefix = {arXiv},
       eprint = {1603.08924},
 primaryClass = {astro-ph.CO},
       adsurl = {https://ui.adsabs.harvard.edu/abs/2017A&A...604A.133M}
}

@ARTICLE{Slepian2017,
       author = {{Slepian}, Zachary and {Eisenstein}, Daniel J. and {Beutler}, Florian and {Chuang}, Chia-Hsun and {Cuesta}, Antonio J. and {Ge}, Jian and {Gil-Mar{\'\i}n}, H{\'e}ctor and {Ho}, Shirley and {Kitaura}, Francisco-Shu and {McBride}, Cameron K. and {Nichol}, Robert C. and {Percival}, Will J. and {Rodr{\'\i}guez-Torres}, Sergio and {Ross}, Ashley J. and {Scoccimarro}, Rom{\'a}n and {Seo}, Hee-Jong and {Tinker}, Jeremy and {Tojeiro}, Rita and {Vargas-Maga{\~n}a}, Mariana},
        title = "{The large-scale three-point correlation function of the SDSS BOSS DR12 CMASS galaxies}",
      journal = {\mnras},
         year = 2017,
        month = jun,
       volume = {468},
       number = {1},
        pages = {1070-1083},
          doi = {10.1093/mnras/stw3234},
archivePrefix = {arXiv},
       eprint = {1512.02231},
 primaryClass = {astro-ph.CO},
       adsurl = {https://ui.adsabs.harvard.edu/abs/2017MNRAS.468.1070S}
}

@ARTICLE{Garcia2022,
       author = {{Garcia}, Karolina and {Slepian}, Zachary},
        title = "{Improving the line of sight for the anisotropic 3-point correlation function of galaxies: Centroid and Unit-Vector-Average methods scaling as $\mathcal{O}(N^2)$}",
      journal = {\mnras},
         year = 2022,
        month = sep,
       volume = {515},
       number = {1},
        pages = {1199-1217},
          doi = {10.1093/mnras/stac1540},
       adsurl = {https://ui.adsabs.harvard.edu/abs/2022MNRAS.515.1199G}
}

@ARTICLE{Sugiyama2021,
       author = {{Sugiyama}, Naonori S. and {Saito}, Shun and {Beutler}, Florian and {Seo}, Hee-Jong},
        title = "{Towards a self-consistent analysis of the anisotropic galaxy two- and three-point correlation functions on large scales: application to mock galaxy catalogues}",
      journal = {\mnras},
         year = 2021,
        month = feb,
       volume = {501},
       number = {2},
        pages = {2862-2896},
          doi = {10.1093/mnras/staa3725},
archivePrefix = {arXiv},
       eprint = {2010.06179},
 primaryClass = {astro-ph.CO},
       adsurl = {https://ui.adsabs.harvard.edu/abs/2021MNRAS.501.2862S}
}

@ARTICLE{Veropalumbo2021,
       author = {{Veropalumbo}, Alfonso and {S{\'a}ez Casares}, I{\~n}igo and {Branchini}, Enzo and {Granett}, Benjamin R. and {Guzzo}, Luigi and {Marulli}, Federico and {Moresco}, Michele and {Moscardini}, Lauro and {Pezzotta}, Andrea and {de la Torre}, Sylvain},
        title = "{A joint 2- and 3-point clustering analysis of the VIPERS PDR2 catalogue at z $\sim$ 1: breaking the degeneracy of cosmological parameters}",
      journal = {\mnras},
         year = 2021,
        month = oct,
       volume = {507},
       number = {1},
        pages = {1184-1201},
          doi = {10.1093/mnras/stab2205},
archivePrefix = {arXiv},
       eprint = {2106.12581},
 primaryClass = {astro-ph.CO},
       adsurl = {https://ui.adsabs.harvard.edu/abs/2021MNRAS.507.1184V}
}

@ARTICLE{2021Moresco,
       author = {{Moresco}, Michele and {Veropalumbo}, Alfonso and {Marulli}, Federico and {Moscardini}, Lauro and {Cimatti}, Andrea},
        title = "{C$^{3}$: Cluster Clustering Cosmology. II. First Detection of the Baryon Acoustic Oscillations Peak in the Three-point Correlation Function of Galaxy Clusters}",
      journal = {\apj},
         year = 2021,
        month = oct,
       volume = {919},
       number = {2},
          eid = {144},
        pages = {144},
          doi = {10.3847/1538-4357/ac10c9},
archivePrefix = {arXiv},
       eprint = {2011.04665},
 primaryClass = {astro-ph.CO},
       adsurl = {https://ui.adsabs.harvard.edu/abs/2021ApJ...919..144M}
}

@ARTICLE{Veropalumbo2022,
       author = {{Veropalumbo}, A. and {Binetti}, A. and {Branchini}, E. and {Moresco}, M. and {Monaco}, P. and {Oddo}, A. and {S{\'a}nchez}, A.~G. and {Sefusatti}, E.},
        title = "{The halo 3-point correlation function: a methodological analysis}",
      journal = {\jcap},
         year = 2022,
        month = sep,
       volume = {2022},
       number = {9},
          eid = {033},
        pages = {033},
          doi = {10.1088/1475-7516/2022/09/033},
archivePrefix = {arXiv},
       eprint = {2206.00672},
 primaryClass = {astro-ph.CO},
       adsurl = {https://ui.adsabs.harvard.edu/abs/2022JCAP...09..033V}
}

@ARTICLE{Aviles2023,
       author = {{Aviles}, Alejandro and {Niz}, Gustavo},
        title = "{Galaxy three-point correlation function in modified gravity}",
      journal = {\prd},
         year = 2023,
        month = mar,
       volume = {107},
       number = {6},
          eid = {063525},
        pages = {063525},
          doi = {10.1103/PhysRevD.107.063525},
archivePrefix = {arXiv},
       eprint = {2301.07240},
 primaryClass = {astro-ph.CO},
       adsurl = {https://ui.adsabs.harvard.edu/abs/2023PhRvD.107f3525A}
}

@ARTICLE{Guidi2023,
       author = {{Guidi}, M. and {Veropalumbo}, A. and {Branchini}, E. and {Eggemeier}, A. and {Carbone}, C.},
        title = "{Modelling the next-to-leading order matter three-point correlation function using FFTLog}",
      journal = {\jcap},
         year = 2023,
        month = aug,
       volume = {2023},
       number = {8},
          eid = {066},
        pages = {066},
          doi = {10.1088/1475-7516/2023/08/066},
archivePrefix = {arXiv},
       eprint = {2212.07382},
 primaryClass = {astro-ph.CO},
       adsurl = {https://ui.adsabs.harvard.edu/abs/2023JCAP...08..066G}
}

@ARTICLE{Farina2026,
       author = {{Farina}, A. and {Veropalumbo}, A. and {Branchini}, E. and {Guidi}, M.},
        title = "{Modeling and measuring the anisotropic halo 3-point correlation function: a coordinated study}",
      journal = {\jcap},
         year = 2026,
        month = feb,
       volume = {2026},
       number = {2},
          eid = {028},
        pages = {028},
          doi = {10.1088/1475-7516/2026/02/028},
archivePrefix = {arXiv},
       eprint = {2408.03036},
 primaryClass = {astro-ph.CO},
       adsurl = {https://ui.adsabs.harvard.edu/abs/2026JCAP...02..028F}
}

@ARTICLE{Kamalinejad2026,
       author = {{Kamalinejad}, Farshad and {Slepian}, Zachary and {Krolewski}, Alex and {Greco}, Alessandro and {Ortol{\'a} Leonard}, William and {Chellino}, Jessica and {Reinhard}, Matthew and {Fern{\'a}ndez-Garc{\'\i}a}, Elena and {Prada}, Francisco and {Aguilar}, J. and {Ahlen}, S. and {Anand}, A. and {Bebek}, C. and {Bianchi}, D. and {Brooks}, D. and {Claybaugh}, T. and {Cuceu}, A. and {Dawson}, K.~S. and {de la Macorra}, A. and {Demina}, R. and {Doel}, P. and {Edelstein}, J. and {Forero-Romero}, J.~E. and {Gazta{\~n}aga}, E. and {Gontcho}, S. Gontcho A and {Gutierrez}, G. and {Herrera-Alcantar}, H.~K. and {Honscheid}, K. and {Howlett}, C. and {Huterer}, D. and {Ishak}, M. and {Joyce}, R. and {Juneau}, S. and {Kirkby}, D. and {Kisner}, T. and {Kremin}, A. and {Lahav}, O. and {Lamman}, C. and {Landriau}, M. and {Le Guillou}, L. and {Manera}, M. and {Meisner}, A. and {Miquel}, R. and {Newman}, J.~A. and {Percival}, W.~J. and {Poppett}, C. and {P{\'e}rez-R{\`a}fols}, I. and {Samushia}, L. and {Sanchez}, E. and {Schlegel}, D. and {Schubnell}, M. and {Seo}, H. and {Silber}, J. and {Sprayberry}, D. and {Tarl{\'e}}, G. and {Weaver}, B.~A. and {Zhao}, C. and {Zou}, H.},
        title = "{First Detection of the Baryon Acoustic Oscillation (BAO) Feature in the 3-Point Correlation Function of DESI DR1 Luminous Red Galaxies}",
      journal = {arXiv e-prints},
         year = 2026,
        month = feb,
          eid = {arXiv:2602.16134},
        pages = {arXiv:2602.16134},
          doi = {10.48550/arXiv.2602.16134},
archivePrefix = {arXiv},
       eprint = {2602.16134},
 primaryClass = {astro-ph.CO},
       adsurl = {https://ui.adsabs.harvard.edu/abs/2026arXiv260216134K}
}

@ARTICLE{Fry1978,
       author = {{Fry}, J.~N. and {Peebles}, P.~J.~E.},
        title = "{Statistical analysis of catalogs of extragalactic objects. IX. The four-point galaxy correlation function.}",
      journal = {\apj},
         year = 1978,
        month = apr,
       volume = {221},
        pages = {19-33},
          doi = {10.1086/156001},
       adsurl = {https://ui.adsabs.harvard.edu/abs/1978ApJ...221...19F}
}

@ARTICLE{Fry1983,
       author = {{Fry}, J.~N.},
        title = "{A new evaluation of the four-point galaxy correlation function amplitudes}",
      journal = {\apj},
         year = 1983,
        month = apr,
       volume = {267},
        pages = {483-487},
          doi = {10.1086/160885},
       adsurl = {https://ui.adsabs.harvard.edu/abs/1983ApJ...267..483F}
}

@ARTICLE{Ortola2025,
       author = {{Ortol{\'a} Leonard}, William and {Slepian}, Zachary and {Hou}, Jiamin},
        title = "{A model for the redshift-space galaxy 4-point correlation function}",
      journal = {\jcap},
         year = 2025,
        month = jan,
       volume = {2025},
       number = {1},
          eid = {090},
        pages = {090},
          doi = {10.1088/1475-7516/2025/01/090},
archivePrefix = {arXiv},
       eprint = {2402.15510},
 primaryClass = {astro-ph.CO},
       adsurl = {https://ui.adsabs.harvard.edu/abs/2025JCAP...01..090O}
}

@ARTICLE{Tinker2008,
       author = {{Tinker}, Jeremy and {Kravtsov}, Andrey V. and {Klypin}, Anatoly and {Abazajian}, Kevork and {Warren}, Michael and {Yepes}, Gustavo and {Gottl{\"o}ber}, Stefan and {Holz}, Daniel E.},
        title = "{Toward a Halo Mass Function for Precision Cosmology: The Limits of Universality}",
      journal = {\apj},
         year = 2008,
        month = dec,
       volume = {688},
       number = {2},
        pages = {709-728},
          doi = {10.1086/591439},
archivePrefix = {arXiv},
       eprint = {0803.2706},
 primaryClass = {astro-ph},
       adsurl = {https://ui.adsabs.harvard.edu/abs/2008ApJ...688..709T}
}

@ARTICLE{Tinker2010,
       author = {{Tinker}, Jeremy L. and {Robertson}, Brant E. and {Kravtsov}, Andrey V. and {Klypin}, Anatoly and {Warren}, Michael S. and {Yepes}, Gustavo and {Gottl{\"o}ber}, Stefan},
        title = "{The Large-scale Bias of Dark Matter Halos: Numerical Calibration and Model Tests}",
      journal = {\apj},
         year = 2010,
        month = dec,
       volume = {724},
       number = {2},
        pages = {878-886},
          doi = {10.1088/0004-637X/724/2/878},
archivePrefix = {arXiv},
       eprint = {1001.3162},
 primaryClass = {astro-ph.CO},
       adsurl = {https://ui.adsabs.harvard.edu/abs/2010ApJ...724..878T}
}

\end{document}